 \documentclass[11pt]{article}

\usepackage{amsmath,amsfonts,amssymb}

 \usepackage[numbers,sort&compress]{natbib}

\usepackage{hyperref}
\hypersetup{
   colorlinks   =  true,
    linkcolor    = blue,
    citecolor    = red,
     urlcolor	=blue,
}

\renewcommand{\theequation}{\arabic{section}.\arabic{equation}}

\newcommand\be{\begin{equation}}
\newcommand\ee{\end{equation}}
\newcommand\bea{\begin{eqnarray}}
\newcommand\eea{\end{eqnarray}}

\usepackage{lmodern}
\usepackage[english]{babel}
\usepackage[utf8]{inputenc}
\usepackage[T1]{fontenc}

 \usepackage{pgf,tikz,pgfplots}
\pgfplotsset{compat=1.15}
\usetikzlibrary{matrix,arrows}
\usepackage{graphicx}
\usepackage{adjustbox}
\usepackage{subcaption,booktabs}
\graphicspath{{graficos/}}

\usepackage{amsmath}
\usepackage{amsthm}
 \usepackage{physics}
\usepackage{amssymb}
\usepackage{stmaryrd}
\usepackage{bm}
\usepackage{doi}

\providecommand{\abs}[1]{\lvert#1\rvert} 
\newcommand{\set}[1]{\left\{ #1 \right\}} 

\newcommand{\R}{\mathbb{R}} 
\newcommand{\Z}{\mathbb{Z}} 

\newcommand{\cH}{\mathcal{H}} 
\newcommand{\T}{\mathbf{T}} 

\newcommand{\conf}{\mathfrak{conf}} 
\newcommand{\E}{\mathbf{E}} 
\newcommand{\AdS}{\mathbf{A}\mathbf{d}\mathbf{S}} 
\newcommand{\dS}{\mathbf{d}\mathbf{S}} 
\newcommand{\M}{\mathbf{M}} 
\newcommand{\NH}{\mathbf{N}\mathbf{H}} 
\newcommand{\G}{\mathbf{G}} 

\newcommand{\D}{\mathbf{D}}

\renewcommand{\P}{\mathbf{P}}
\renewcommand{\G}{\mathbf{G}}
\newcommand{\J}{\mathbf{J}_{12}}

\newcommand{\h}{\mathfrak{h}}
\newcommand{\p}{\mathfrak{p}}

\renewcommand{\sl}{\mathfrak{s}\mathfrak{l}}
\newcommand{\so}{\mathfrak{s}\mathfrak{o}}
\newcommand{\iso}{\mathfrak{i}\mathfrak{s}\mathfrak{o}}
\newcommand{\iiso}{\mathfrak{i}\mathfrak{i}\mathfrak{s}\mathfrak{o}}

\renewcommand{\k}{\bm{\kappa}}

\newcommand{\kk}{\kappa}

\newcommand{\diag}{\mathrm{diag}}
\newcommand{\X}{\mathbf{X}} 
\newcommand{\x}{\mathbf{x}} 

\renewcommand{\H}{\mathbf{H}}

\newcommand{\Cos}{\,\mathrm{C}}
\newcommand{\Sin}{\,\mathrm{S}}
\newcommand{\Tan}{\,\mathrm{T}}
\newcommand{\Ver}{\mathrm{V}}

\newcommand{\End}{\mathrm{End}}

\renewcommand{\S}{\mathbf{S}}

\newcommand{\Iq}{\mathbb{I}}

\newcommand{\mm}{\mathbb{\mu}}

\theoremstyle{definition}

\theoremstyle{plain}

\theoremstyle{remark}

\numberwithin{equation}{section} 

\begin{document}

\
  \vskip0.5cm

  \begin{center}

 \noindent
 {\Large \bf
 Lie--Hamilton systems on Riemannian and Lorentzian spaces \\ [5pt] from conformal transformations and some of their applications}

   \end{center}

\medskip

\begin{center}

{\sc  Rutwig Campoamor-Stursberg$^{1,2}$, Oscar Carballal$^{2}$\\[2pt] and Francisco J.~Herranz$^{3}$}

\end{center}

\medskip

 \noindent
$^1$ Instituto de Matem\'atica Interdisciplinar, Universidad Complutense de Madrid, E-28040 Madrid,  Spain

\noindent
$^2$ Departamento de \'Algebra, Geometr\'{\i}a y Topolog\'{\i}a,  Facultad de Ciencias
Matem\'aticas, Universidad Complutense de Madrid, Plaza de Ciencias 3, E-28040 Madrid, Spain

\noindent
{$^3$ Departamento de F\'isica, Universidad de Burgos,
E-09001 Burgos, Spain}

  \medskip

\noindent  E-mail: {\small
 \href{mailto:rutwig@ucm.es}{\texttt{rutwig@ucm.es}}, \href{mailto:oscarbal@ucm.es}{\texttt{oscarbal@ucm.es}}, \href{mailto:fjherranz@ubu.es}{\texttt{fjherranz@ubu.es}}
}

\medskip

\begin{abstract}
\noindent
We propose a generalization of two classes of Lie--Hamilton systems on the Euclidean plane to two-dimensional curved spaces, leading to novel Lie--Hamilton systems on Riemannian spaces (flat $2$-torus, product of hyperbolic lines, sphere and hyperbolic plane), pseudo-Riemannian spaces (anti-de Sitter, de Sitter, and Minkowski spacetimes), as well as to semi-Riemannian spaces (Newtonian or non-relativistic spacetimes). The vector fields, Hamiltonian functions, symplectic form and constants of the motion of the Euclidean classes are recovered by a contraction process. The construction is based on the structure of certain subalgebras of the so-called conformal algebras of the two-dimensional Cayley--Klein spaces. These curved Lie--Hamilton classes allow us to generalize naturally the Riccati, Kummer--Schwarz and Ermakov equations on the Euclidean plane to curved spaces, covering both the Riemannian and Lorentzian possibilities, and where the curvature  can be considered as an integrable deformation parameter of the initial Euclidean system.

\end{abstract}
\medskip
\medskip

\medskip


\smallskip
\noindent
\textbf{PACS 2010 codes}:  {02.20.Sv, 02.30.Hq, 02.40.Dr, 02.40.Ky}

 \newpage

\tableofcontents

\newpage


\section{Introduction}
\label{s1}
A \textit{Lie system} is a time-dependent system of first-order ordinary differential equations (ODEs in short) in normal form that admits a \textit{superposition rule}, i.e., a time-independent map through which the general solution of the system is expressed in terms of a generic finite family of particular solutions and some constants related to the initial conditions \cite{Lie1893,Vessiot1895,Shnider1984,Bountis1986,Lucas2020}. The fundamental Lie--Scheffers theorem states that a Lie system is equivalent to a time-dependent vector field that can be seen as a curve taking values in a finite-dimensional Lie algebra of vector fields, a so-called \textit{Vessiot--Guldberg Lie algebra} (VG Lie algebra in short) \cite{Carinena2007}.

A  \textit{Lie--Hamilton system} (LH system in short) is a Lie system whose VG Lie algebra is formed by Hamiltonian vector fields with respect to a certain symplectic structure. These enriched Lie systems have been widely studied and classified under local diffeomorphisms on the real plane (see \cite{Ballesteros2015,Blasco2015} and references therein). One of the most remarkable properties of LH systems is that they allow an algorithmic computation of time-independent constants of the motion through the so-called \textit{coalgebra formalism}, hence eventually simplifying the deduction of a superposition rule \cite{Ballesteros2013}.

In order to explicitly state our main objectives,   let us consider the so-called \textit{Ermakov equation}~\cite{Ermakov1880} on the Euclidean line $\E^{1} := \R$ (see also~\cite{Leach1991,Maamache1995,Carinena2008,Leach2008} and references therein), given by
\begin{equation}
\dv[2]{u}{t} = - \Omega^{2}(t) u + \frac{c}{u^{3}}, \qquad c \in \R,\nonumber
\end{equation}
where $\Omega(t)$ is any $t$-dependent real or pure imaginary function. Recall that this equation is also known as the Milne--Pinney    equation~\cite{Milne1930,Pinney1950}  and that it is equivalent to the following first-order system of ODEs on the Euclidean plane $\E^{2} := \R^{2}$:
\begin{equation}
\dv{u}{t} = v, \qquad \dv{v}{t} = - \Omega^{2}(t) u + \frac{c}{u^{3}}.
\label{eq:Er_Euc:system}
\end{equation}
Hence this  system is associated with the $t$-dependent vector field $\X: \R \times \E^{2} \to T\E^{2}$ given by
\begin{equation}
 \X = \X_{3} + \Omega^{2}(t) \X_{1},
\label{eq:Er_Euc:tvf}
\end{equation}
where the vector fields
\begin{equation}
\X_{1} = - u \pdv{v}, \qquad \X_{2}= \frac{1}{2} \left( v \pdv{v} - u \pdv{u} \right), \qquad \X_{3} = v \pdv{u} + \frac{c}{u^{3}} \pdv{v}
\label{eq:Er_Euc:vf}
\end{equation}
fulfil the commutation relations
\begin{equation}
[\X_{1}, \X_{2}] = \X_{1}, \qquad [\X_{1}, \X_{3}] = 2 \X_{2}, \qquad [\X_{2}, \X_{3}] = \X_{3} .
\label{sl2}
\end{equation}
Therefore they span a VG Lie algebra  $V^X$ of $\X$ (\ref{eq:Er_Euc:tvf})  isomorphic to $\sl(2, \R) \simeq \so(2, 1)$ for any value of the constant $c$~\cite{Ballesteros2013,Ballesteros2015,Blasco2015}.
Hence,  the $t$-dependent vector field $\X$   (or its associated equation (\ref{eq:Er_Euc:system})) is a Lie system (see~\cite{Lucas2020} and references therein).

In addition, the vector fields \eqref{eq:Er_Euc:vf} are Hamiltonian vector fields relative to the canonical symplectic form
\be
\omega = \dd u \wedge \dd v
\label{canE}
\ee
 on $\E^{2}$, as they satisfy the  invariance condition
\be
{\cal L}_{\mathbf{X}_i}\omega =0, \qquad 1 \leq i \leq 3.
\label{LieDer}
\ee
In order to clarify the terminology used throughout the paper, we observe that, as we  always deal with vector fields and a symplectic form $\omega$, the condition on the vanishing of the Lie derivative with respect to  $\omega$ implies that the vector field is locally Hamiltonian (also called symplectic in the literature, see for instance~\cite{Olver}). In the particular case of the Ermakov system, as the domain of the latter is contractible, it can be inferred that the Hamiltonian functions are globally defined.
 
The associated Hamiltonian functions of the vector fields (\ref{eq:Er_Euc:vf}) are then determined by the inner product
 \be
 \iota_{\X_{i}} \omega = \dd h_{i}, \qquad 1 \leq i \leq 3,
 \label{condition}
 \ee
  turning out to be
\begin{equation}
h_{1} = \frac{1}{2} u^{2}, \qquad h_{2} = -\frac{1}{2} uv,  \qquad h_{3} = \frac{1}{2} \left( v^{2} + \frac{c}{u^{2}} \right).
\label{eq:Er_Euc:Ham}
\end{equation}
 The Poisson bracket $\set{\cdot, \cdot}_{\omega}$, induced by   $\omega$, of the above functions leads to
\begin{equation}
 \set{h_{1}, h_{2}}_{\omega} = - h_{1}, \qquad \set{h_{1}, h_{3}}_{\omega} = - 2 h_{2}, \qquad \set{h_{2}, h_{3}}_{\omega} = - h_{3},
 \label{sl2h}
\end{equation}
so that they span a Lie--Hamilton  algebra (LH algebra in short) $\cH_{\omega} \simeq \sl(2, \R)$.  The Casimir element $\mathcal{C}$, that Poisson commutes with $h_i$, reads
\be
\mathcal{C}= h_1 h_3 - h_2^2,
\label{cas2}
\ee
which under the functional realization (\ref{eq:Er_Euc:Ham}) gives the constant
\be
\mathcal{C}=c/4.
\label{casconstant}
\ee
Consequently, the $t$-dependent vector field $\X$ (\ref{eq:Er_Euc:tvf}) not only determines a  Lie system but also a LH system~\cite{Lucas2020}.

It is worth noting  that  if we identify $(u,v)$ with the usual canonical variables $(q,p)$ of $T^{*}\R$ in classical mechanics,  the  Ermakov system (\ref{eq:Er_Euc:system}) can also be obtained from the Hamilton equations determined by the $t$-dependent Hamiltonian given by (see (\ref{eq:Er_Euc:tvf}))
\be
h=  h_3 +   \Omega^{2}(t) h_1   =    \frac 12   p^2+    \frac 12   \Omega^{2}(t) q^2 +   \frac c{2q^2}  .\nonumber
\ee
In this form,  $h$  can be interpreted as the one-dimensional (1D) $t$-dependent   counterpart~\cite{Ballesteros2013,Ballesteros2015,Blasco2015} of the Smorodinsky--Winternitz oscillator \cite{Fris1965}, with     time-dependent frequency  $ \Omega(t)$,  unit mass  and with a  Rosocha\-tius or Winternitz potential depending on the constant $c$; note that the latter  is  in fact a centrifugal barrier whenever $c>0$ (see~\cite{Ballesteros2013b} and references therein).

Consequently, although the value of the parameter $c$ does not interfere in the algebraic part of the LH structure associated with the Ermakov equation (\ref{eq:Er_Euc:system}), the sign of $c$ (\ref{casconstant}) does matter. This leads to three types  of submanifolds determined by the surfaces with constant value    of the Casimir $\mathcal{C}$ for the Poisson structure of $\sl(2, \R)$~\cite{Ballesteros2018}. This result agrees perfectly  with the classification  of LH systems on the Euclidean plane~\cite{Ballesteros2015,Blasco2015}. Explicitly, there are three   different classes of non-diffeomorphic $\sl(2, \R)$-LH systems on   $\E^{2} $, which are   distinguished among them  by    the value of  $c$:   class  {\rm P}$_2$ for $c>0$;  class {\rm I}$_4$ for $c<0$; and class {\rm I}$_5$ for $c=0$.  This means that there does not exist any  local diffeomorphism mapping one class into another, so that  any   LH system related to a VG Lie algebra of Hamiltonian vector fields isomorphic to $\sl(2, \R)$ must be, up to a  local change of coordinates, of the form (\ref{eq:Er_Euc:system}) for a positive, zero or negative value of $c$.

The resulting (Hamiltonian) vector fields corresponding to one representative from each of these three classes are displayed  in Table~\ref{table1} in Cartesian coordinates $(x,y)$, along with its associated symplectic form and the constant value of the Casimir (\ref{cas2}); note that these   variables  are different from $(u,v)$   such   that  $\omega$ now becomes  a non-canonical symplectic form.  The vector fields $\X_i$ obey the commutation relations (\ref{sl2}), while the Hamiltonian functions $h_i$ fulfil (\ref{sl2h}) for the three classes. Recall that, in addition to the Ermakov or Milne--Pinney equation, it has  been shown in~\cite{Ballesteros2015,Blasco2015} that $\sl(2, \R)$-LH systems in  Table~\ref{table1}   also  comprise several types of Riccati equations~\cite{Mariton1985, Campos1997, Zoladek2000, Egorov2007, Wilczynski2008, Suazo2011, Ortega2012,Schuch2012,Suazo2014, Estevez2016, Grundland2017}  (see also  references therein)
 and second-order Kummer--Schwarz equations~\cite{Berkovich2007,Carinena2012,deLucas2013}.


\begin{table}[t]
\centering
{\small
\caption{ \small{Classes P$_{2}$, I$_{4}$ and I$_{5}$ of LH systems on the plane \cite{Ballesteros2015,Blasco2015}. The VG Lie algebra spanned by the vector fields $\X_{i}$ $(1 \leq i \leq 3)$ is isomorphic to $\sl(2, \R) \simeq \so(2, 1)$ in the three cases, while the Hamiltonian functions $h_{i}$   span a LH algebra isomorphic to $\sl(2, \R) $ with respect to  the  symplectic form $\omega$.}}
\begin{adjustbox}{max width=\textwidth}
\begin{tabular}{llllll}
\hline

\hline
\\[-6pt]
Class & Vector fields $\X_i$ & Hamiltonian functions $h_{i}$& $\omega$ &Domain& Casimir \\[4pt]
\hline \rule{0pt}{1.5\normalbaselineskip}
P$_{2}$ & $ \displaystyle \partial_{x},\  x \partial_{x} + y \partial_{y},\   (x^{2} - y^{2}) \partial_{x} + 2xy \partial_{y}$ & $ \displaystyle -\frac{1}{y},\   - \frac{x}{y},\   - \frac{x^{2} + y^{2}}{y}$ & $ \displaystyle \frac{\dd x \wedge \dd y}{y^{2}}$ &$\R^2_{y\ne 0}$ &\ \ $1$  \\ \rule{0pt}{1.5\normalbaselineskip}
I$_{4}$ & $ \displaystyle \partial_{x} + \partial_{y}, \   x \partial_{x} + y \partial_{y},\   x^{2} \partial_{x} + y^{2} \partial_{y}$ & $ \displaystyle \frac{1}{x-y}, \  \frac{x+y}{2(x-y)},\   \frac{xy}{x-y}$ & $ \displaystyle \frac{\dd x \wedge \dd y}{(x-y)^{2}}$ &$\R^2_{x\ne y}$&$\displaystyle -\frac 14$ \\\rule{0pt}{1.5\normalbaselineskip}
I$_{5}$ & $ \displaystyle \partial_{x},\    x \partial_{x} + \frac 12y \partial_{y},\   x^{2} \partial_{x} + xy \partial_{y}$ &$ \displaystyle - \frac{1}{2y^{2}}, \  - \frac{x}{2y^{2}}, \  - \frac{x^{2}}{2y^{2}}$ & $ \displaystyle \frac{\dd x \wedge \dd y}{y^{3}}$  &$\R^2_{y\ne 0}$ &\ \  $0$ \\[10pt]
\hline

\hline

\end{tabular}
\end{adjustbox}
\label{table1}
}
\end{table}


 The striking point now is that the vector fields of class  {\rm P}$_2$  expressed exactly in the form of Table~\ref{table1} admit a natural geometric interpretation as conformal symmetries of  $\E^{2} $. In particular, they span a Lie subalgebra of the conformal Euclidean algebra $ \mathfrak{so}(3,1)$ such that   the vector field $\X_1$ corresponds to the generator of
   translations along the axis $x$,  $\X_2$  gives rise to dilations and $\X_3$ provides specific conformal transformations again related to the axis $x$.   Likewise, the vector fields of class  {\rm I}$_4$ in Table~\ref{table1}   can  be directly interpreted  as conformal transformations of  the space $\E^{1} \times \E^{1} $:  $\X_1$ as the translation, $\X_2$  as the dilation and  $\X_3$ as the conformal transformation. However, the remaining class {\rm I}$_5$ does not allow a conformal interpretation.

    These ideas naturally suggest to construct the `curved' counterparts of the classes {\rm P}$_2$ and {\rm I}$_4$ by making use of the known conformal symmetries of 2D spaces of constant curvature~\cite{Herranz2002,Herranz2002b}. Our procedure will make use of a Cayley--Klein (CK) setting that requires to deal with two explicit real (graded) contraction parameters $\kk_1$ and $\kk_2$. The former corresponds to the constant (Gaussian) curvature of the space, while the latter determines the signature, so the nine 2D CK spaces are collectively denoted by $\S^{2}_{[\kappa_{1}],\kappa_{2}}$. The three 1D CK spaces $\S^{1}_{[\kappa]}$ are obtained as suitable  submanifolds of some 2D CK spaces.   In this way, we will obtain unified expressions that generalize those for  {\rm P}$_2$ and {\rm I}$_4$ presented in Table~\ref{table1} to the curved spaces $\S^{2}_{[\kappa_{1}],\kappa_{2}}$ and $\S^{1}_{[\kappa]} \times \S^{1}_{[\kappa]}$, respectively. In particular, our novel results will cover LH systems on   the three classical Riemannian spaces (sphere, Euclidean and hyperbolic), the three Lorentzian (Minkowski and (anti-)de Sitter) and the three Newtonian (Galilei and Newton--Hooke) spacetimes of constant curvature, as well as on the flat $2$-torus and the product of two hyperbolic lines. We recall that a similar approach was already developed in~\cite{Tobolski2017} for the class {\rm P}$_1$ of LH systems, but in that case it was based on the isometries of  2D CK spaces. Similarly, examples of Lie systems admitting a VG Lie algebra of conformal vector fields can also be found in \cite{Blasco2015,Grundland2017,Anderson1981} and, for the same reasons pointed out in \cite{Tobolski2017}, these results cannot be applied to the new Lie systems we propose in this work.

      In addition, in the framework of Lie systems on homogeneous spaces its is worth recalling that   in~\cite{Shnider1984,SW84} an `almost' complete classification of superposition rules for {\em complex} Lie systems with primitive transitive VG Lie algebras of  vector fields on homogeneous spaces has been achieved. However, we stress that there are not many results for Lie systems endowed with   {\em real} VG Lie algebras on homogeneous spaces, which constitute a much more difficult task (cf.~\cite{Shnider1984,SW84}). In this respect, some results on spheres can be found  in~\cite{Brockett}.

Consequently, the construction developed in~\cite{Tobolski2017} constitutes, in fact,  the first known example of LH systems on curved Riemannian, Lorenzian and Newtonian spaces, and a superposition rule for all these nine curved systems was obtained in a very geometrical way through the trigonometry of the 2D CK spaces \cite{Herranz2000}. Nevertheless, no relevant applications were found, as the Euclidean P$_{1}$-LH class does not possess  any remarkable system appart from the complex Bernoulli equation  \cite{Blasco2015}. This last point was the main motivation of our novel conformal-based approach, as the Euclidean P$_{2}$ and I$_{4}$-LH classes comprise several notable applications, as pointed out before.

This paper is structured as follows. After explicitly setting the conformal-based motivation of our work, in Section~\ref{s2} we review the conformal symmetries of the 2D CK spaces $\S^{2}_{[\kappa_{1}],\kappa_{2}}$ introduced in \cite{Herranz2002}, from which the conformal symmetries of the 1D CK spaces $\S^{1}_{[\kappa]}$ are deduced. In Section~\ref{s3} we construct the curved counterpart of the Euclidean I$_{4}$-LH class  (see Table~\ref{table1}) on the spaces $\S^{1}_{[\kappa]} \times \S^{1}_{[\kappa]}$, which we call the \textit{curved I$_{4}$-LH class}. Their $t$-independent constants of the motion are computed in Subsection~\ref{s31}, and a  superposition rule for this curved class is derived in Subsection~\ref{s32}. In Section~\ref{s4}, some relevant applications of the curved I$_{4}$-LH class are given, generalizing to $\S^{1}_{[\kappa]} \times \S^{1}_{[\kappa]}$ well-known LH systems on the Euclidean plane: curved coupled Riccati equations, a curved split-complex Riccati equation, a curved diffusion Riccati system, a curved Kummer--Schwarz equation and a curved Ermakov equation.   The same ansatz is applied in Section~\ref{s5}, where we construct the \textit{curved P$_{2}$-LH class} on the 2D CK spaces $\S^{2}_{[\kappa_{1}],\kappa_{2}}$, from which the Euclidean P$_{2}$-LH class (see Table~\ref{table1}) is recovered after contraction. Their $t$-independent constants of motion are computed  in Subsection~\ref{s51}, and a superposition rule on those 2D CK spaces $\S^{2}_{[\kappa_{1}],\kappa_{2}}$ with some $\kappa_{i} = 0$ is obtained in Subsection~\ref{s52}, as the computations in the general case are very cumbersome. In Section~\ref{s6} we give remarkable applications of the curved P$_{2}$-LH class: a curved complex Riccati equation, a curved Kummer--Schwarz equation and a curved Ermakov equation. Finally, in Section~\ref{s7}, some conclusions and open problems are drawn. In particular, the novel interpretation of the curvature as an integrable deformation parameter of LH systems is addressed to.


\section{Conformal symmetries of Cayley--Klein spaces revisited}
\label{s2}

Let us consider the three-dimensional real Lie algebra $\so(3)$ with commutation relations
\begin{equation}
[J_{12}, P_{1}] = P_{2}, \qquad [J_{12}, P_{2}] = - P_{1}, \qquad [P_{1}, P_{2}] = J_{12}\nonumber
\end{equation}
over a basis $\set{P_{1}, P_{2}, J_{12}}$ and with Casimir given by
\begin{equation}
C = P_{1}^{2} + P_{2}^{2} + J_{12}^{2} .\nonumber
\end{equation}
The automorphisms $\Theta_{0}, \Theta_{01}: \so(3) \to \so(3)$ defined by
\begin{equation}
\begin{array}{lll}
\Theta_{0}(P_{1}) = - P_{1}, &\quad \Theta_{0}(P_{2} )  = - P_{2}, &\quad  \Theta_{0}(J_{12})  = J_{12},  \\
\Theta_{01} (P_{1}) = P_{1},& \quad \Theta_{01} (P_{2})  = - P_{2}, &\quad \Theta_{01}(J_{12})  = - J_{12}
\end{array}
\label{auto}
\end{equation}
generate a $\Z_{2} \times \Z_{2}$ group of commuting involutive automorphisms of $\so(3)$, inducing   the following $\Z_{2} \times \Z_{2}$-grading of $\so(3)$:
\begin{equation}
 \so(3) = E_{(1,0)} \oplus E_{(0,1)} \oplus E_{(1, 1)}, \nonumber
\end{equation}
where $E_{(1,0)} := \langle P_{1} \rangle$, $E_{(0,1)} := \langle J_{12} \rangle$, $E_{(1,1)}:= \langle P_{2} \rangle$  and $E_{(0,0)} := \set{0}$ satisfy $[E_{(\alpha_{1}, \alpha_{2})}, E_{(\beta_{1}, \beta_{2})}] = E_{(\alpha_{1} + \beta_{1}, \alpha_{2} + \beta_{2})}$ for every $\alpha_{1}, \alpha_{2}, \beta_{1}, \beta_{2} \in \Z_{2}$. A particular solution of the set of $\Z_{2} \times \Z_{2}$-graded contractions from $\so(3)$ leads to a two-parametric family of real Lie algebras, denoted by $\so_{\kappa_{1}, \kappa_{2}}(3)$ and given by \cite{Herranz1994}
\begin{equation}
 [J_{12}, P_{1}] = P_{2}, \qquad [J_{12}, P_{2}] = - \kappa_{2} P_{1}, \qquad [P_{1}, P_{2}] = \kappa_{1} J_{12},
 \label{CKalgebra}
\end{equation}
where $\kappa_{1}$ and $\kappa_{2}$ are two real graded contraction parameters that can be reduced to either $+1$, $0$ or $-1$ through a rescaling of the Lie algebra generators. Recall that the vanishment of any $\kappa_{i}$ is equivalent to applying an \textit{Inönu--Wigner contraction} \cite{Herranz1994,Inonu1953}. The corresponding Casimir reads
\begin{equation}
 C_{\k} =  \kappa_{2} P_{1}^{2} + P_{2}^{2} + \kappa_{1} J_{12}^{2}, \qquad \k:= (\kappa_{1}, \kappa_{2}).\nonumber
\end{equation}
The family $\so_{\kappa_{1}, \kappa_{2}}(3)$ contains nine specific Lie algebras (some of them isomorphic). Simple Lie algebras  arise when both $\kappa_{i} \neq 0$ ($\so(3)$ for positive values and $\so(2, 1) \simeq \sl(2, \R)$ otherwise) as well as non-simple ones when one $\kappa_{i} = 0$ (the inhomogenous $\iso(1, 1)$, $\iso(2)$ and $\iiso(1)$, where $\iso(1) \equiv \R$). The remarkable fact is that $\so_{\kappa_{1}, \kappa_{2}}(3)$ contains all the Lie algebras of the motion groups of the so-called $2$D \textit{Cayley--Klein (CK) homogeneous spaces} \cite{Yaglom1979,Gromov1990,Ballesteros1993} and thus $\so_{\kappa_{1}, \kappa_{2}}(3)$ is called a \textit{CK algebra}.

Let us now exhibit the connection between the CK algebras $\so_{\kappa_{1}, \kappa_{2}}(3)$ and the 2D CK homogeneous spaces explicitly. The automorphism $\Theta_{0}$ (\ref{auto}) gives rise to the Cartan decomposition
\begin{equation}
\so_{\kappa_{1}, \kappa_{2}}(3) = \h_{0} \oplus \p_{0}, \qquad \h_{0}:= \langle J_{12} \rangle, \qquad \p_{0} := \langle P_{1}, P_{2} \rangle.
\label{eq:CKAlg:Cartan_desc}
\end{equation}
Consider now the faithful representation of the CK algebra $\Gamma: \so_{\kappa_{1}, \kappa_{2}}(3) \to \End(\R^{3})$ given by
\begin{equation}
\Gamma(J_{12}) = - \kappa_{2} E_{12} + E_{21}, \qquad \Gamma(P_{1}) = - \kappa_{1} E_{01} + E_{10}, \qquad \Gamma(P_{2}) = - \kappa_{1} \kappa_{2} E_{02} + E_{20},
\label{eq:CKAlg:rep}
\end{equation}
where $E_{ij}$ denotes the $3 \times 3$ elementary matrix with a single non-zero entry $1$ at row $i$ and column $j$ ($0 \leq i, j \leq 2$). This representation establishes an isomorphism of Lie algebras between the CK algebra $\so_{\kappa_{1}, \kappa_{2}}(3)$ and the Lie algebra of real $3 \times 3$ matrices $M$ satisfying
\begin{equation}
M^{T} \Iq_{\k} + \Iq_{\k}M = 0, \qquad \Iq_{\k} := \diag (1, \kappa_{1}, \kappa_{1}\kappa_{2}).
\label{isom}
\end{equation}
The elements of the representation $\Gamma$ generated by matrix exponentiation are referred to as the \textit{CK Lie group} $\mathrm{\mathrm{SO}}_{\kappa_{1}, \kappa_{2}}(3)$. In particular, the following one-parameter subgroups of the CK Lie group $\mathrm{\mathrm{SO}}_{\kappa_{1}, \kappa_{2}}(3)$ are obtained from   \eqref{eq:CKAlg:rep}:
\begin{equation}
\begin{split}
\exp(\alpha \Gamma(P_{1})) &= \begin{pmatrix}
\Cos_{\kappa_{1}}(\alpha) & - \kappa_{1} \Sin_{\kappa_{1}}(\alpha) & 0 \\
\Sin_{\kappa_{1}}(\alpha) & \Cos_{\kappa_{1}}(\alpha) & 0 \\
0 & 0 & 1
\end{pmatrix}, \quad
\exp (\gamma \Gamma(J_{12})) = \begin{pmatrix}
1 & 0 & 0 \\
0 & \Cos_{\kappa_{2}}(\gamma) & - \kappa_{2} \Sin_{\kappa_{2}}(\gamma) \\
0 & \Sin_{\kappa_{2}}(\gamma) & \Cos_{\kappa_{2}}(\gamma)
\end{pmatrix}, \\
\exp(\beta \Gamma(P_{2})) &= \begin{pmatrix}
\Cos_{\kappa_{1} \kappa_{2}}(\beta) & 0 & - \kappa_{1} \kappa_{2} \Sin_{\kappa_{1} \kappa_{2}}(\beta) \\
0 & 1 & 0 \\
\Sin_{\kappa_{1} \kappa_{2}}(\beta) & 0 & \Cos_{\kappa_{1} \kappa_{2}}(\beta)
\end{pmatrix}  , \qquad \alpha, \beta, \gamma \in \R,
\end{split}
\label{eq:CKGrp:rep}
\end{equation}
where we have introduced the so-called \textit{$\kappa$-dependent cosine} and \textit{sine} functions defined by (see \cite{Ballesteros1993,Herranz2000,Herranz2002} for details)
\begin{equation}
\begin{split}
\Cos_{\kappa}(u) &:= \sum_{l = 0}^{\infty} (-\kappa)^{l} \frac{u^{2l}}{(2l)!} = \begin{cases}
\cos{\sqrt{\kappa} \, u} & \quad \kappa > 0, \\
\qquad 1 & \quad \kappa = 0, \\
\cosh{\sqrt{-\kappa} \, u} & \quad \kappa < 0 .
\end{cases}  \\
\Sin_{\kappa} (u) &:= \sum_{l = 0}^{\infty} (-\kappa)^{l}  \frac{u^{2l + 1}}{(2l+1)!} = \begin{cases}
\frac{1}{\sqrt{\kappa}}\sin{\sqrt{\kappa} \, u} & \quad \kappa > 0, \\
\qquad u & \quad \kappa = 0, \\
\frac{1}{\sqrt{-\kappa}}\sinh{\sqrt{-\kappa} \, u} & \quad \kappa < 0.
\end{cases}
\end{split}
\label{eq:kdep:cos_sin}
\end{equation}
The above $\kappa$-dependent trigonometric functions give rise to the \textit{$\kappa$-tangent} and the \textit{$\kappa$-versed sine} (or \textit{versine}), which are defined as
\begin{equation}
\Tan_{\kappa}(u) := \frac{\Sin_{\kappa}(u)}{\Cos_{\kappa}(u)}, \qquad \Ver_{\kappa}(u) := \frac{1}{\kappa} \bigl(1 - \Cos_{\kappa}(u) \bigr).
\label{eq:kdep:tan_ver}
\end{equation}
These $\kappa$-functions cover the usual circular $(\kappa >0)$ as well as the hyperbolic $(\kappa < 0)$ trigonometric functions, while the parabolic or Galilean ones $(\kappa = 0)$ are obtained from the contraction $\kappa \to 0$ as
\be
\Cos_{0}(u)= 1, \qquad \Sin_{0}(u) = \Tan_{0}(u) = u,\qquad \Ver_{0}(u) = u^{2} /2.
\label{contract}
\ee
Their main relations, necessary for further computations, are summarized in the Appendix.

Consider now the Lie subgroup $H_{0}:= \mathrm{\mathrm{SO}}_{\kappa_{2}}(2)$  of $\mathrm{\mathrm{SO}}_{\kappa_{1},\kappa_{2}}(3)$ associated with the Lie subalgebra $\h_{0} \subset \so_{\kappa_{1},\kappa_{2}}(3)$ (\ref{eq:CKAlg:Cartan_desc}). The \textit{CK family of $2$D symmetrical homogeneous spaces} is defined by the quotient
\be
 \S^{2}_{[\kappa_{1}], \kappa_{2}} := \mathrm{\mathrm{SO}}_{\kappa_{1},\kappa_{2}}(3) /  {\mathrm{SO}}_{\kappa_{2}}(2).
\label{CKspace}
\ee
It can be shown that the graded   contraction parameter $\kappa_{1}$ becomes the (Gaussian) \textit{curvature} of the space, while $\kappa_{2}$ determines the \textit{signature} of the metric through $\diag(1, \kappa_{2})$.


\subsection{Coordinate systems on the Cayley--Klein spaces}
\label{s21}

The matrix representation \eqref{eq:CKGrp:rep} of $\mathrm{\mathrm{SO}}_{\kappa_{1}, \kappa_{2}}(3)$ shows that every element $g \in \mathrm{\mathrm{SO}}_{\kappa_{1},\kappa_{2}}(3)$ satisfies
$g^{T} \Iq_{\k} g = \Iq_{\k}$,
so we can consider the Lie group action of $\mathrm{\mathrm{SO}}_{\kappa_{1}, \kappa_{2}}(3)$ on $\R^{3}$ as isometries of $\Iq_{\k}$ (\ref{isom}). This action is not transitive, as it preserves the quadratic form induced by $\Iq_{\k}$. Also, $\mathrm{\mathrm{SO}}_{\kappa_{2}}(2) = \langle \exp(\gamma \Gamma(J_{12})) \rangle$ is the isotropy group of the point  $O:=(1,0,0)$, which is thus taken as the \textit{origin} in the space $\S^{2}_{[\kappa_{1}], \kappa_{2}}$. Nevertheless, the action becomes transitive when we restrict ourselves to the orbit of the point $O$; namely, the connected component of the submanifold
\begin{equation}
 \Sigma_{\k} := \set{v := (x^{0}, x^{1}, x^{2}) \in \R^{3} : \Iq_{\k}(v, v) = (x^{0})^{2} + \kappa_{1}(x^{1})^{2} + \kappa_{1} \kappa_{2} (x^{2})^{2} =   1}
 \label{eq:CK2:subman}
\end{equation}
containing the origin $O$, thus allowing us to identify the space $\S^{2}_{[\kappa_{1}], \kappa_{2}}$  (\ref{CKspace}) with the latter orbit.
The coordinates $(x^{0}, x^{1}, x^{2})$ on $\R^{3}$ satisfying the constraint \eqref{eq:CK2:subman} on $\Sigma_{\k}$ are called \textit{ambient} or \textit{Weierstrass} coordinates. In terms of these variables, the metric $\dd s_{\k}^{2}$ of $\S^{2}_{[\kappa_{1}], \kappa_{2}}$ comes from the flat ambient metric in $\R^{3}$ divided by  the curvature $\kappa_{1}$ and restricted to $\Sigma_{\k}$:
\begin{equation}
\dd s_{\k}^{2} = \frac{1}{\kappa_{1}} (  (\dd x^{0})^{2}  + \kappa_{1}(\dd x^{1})^{2} + \kappa_{1} \kappa_{2} (\dd x^{2})^{2}) \Big\vert_{\Sigma_{\k}} = \frac{\kappa_{1} (x^{1} \dd x^{1} + \kappa_{2} x^{2} \dd x^{2})^{2}}{1 - \kappa_{1} (x^{1})^{2} - \kappa_{1} \kappa_{2}(x^{2})^{2}} + (\dd x^{1})^{2} + \kappa_{2} (\dd x^{2})^{2} .
\label{eq:CK2:metric_amb}
\end{equation}

\begin{figure}[t]
\centering
\includegraphics[width = 0.45\textwidth]{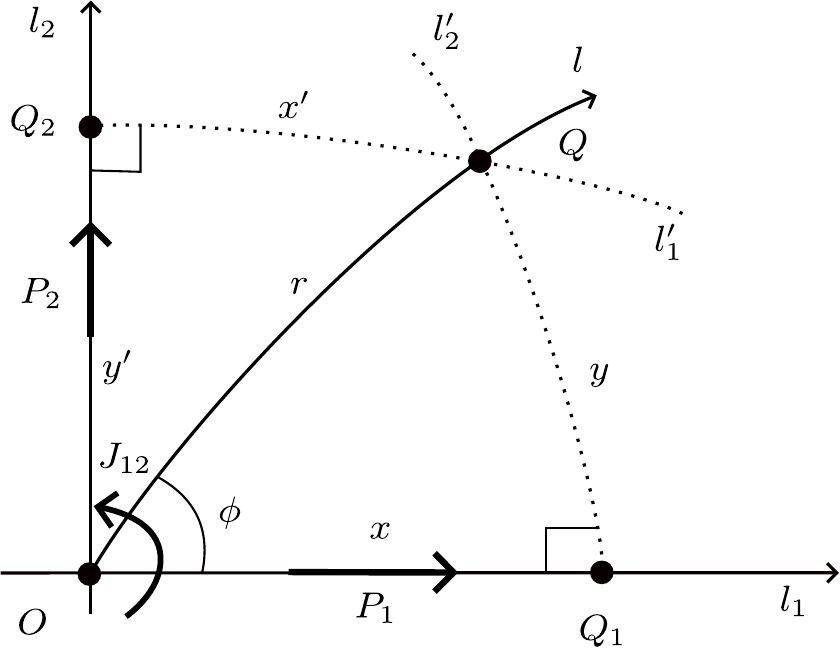}
\caption{\small Isometry infinitesimal generators $\set{J_{12},P_1,P_2}$ and  geodesic coordinates   $(x,y)$, $(x',y')$ and $(r, \phi)$     (\ref{eq:CK2:coord}) of a   point  $Q=(x^0,x^1,x^2)$   on the 2D CK space $\S^{2}_{[\kappa_{1}], \kappa_{2}}$  (\ref{CKspace}).}
\label{fig:coord_curved}
\end{figure}

Therefore,  the Lie group $\mathrm{\mathrm{SO}}_{\kappa_{1}, \kappa_{2}}(3)$ is the isometry group of the space $\S^{2}_{[\kappa_{1}], \kappa_{2}}$ in such a manner that $J_{12}$ is a rotation generator, while $P_{1}$ and $P_{2}$ are translations moving the origin $O$ along two geodesics $l_{1}$ and $l_{2}$ orthogonal at $O$, respectively (see Figure~\ref{fig:coord_curved}). We next  introduce   three relevant kinds of geodesic coordinate systems   of a point $Q := (x^{0}, x^{1}, x^{2})$ in   $\S^{2}_{[\kappa_{1}], \kappa_{2}}$  (\ref{CKspace}):  \textit{geodesic parallel}  coordinates of type I $(x,y)$  and type II $(x',y')$,  and   \textit{geodesic polar} coordinates $(r, \phi)$. These are defined through the action of the one-parametric groups \eqref{eq:CKGrp:rep} on the origin $O$ as follows~\cite{Herranz2002}:
\[ Q = \exp(x\Gamma(P_{1}))(\exp(y \Gamma(P_{2}))O =\exp(y' \Gamma(P_{2})) \exp(x' \Gamma(P_{1})) O  = \exp(\phi \Gamma(J_{12})) \exp(r \Gamma(P_{1})) O, \]
obtaining
\begin{equation}
\begin{split}
x^{0} &= \Cos_{\kappa_{1}}(x) \Cos_{\kappa_{1} \kappa_{2}}(y) = \Cos_{\kappa_{1}}(x') \Cos_{\kappa_{1}, \kappa_{2}} (y') =  \Cos_{\kappa_{1}}(r), \\
x^{1} & = \Sin_{\kappa_{1}}(x) \Cos_{\kappa_{1} \kappa_{2}}(y) =  \Sin_{\kappa_{1}}(x') = \Sin_{\kappa_{1}}(r) \Cos_{\kappa_{2}}(\phi), \\
x^{2} & = \Sin_{\kappa_{1} \kappa_{2}}(y)  = \Cos_{\kappa_{1}}(x') \Sin_{\kappa_{1} \kappa_{2}}(y') = \Sin_{\kappa_{1}}(r) \Sin_{\kappa_{2}}(\phi).
\end{split}
\label{eq:CK2:coord}
\end{equation}
In these coordinates, the metric \eqref{eq:CK2:metric_amb}  adopts the form

\begin{equation}
\dd s_{\k}^{2} = \Cos_{\kappa_{1} \kappa_{2}}^{2}(y) \dd x^{2} + \kappa_{2} \dd y^{2} = \dd x'^{2} + \kappa_{2} \Cos_{\kappa_{1}}^{2}(x') \dd y'^{2} =  \dd r^{2} + \kappa_{2} \Sin_{\kappa_{1}}^{2}(r) \dd \phi^{2}.
\label{eq:CK2:metric_curv}
\end{equation}


As Figure~\ref{fig:coord_curved} shows, the variable $r$ is the distance between the origin $O$ and the point $Q$ measured along the geodesic $l$ joining both points, while $\phi$ is the angle formed by the geodesics $l$ and $l_{1}$. Consider the intersection point  $Q_{1}$ of $l_{1}$ with its orthogonal geodesic $l_{2}'$ through $Q$, so $x$ is the geodesic distance between $O$ and $Q_{1}$ measured along $l_{1}$ and $y$ is the geodesic distance between $Q_{1}$ and $Q$ measured along $l_{2}'$. Similarly, if $Q_{2}$ is the intersection point of $l_{2}$ with its orthogonal geodesic $l_{1}'$, then $x'$ is the geodesic distance between $Q$ and $Q_{2}$ measured along $l_{1}'$ and $y'$ is the geodesic distance beween $O$  and $Q_{2}$ measured along $l_{2}$.
 Note that  $(x',y') \neq (x,y)$ if the curvature $\kappa_{1} \neq 0$, while on the flat Euclidean plane $\E^{2}\equiv \S^{2}_{[0], +}$  we recover the usual Cartesian coordinates $(x,y) = (x',y')$ and the polar ones $(r, \phi)$.

\begin{figure}[t]
\centering
\includegraphics[width = 0.8 \textwidth]{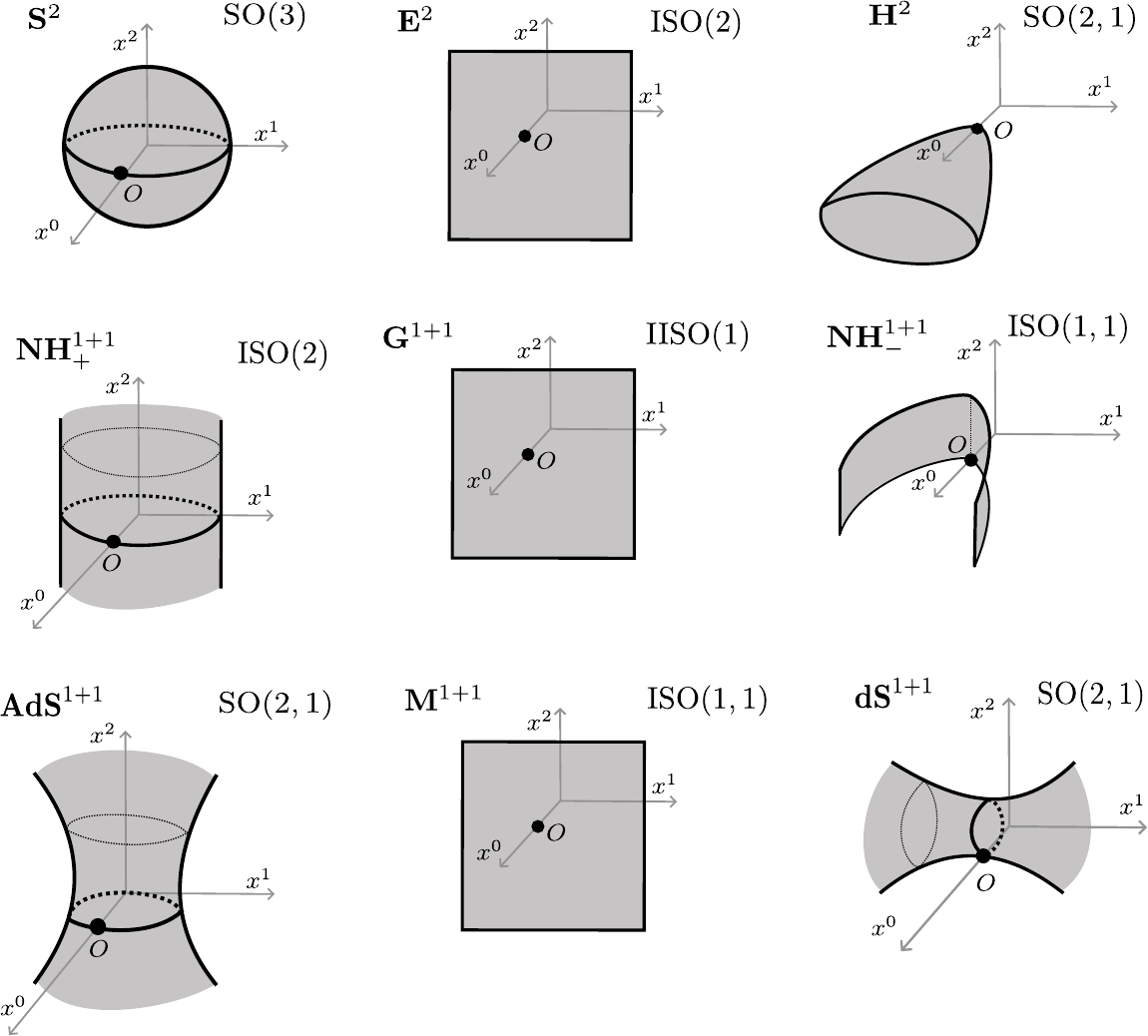}
\caption{{\small The  nine 2D CK spaces $\S^{2}_{[\kappa_{1}], \kappa_{2}}$  (\ref{CKspace}) in Weierstrass coordinates $(x^{0}, x^{1}, x^{2})$ (\ref{eq:CK2:subman}) together with their correspondent 2D CK motion group $\mathrm{SO}_{\kappa_{1}, \kappa_{2}}(3)$ according to the `normalized' values of the contraction parameters $\kappa_{i} \in \set{-1,0,1}$.}}
\label{fig:ambientmodel2D}
\end{figure}

Summing up, depending on the values of the parameters $\kappa_{i}$, the CK family $\S^{2}_{[\kappa_{1}], \kappa_{2}}$  (\ref{CKspace}) comprises \textit{nine} specific 2D symmetrical homogeneous spaces, which are classified intro three types according to the value of the signature parameter $\kappa_{2}$~\cite{Herranz2002}, which are displayed by rows in   Figure~\ref{fig:ambientmodel2D}:
\begin{itemize}
\item \textit{Riemannian spaces} for $\kappa_{2} > 0$. In this case the isotropy subgroup ${\mathrm{SO}}_{\kappa_{2}}(2)\equiv {\mathrm{SO}} (2)$ and we find:  the   {\em sphere} $\S^{2}$   ($\kappa_{1} > 0$);
the {\em Euclidean plane}  $\E^{2}$ ($\kappa_{1} = 0$) with $x^{0} = + 1$; and
 the \textit{hyperbolic space}  $\H^{2}$ ($\kappa_{1} < 0$) as the part with $x^{0} \geq 1$ of a two-sheeted hyperboloid.

 \item\textit{Pseudo-Riemannian spaces} or \textit{Lorentzian spacetimes} for $\kappa_{2} <0$.
 Now ${\mathrm{SO}}_{\kappa_{2}}(2)\equiv {\mathrm{SO}} (1,1)$ is the Lorentz subgroup and
 these spaces admit a kinematical intertepretation as $(1+1)$D spacetimes. Thus we obtain:  the {co-hyperbolic} or \textit{anti-de Sitter space} $\AdS^{1+1}$ ($\kappa_{1} > 0$);
 the \textit{Minkowskian plane} $\M^{1+1}$ ($\kappa_{1} = 0$); and  
  the  {doubly-hyperbolic} or \textit{de Sitter space} $\dS^{1+1}$   ($\kappa_{1} < 0$). In all these cases, the generators $J_{12}$, $P_{1}$ and $P_{2}$ of  $\so_{\kappa_{1},\kappa_{2}}(3)$ correspond to the infinitesimal generators of boosts, time translations and spatial translations, respectively. Moreover, from a physical viewpoint the $\kappa_{i}$ parameters  are related to the cosmological constant $\Lambda$ and the speed of light $c$ through
\be
 \kappa_{1} = - \Lambda, \qquad \kappa_{2} = -1 / c^{2}.
 \label{luz}
 \ee
The geodesic parallel coordinates of type I $(x,y)$ are just the time $t$ and spatial $y$ ones.

\item\textit{Semi-Riemannian spaces} or \textit{Newtonian spacetimes} for $\kappa_{2} = 0$. Here ${\mathrm{SO}}_{\kappa_{2}}(2)\equiv {\mathrm{ISO}} (1)\equiv \R$ and the three spaces can be interpreted as non-relativistic spacetimes with $c  = \infty$ (\ref{luz}).
These are:   the  {co-Euclidean} or \textit{oscillating Newton--Hooke (NH) space} $\NH^{1+1}_{+}$ ($\kappa_{1} > 0$); the  \textit{Galilean   plane} $\G^{1+1}$ ($\kappa_{1} = 0$); and the  {co-Minkowskian} or \textit{expanding NH space} $\NH_{-}^{1+1}$ ($\kappa_{1} < 0$). The metric \eqref{eq:CK2:metric_curv} is degenerate and its kernel gives rise to an integrable foliation of $\S^{2}_{[\kappa_{1}], 0}$ which is invariant under the action of the CK group $\mathrm{\mathrm{SO}}_{\kappa_{1},0}(3)$.
A    {subsidiary metric} $\dd s_{\k}^{\prime 2}$   appears being well-defined when restricted to each leaf of the foliation. In  geodesic parallel coordinates of type I $(x,y)$ the metrics read
\be
\dd s_{\k}^{2} = \dd x^{2}, \qquad \dd s_{\k}^{\prime 2} = \dd y^{2} \quad \text{on} \quad \text{$x = \text{constant}$}.
\label{metricN}
\ee
The main metric $\dd s_{\k}^{2}$ provides `absolute-time' $t$, the leaves of the invariant foliation are the `absolute-space' at $t = t_{0}$ and $\dd s^{\prime 2}_{\k}$ is the subsidiary spatial metric defined on each leaf.
\end{itemize}


\begin{figure}[t]
\centering
\includegraphics[width = 0.8\textwidth]{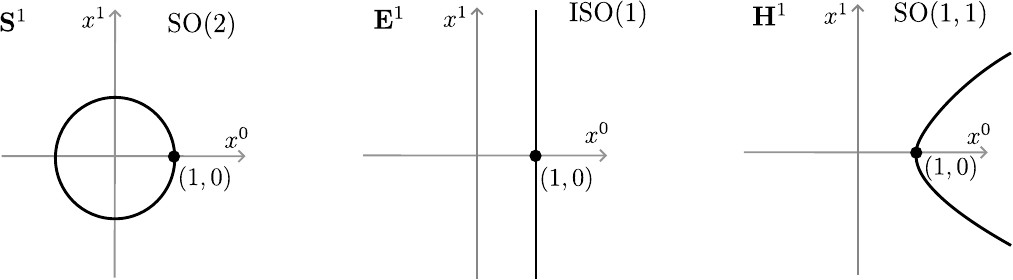}
\caption{{\small The  three 1D CK spaces $\S^{2}_{[\kappa]}$ in Weierstrass coordinates $(x^{0}, x^{1})$ (\ref{eq:CK1:subman})  together with their corresponding 1D CK motion group $\mathrm{SO}_{\kappa}(2)$ according to the `normalized' values of the contraction parameter $\kappa \in \set{-1,0,1}$.}}
\label{fig:ambientmodel1D}
\end{figure}

Consider now the projection
\begin{equation}
\R^{3} \ni (x^{0}, x^{1}, x^{2}) \mapsto (x^{0}, x^{1}) \in \R^{2},
\label{eq:CK1:proj}
\end{equation}
so the image of the submanifold $\Sigma_{\k}$ \eqref{eq:CK2:subman} under this mapping is
\begin{equation}
\Sigma_{\kappa} := \set{(x^{0}, x^{1}) \in \R^{2}: (x^{0})^{2} + \kappa (x^{1})^{2} = 1},
\label{eq:CK1:subman}
\end{equation}
where we have set $\kappa := \kappa_{1}$. The image of the origin $O$   of $\S^{2}_{[\kappa_{1}], \kappa_{2}}$ under the projection is the point $(1, 0)$, so we call  {$1$D CK space} $\S^{1}_{[\kappa]}$  the connected component of $\Sigma_{\kappa}$ containing the point $(1, 0)$. For $\kappa > 0$ we obtain the 1D sphere $\S^{1}$, while the case $\kappa <0$ gives rise to the hyperbolic line $\H^{1}$, which is the branch of the hyperbola $(x^{0})^{2} -(x^{1})^{2} = 1$ with $x^{0} \geq 1$. The Euclidean line $\E^{1}$, identified with the line $x^{0} = 1$ in $\R^{2}$, appears after the contraction $\kappa = 0$ as depicted in   Figure~\ref{fig:ambientmodel1D}.

Analogously to the 2D case, the coordinates $(x^{0}, x^{1})$ on $\R^{2}$ satisfying the constraint \eqref{eq:CK1:subman}   are called \textit{ambient} or \textit{Weierstrass} coordinates. The single coordinate system $(x)$ on $\S^{1}_{[\kappa]}$ is given by
\be
x^{0} = \Cos_{\kappa}(x), \qquad x^{1} = \Sin_{\kappa}(x),
\label{CK1}
\ee
and it can be obtained from the coordinates \eqref{eq:CK2:coord} of $\S^{2}_{[\kappa_{1}], \kappa_{2}}$ by setting $y = 0$, $\kappa_{1} \equiv \kappa$ and $\kappa_{2} \equiv 0$.  Thus, the 1D CK space $\S^{1}_{[\kappa]}$ is canonically identified with the submanifold $y = 0$ of the 2D CK space $\S^{2}_{[\kappa], 0}$. Consequently, the (Riemannian) metric $\dd s_{\kappa}^{2}$ of $\S^{1}_{[\kappa]}$ reads from \eqref{eq:CK2:metric_curv} as
\begin{equation}
 \dd s_{\kappa}^{2} = \dd x^{2}.
\label{eq:CK1:metric}
\end{equation}
Note that in these spaces the parameter $\kappa$ cannot be interpreted as the Gaussian curvature, as the three 1D spaces are flat. Nevertheless, there are some differences better described after the value of $\kappa$. The associated 1D CK groups $\mathrm{\mathrm{SO}}_{\kappa}(2)$ are given by
\[ \mathrm{\mathrm{SO}}_{\kappa} (2) =  \set{\begin{pmatrix}
\Cos_{\kappa}(\alpha) & - \kappa \Sin_{\kappa}(\alpha) \\
\Sin_{\kappa}(\alpha) & \Cos_{\kappa}(\alpha)
\end{pmatrix} : \alpha \in \R},\]
which correspond to the projection \eqref{eq:CK1:proj} to $\R^{2}$ of the one-parameter subgroup \eqref{eq:CKGrp:rep} of $\mathrm{\mathrm{SO}}_{\kappa, 0}(3)$ generated by $P_{1}$. If $\Iq_{\kappa} := \diag(1, \kappa)$ denotes the quadratic form inducing the submanifold $\Sigma_{\kappa}$ \eqref{eq:CK1:subman}, then every element $g \in \mathrm{\mathrm{SO}}_{\kappa}(2)$ satisfies
$g^{T} \Iq_{\kappa} g = \Iq_{\kappa}$,
showing that $\mathrm{\mathrm{SO}}_{\kappa}(2)$ is a group of isometries of the 1D CK space $\S^{1}_{[\kappa]}$. After the values $\kappa = +1, \kappa = 0$ and $\kappa = -1$ we recover from $\mathrm{\mathrm{SO}}_{\kappa}(2)$ the well-known groups of isometries $\mathrm{\mathrm{SO}}(2)$, $\mathrm{ISO}(1) \equiv \R$ and $\mathrm{\mathrm{SO}}(1, 1)$ of $\S^{1}$, $\E^{1}$ and $\H^{1}$, respectively.


\subsection{Conformal Cayley--Klein algebras}
\label{s22}

 Infinitesimal conformal symmetries of the nine 2D CK spaces $\S^{2}_{[\kappa_{1}], \kappa_{2}}$ were obtained in \cite{Herranz2002} by looking for one-parameter subgroups of cycle-preserving transformations (cycles are lines with constant geodesic curvature), while their corresponding conformal Lie groups and compactification were constructed in~\cite{Herranz2002b}.    This global approach requires to consider the three sets of geodesic coordinates (\ref{eq:CK2:coord}) and then to express all the conformal Lie generators in a common set. The resulting  conformal symmetries span a  6D real Lie algebra $\conf_{\kappa_{1}, \kappa_{2}}$, known as the \textit{conformal algebra} of $\S^{2}_{[\kappa_{1}], \kappa_{2}}$. Let us summarize the main results necessary for this work.

We consider the  basis $\set{P_{i}, J_{12}, G_{i}, D: 1 \leq i \leq 2}$   corresponding to the generators of translations along the basic geodesic $l_i$, rotations around the origin (see Figure~\ref{fig:coord_curved}), specific conformal transformations related to the geodesic $l_i$ and dilations, respectively.  The commutation relations of $\conf_{\kappa_{1}, \kappa_{2}}$ over such a basis  are given by
\be
\begin{array}{lll}
[J_{12}, P_{1}] = P_{2}, &\quad [J_{12}, P_{2}] = - \kappa_{2} P_{1}, & \quad[P_{1}, P_{2}] = \kappa_{1} J_{12}, \\[2pt]
[J_{12}, G_{1}] = G_{2},  & \quad [J, G_{2}] = - \kappa_{2} G_{1},  & \quad [G_{1}, G_{2}] = 0, \\[2pt]
[D, P_{i}] = P_{i} + \kappa_{1} G_{i},  &\quad [D, G_{i}] = -G_{i},  &\quad [D, J_{12}] = 0,\\[2pt]
[P_{1}, G_{1}] =  D,  &\quad [P_{2},G_{2}] =  \kappa_{2} D,  &\quad \\[2pt]
[P_{1}, G_{2}]  = - J_{12},  &\quad [P_{2}, G_{1}] = J_{12}. &\quad
\end{array}
\label{cCK}
\ee
Recall that for the six kinematical cases with $\kk_2\le 0$  (see Figure~\ref{fig:ambientmodel2D}), $P_1$ is the time-like translation generator, $P_2$ is the space-like one and $J_{12}$ is just the boost. Hence, $G_1$ and $G_2$ can be regarded as  time- and space-like specific conformal transformations.

As some relevant Lie subalgebras of $\conf_{\kappa_{1}, \kappa_{2}}$, observe that: (i)  the isometries $\set{P_{1}, P_2, J_{12}}$ span the CK algebra $\so_{\kappa_{1}, \kappa_{2}}(3)$ (\ref{CKalgebra}) as expected; (ii)  the set $\set{P_{1}, P_2,J_{12},D}$ (isometries plus dilations) only closes on a   subalgebra in the flat spaces with $\kk_1=0$, the so-called   Weyl subalgebra;  (iii)~the set $\set{G_{1}, G_2, J_{12}}$ spans a subalgebra isomorphic to the inhomogenous $\so_{0, \kappa_{2}}(3)\simeq \iso_{\kk_2}(2)$;   (iv)    $\set{P_1,G_{1}, D}$ provides  a subalgebra isomorphic to $\so(2, 1)$; and (v) $\set{P_2,G_{2}, D}$ again leads  to a subalgebra isomorphic to  $\so(2, 1)$ whenever $\kk_2\ne 0$, but to Poincar\'e  $\iso(1, 1)$ for $\kk_2=0$.

The conformal CK algebra $\conf_{\kappa_{1}, \kappa_{2}}$  (\ref{cCK}) covers nine specific conformal Lie algebras which are isomorphic to three cases according to the value of the signature parameter $\kk_2$ (regardless of the curvature value $\kk_1$):
\begin{itemize}
\item $\so(3,1)$, isomorphic to the 3D hyperbolic algebra or to the $(2+1)$D de Sitter algebra, for the three 2D Riemannian spaces with $\kk_2>0$.

\item $\iso(2,1)$, isomorphic to   the $(2+1)$D  Poincar\'e algebra, for the three $(1+1)$D non-relativistic or Newtonian spacetimes   with $\kk_2=0$.

\item $\so(2,2)$, isomorphic to   the $(2+1)$D  anti-de Sitter  algebra, for the three $(2+1)$D relativistic or Lorentzian spacetimes   with $\kk_2<0$.

\end{itemize}

A differential realization of $\conf_{\kappa_{1}, \kappa_{2}}$ (\ref{cCK}) can be derived in any of the  three sets of geodesic coordinates (\ref{eq:CK2:coord}). Hereafter  we will consider geodesic parallel coordinates of type I $(x,y)$ that give rise to the following vector fields~\cite{Herranz2002}
\begin{equation}
\begin{split}
&\P_{1} = - \pdv{x} ,\qquad  \P_{2} = - \kappa_{1} \kappa_{2} \Sin_{\kappa_{1}}(x) \Tan_{\kappa_{1} \kappa_{2}}(y) \pdv{x}- \Cos_{\kappa_{1}}(x) \pdv{y} ,\\
& \J= \kappa_{2} \Cos_{\kappa_{1}}(x) \Tan_{\kappa_{1} \kappa_{2}}(y) \pdv{x}- \Sin_{\kappa_{1}}(x) \pdv{y},\\
 &\G_{1} = \frac{1}{\Cos_{\kappa_{1} \kappa_{2}}(y)}\, \bigl(\Ver_{\kappa_{1}}(x) - \kappa_{2} \Ver_{\kappa_{1} \kappa_{2}}(y) \bigr) \pdv{x} + \Sin_{\kappa_{1}}(x) \Sin_{\kappa_{1} \kappa_{2}}(y) \pdv{y} ,\\
&\G_{2} = \kappa_{2} \Sin_{\kappa_{1}}(x) \Tan_{\kappa_{1}\kappa_{2}}(y) \pdv{x} - \bigl(\Ver_{\kappa_{1}}(x) - \kappa_{2} \Ver_{\kappa_{1} \kappa_{2}}(y) \bigr) \pdv{y},\\
& \D = - \frac{\Sin_{\kappa_{1}}(x)}{\Cos_{\kappa_{1} \kappa_{2}}(y)} \pdv{x} - \Cos_{\kappa_{1}}(x) \Sin_{\kappa_{1} \kappa_{2}}(y) \pdv{y} ,
\end{split}
\label{eq:Conf:specific}
\end{equation}
 which are always well-defined for any value of $\kappa_{1}$ and $\kappa_{2}$. The relations  (\ref{eq:CK2:coord}) allow one to express these vector fields in the  remaining two sets  of  geodesic coordinates.

 It is worth observing that in the six kinematical spaces with $\kk_2\le 0$, the  expressions (\ref{eq:Conf:specific}) will be well adapted to construct `time-like' curved LH systems,  since the time translation generator is simply $\P_{1} = - \partial_{x}$, which is our aim here.  Nevertheless, it would also be  possible to deduce `space-like' curved LH systems, but for this it would be necessary   to choose geodesic parallel coordinates of type II $(x',y')$ such that the spatial translation generator becomes $\P_{2} = - \partial_{y'}$. Note that the dilation vector field takes the simplest form in geodesic polar coordinates: $\D=- \Sin_{\kappa_{1}}(r)\partial_r$.

In addition, it can  be  easily shown that the vector fields  (\ref{eq:Conf:specific}) satisfy the conformal Killing equations for the CK metric $g_1$ (\ref{eq:CK2:metric_curv}), that is, the Lie derivative must be $\mathcal{L}_\X g_1 = \mu_\X g_1$ where  $\mu_\X$ is the conformal factor of the vector field $\X$. Explicitly, the conformal factors in the geodesic $(x,y)$ and ambient $(x^0,x^1,x^2)$ coordinates are found to be
\begin{equation}
\begin{split}
&\mu_{\P_1}= \mu_{\P_2}= \mu_{ \J }= 0,\qquad \mu_\D= -2  \Cos_{\kappa_{1}}(x)\Cos_{\kappa_{1} \kappa_{2}}(y) \equiv -2 x^0, \\
& \mu_{\G_1}= 2\Sin_{\kappa_{1}}(x)\Cos_{\kappa_{1} \kappa_{2}}(y)  \equiv 2 x^1, \qquad \mu_{\G_2}=2 \kk_2\Sin_{\kappa_{1} \kappa_{2}}(y)  \equiv 2 \kk_2 x^2.
\end{split}
\label{factors}
\end{equation}

In order to illustrate the `intrinsic' contraction-scheme of the CK approach, let us write the resulting flat contraction $\kk_1=0$ of the above vector fields. By taking into account (\ref{contract}) we directly obtain that the conformal vector fields  (\ref{eq:Conf:specific})  for the three spaces $\S^{2}_{[0], \kappa_{2}}$ are
\begin{equation}
\begin{split}
&\P_{1} = - \pdv{x} ,\qquad  \P_{2} = -  \pdv{y} , \qquad  \J= \kappa_{2}\, y \pdv{x}-x \pdv{y},\qquad \D = - x \pdv{x} - y \pdv{y} ,\\
 &\G_{1} =  \frac 12 \bigl( x^2 - \kappa_{2} \, y^2 \bigr) \pdv{x} + xy \pdv{y} ,\qquad
 \G_{2} = \kappa_{2} \, x y\pdv{x} - \frac 12 \bigl(x^2 - \kappa_{2} \, y^2 \bigr) \pdv{y},
\end{split}
\label{eq:Conf:specificB}
\end{equation}
corresponding to the Euclidean $\E^{2} \equiv \S^{2}_{[0], +}$, Galilean $\G^{1+1} \equiv \S^{2}_{[0], 0}$ and  Minkowskian $\M^{1+1} \equiv \S^{2}_{[0], -}$ spaces. The conformal factors (\ref{factors}) reduce to $\mu_\D= -2$, $\mu_{\G_1}= 2x$ and $\mu_{\G_2}=2 \kk_2\, y$.

As a byproduct of the above results, we remark that the conformal algebra $\conf_{\kappa}$ of the 1D CK space $\S^{1}_{[\kappa]}$ (\ref{CK1}) (see Figure~\ref{fig:ambientmodel1D}) can be obtained in straightforward manner from $\conf_{\kappa_{1}, \kappa_{2}}$ by the restriction of its vector fields (\ref{eq:Conf:specific}) to the submanifold $y = 0$ of $\S^{2}_{[\kappa], 0}$, such that $\kk\equiv \kk_1$ and $\kk_2=0$. Therefore, $\conf_{\kappa}$  turns out to be a 3D Lie algebra spanned by the  vector fields   $\{\P_1,\G_1,\D\}$ given by
\begin{equation}
\P_{1} = - \pdv{x}, \qquad \G_{1} = \Ver_{\kappa}(x) \pdv{x}, \qquad \D = - \Sin_{\kappa}(x) \pdv{x},
\label{eq:Conf:CK1}
\end{equation}
whose geometrical meaning is  preserved, since they are still   a translation, a specific conformal transformation and a dilation on  $\S^{1}_{[\kappa]}$, respectively. When $\kk=0$, they reduce to
\be
\P_{1} = -\pdv{x} , \qquad  \G_{1} =\frac 12 x^2\pdv{x},\qquad \D=-x\pdv{x} .
 \label{eq:I4_Euc:vf_curved}
\ee
Finally, the  Lie brackets for (\ref{eq:Conf:CK1}) are
\be
 [\D, \P_{1}] = \P_{1} + \kappa \G_{1}, \qquad [\D, \G_{1}] = - \G_{1}, \qquad [\P_{1}, \G_{1}] = \D,
 \label{braCK1}
 \ee
so $\conf_{\kappa} \simeq \so(2, 1)\simeq \sl(2, \R)$ regardless of the value of $\kappa$, as expected.


 \section{The class I$_{4}$ of Lie--Hamilton systems on curved spaces}
 \label{s3}

 We now have  all the geometrical ingredients to construct new LH systems on   2D   spaces  from their conformal symmetries.  Let us start with the class I$_{4}$   of LH systems on the Euclidean plane $\E^{2}$ spanned by  the vector fields given in Table~\ref{table1}, that is,
\begin{equation}
 \X_{1} = \pdv{x} + \pdv{y}, \qquad \X_{2} = x \pdv{x} + y \pdv{y}, \qquad \X_{3} = x^{2} \pdv{x} + y^{2} \pdv{y} ,
\label{eq:I4_Euc:vf}
\end{equation}
which obey the commutation relations (\ref{sl2}) of $\so(2, 1)\simeq \sl(2, \R)$. According to the previous section, it emerges that such vector fields are just the sum of two conformal realizations  (\ref{eq:I4_Euc:vf_curved}) on the Euclidean line $\E^{1}\equiv \S^{1}_{[0]}$ and,  therefore, defined on $\E^{1}\times \E^{1}$ with coordinates $(x,y)$. If we  restrict ourselves to  the first coordinate $x$, we find that  the relationships between $\X_i$ (\ref{eq:I4_Euc:vf})  and the Euclidean conformal vector fields  (\ref{eq:I4_Euc:vf_curved})
read: $\X_1= -\P_1$, $\X_2=-\D$ and $\X_3=2\G_1$.

 Consequently,  a natural generalization  of the Euclidean class  I$_{4}$ (\ref{eq:I4_Euc:vf}) to curved spaces is provided by   the sum of   two realizations of $\so(2, 1)$ on the 1D CK space $\S^{1}_{[\kappa]}$, obtained by replacing the conformal symmetries \eqref{eq:I4_Euc:vf_curved} of $\E^{1}\equiv \S^{1}_{[0]}$ by their curved counterparts   (\ref{eq:Conf:CK1}) of  $\S^{1}_{[\kappa]}$. This yields   the following vector fields on $\S^{1}_{[\kappa]} \times \S^{1}_{[\kappa]}$:
\begin{equation}
\X_{\kappa, 1} = \pdv{x} + \pdv{y}, \qquad \X_{\kappa, 2} = \Sin_{\kappa}(x) \pdv{x} + \Sin_{\kappa}(y) \pdv{y}, \qquad \X_{\kappa, 3} = 2 \Ver_{\kappa}(x) \pdv{x} + 2 \Ver_{\kappa}(y) \pdv{y},
\label{eq:I4_Curv:vf}
\end{equation}
and their commutation relations are  given by (see (\ref{braCK1}))
\begin{equation}
  [\X_{\kappa, 1}, \X_{\kappa, 2}] = \X_{\kappa, 1} - \tfrac{1}{2} \kappa\, \X_{\kappa, 3}, \qquad [\X_{\kappa,1}, \X_{\kappa,3}] = 2 \X_{\kappa, 2}, \qquad [\X_{\kappa,2}, \X_{\kappa,3}] = \X_{\kappa,3} .
 \label{eq:I4_Curv:comm}
 \end{equation}
  The $t$-dependent vector field
 \begin{equation}
 \X_{\kappa} := b_{1}(t) \X_{\kappa, 1} + b_{2}(t) \X_{\kappa,2} + b_{3}(t) \X_{\kappa, 3},
 \label{eq:I4_Curv:tvf}
 \end{equation}
 where $b_{i}(t)$ $(1 \leq i \leq 3)$ are   arbitrary $t$-dependent functions, is associated with the first-order system of ODEs on $\S^{1}_{[\kappa]} \times \S^{1}_{[\kappa]}$ given by
 \begin{equation}
 \dv{x}{t} = b_{1}(t) + b_{2}(t)\Sin_{\kappa}(x) + 2 b_{3}(t)\Ver_{\kappa}(x), \qquad \dv{y}{t} = b_{1}(t) + b_{2}(t)\Sin_{\kappa}(y) + 2 b_{3}(t)\Ver_{\kappa}(y).
 \label{eq:I4_Curv:system}
 \end{equation}
 Notice that the  contraction $\kk=0$ leads to  the sytem
  \begin{equation}
 \dv{x}{t} = b_{1}(t) + b_{2}(t)x+   b_{3}(t)x^2, \qquad \dv{y}{t} = b_{1}(t) + b_{2}(t)y +   b_{3}(t)y^2 ,
 \label{flatriccati}
 \end{equation}
known as    coupled Riccati equations, considered in \cite{Mariton1985}, and which is a particular case of the systems of Riccati equations studied in \cite{Ballesteros2013}. Thus we call the system (\ref{eq:I4_Curv:system}),   the \textit{curved coupled Riccati equations}.

  From  \eqref{eq:I4_Curv:comm} we see that $\X_{\kappa}$ is a Lie system with VG Lie algebra  $V^{X_{\kk}}  \simeq \so(2, 1)$ spanned by the vector fields \eqref{eq:I4_Curv:vf}. Furthermore, they are Hamiltonian vector fields relative to the symplectic form
 \begin{equation}
 \omega_{\kappa} = \frac{1}{4 \Sin_{\kappa}^{2} \bigl( \frac{1}{2} (x-y) \bigr)} \,\dd x \wedge \dd y,
 \label{eq:I4_Curv:symplectic}
 \end{equation}
so fulfilling (\ref{LieDer}),  and their associated Hamiltonian functions, determined by the condition  (\ref{condition}), turn out to be
 \begin{equation}
 h_{\kappa, 1} = \frac{1}{2\Tan_{\kappa}\bigl( \frac{1}{2} (x-y) \bigr)}, \qquad h_{\kappa, 2} = \frac{\Sin_{\kappa} \bigl( \frac{1}{2} (x+y) \bigr)}{2 \Sin_{\kappa} \bigl(\frac{1}{2}(x-y) \bigr)}, \qquad h_{\kappa,3} = \frac{2 \Sin_{\kappa} \bigl(\frac 12 x\bigr)\Sin_{\kappa} \bigl(\frac 12 y \bigr)}{\Sin_{\kappa} \bigl( \frac{1}{2} (x-y) \bigr)}.
 \label{eq:I4_Curv:Ham}
 \end{equation}
 With respect to the Poisson bracket $\set{\cdot, \cdot}_{\omega_{\kappa}}$ induced by the symplectic structure \eqref{eq:I4_Curv:symplectic}, the brackets of the  Hamiltonian functions \eqref{eq:I4_Curv:Ham} are
 \begin{equation}
 \set{h_{\kappa,1}, h_{\kappa, 2}}_{\omega_{\kappa}} = - h_{\kappa, 1} + \tfrac{1}{2} \kappa\, h_{\kappa, 3}, \qquad \set{h_{\kappa, 1}, h_{\kappa,3}}_{\omega_{\kappa}} = - 2 h_{\kappa,2}, \qquad \set{h_{\kappa, 2}, h_{\kappa, 3}}_{\omega_{\kappa}} = - h_{\kappa, 3},
 \label{eq:I4_Curv:brackets}
 \end{equation}
 showing that $\X_{\kappa}$ is a LH system on $\S^{1}_{[\kappa]} \times \S^{1}_{[\kappa]}$ with LH algebra $\cH_{\omega_{\kappa}} \simeq \so(2, 1)$ spanned by the Hamiltonian functions \eqref{eq:I4_Curv:Ham}.

Observe that the product manifold $\S^{1}_{[\kappa]} \times \S^{1}_{[\kappa]}$ is naturally endowed with the product Riemannian metric $\dd x^{2} + \dd y^{2}$, (which is the product metric \eqref{eq:CK1:metric}) showing that the 2D space $\S^{1}_{[\kappa]} \times \S^{1}_{[\kappa]}$ is flat. In particular, for $\kappa > 0$ we obtain the flat $2$-torus $\T^{2} = \S^{1} \times \S^{1}$, while the case $\kappa <0$ gives rise to the product of two hyperbolic lines $\H^{1} \times \H^{1}$. The Euclidean plane $\E^{2}$,  corresponding to $\kappa  = 0$,  is the only 2D CK space which is a product of 1D CK spaces.

  Hence, we have obtained a LH class on the space $\S^{1}_{[\kappa]} \times \S^{1}_{[\kappa]}$ which generalizes the Euclidean  class I$_{4}$  to $\S^{1}_{[\kappa]} \times \S^{1}_{[\kappa]}$, as the vector fields,   symplectic form and   Hamiltonian functions of the Euclidean I$_{4}$-LH class, shown in Table~\ref{table1}, are recovered after the contraction $\kappa = 0$ from \eqref{eq:I4_Curv:vf}, \eqref{eq:I4_Curv:symplectic} and \eqref{eq:I4_Curv:Ham}, respectively. We call such a LH class on $\S^{1}_{[\kappa]} \times \S^{1}_{[\kappa]}$ the \textit{curved I$_{4}$-LH class}, which is explicitly described in Table~\ref{table2} for each   of the three spaces  $\S^{1}_{[\kappa]} \times \S^{1}_{[\kappa]}$. Note that the domain of the coordinates $(x,y)$ requires the condition $x\ne y$ in the three spaces, ensuring  a well-defined symplectic form and Hamiltonian functions.


\begin{table}[t]
{\small
 \noindent
\caption{{\small The curved I$_{4}$-LH class on the three flat spaces $\S^{1}_{[\kappa]} \times \S^{1}_{[\kappa]}$ according to the `normalized' values $\kappa \in \set{-1, 0, 1}$.  For each space   it is shown, in coordinates $(x,y)$,  its Riemannian metric, domain of the variables, VG Lie algebra $V^{X_{\kappa}}$ \eqref{eq:I4_Curv:comm} with vector fields $\X_{\kappa, i}$ \eqref{eq:I4_Curv:vf}, LH algebra $\cH_{\omega_{\kappa}}$ \eqref{eq:I4_Curv:brackets} with Hamiltonian functions $h_{\kappa, i}$ \eqref{eq:I4_Curv:Ham} and   symplectic form $\omega_{\kappa}$ \eqref{eq:I4_Curv:symplectic}. For the sake of clarity, we drop the index $\kappa$.}}
\label{table2}
\medskip
\noindent\hfill
\begin{tabular}{lll}
\hline

\hline
\\[-6pt]
$\bullet$ Flat $2$-torus $\T^{2} = \S^{1} \times \S^{1}$& $\bullet$ Euclidean plane $\E^{2}$
  &$\bullet$ Product of hyperbolic lines $\H^{1} \times \H^{1}$  \\[4pt]
$\dd s^2= \dd x^{2} + \dd y^{2}$ &
 $\dd s^2= \dd x^{2} + \dd y^{2}$ &
$\dd s^2= \dd x^{2} + \dd y^{2}$ \\[2pt]
$x \in (-\pi,\pi],\  y\in (-\pi,\pi],\ x\ne y$&
$x \in \R,\ y \in \R,\ x\ne y$&
$x \in \R,\ y \in \R,\ x\ne y$
\\[5pt]

$V^{X} \simeq \so(2, 1)$ & $V^{X} \simeq \so(2, 1)$ & $V^{X} \simeq \so(2, 1)$ \\[2pt]
$\X_1=\partial_{x} + \partial_{y}$ & $\X_1=\partial_{x} + \partial_{y}$ & $\X_1=\partial_{x} + \partial_{y}$\\[2pt]
$ \X_{2} = \sin x \, \partial_{x} + \sin y \, \partial_{y}$ & $ \X_{2} = x \partial_{x} + y \partial_{y} $ & $ \X_{2} = \sinh x \, \partial_{x} + \sinh y \, \partial_{y}$ \\[2pt]
$\X_{3}=2 \bigl((1 - \cos x) \partial_{x} + (1 - \cos y) \partial_{y} \bigr)$ & $\X_{3} = x^{2} \partial_{x} + y^{2} \partial_{y}$ & $\X_{3} = 2\bigl((\cosh x -1) \partial_{x} + (\cosh y - 1) \partial_{y}\bigr) $
\\[5pt]

$\cH_{\omega} \simeq \so(2, 1)$ & $\cH_{\omega} \simeq \so(2, 1)$ & $\cH_{\omega} \simeq \so(2, 1)$ \\[3pt]
$\displaystyle  h_{1}= \frac{1}{2\tan (\frac{1}{2} (x-y)) } $ & $ \displaystyle  h_{1}=  \frac{1}{x-y} $ & $\displaystyle   h_{1}=\frac{1}{2 \tanh (\frac{1}{2} (x-y)) }  $\\[12pt]
$ \displaystyle  h_{2}= \frac{\sin ( \frac{1}{2} (x+y) )}{2 \sin (\frac{1}{2} (x-y))}$ & $\displaystyle h_{2} = \frac{x+y}{2(x-y)} $ & $ \displaystyle  h_{2}= \frac{\sinh (\frac{1}{2} (x+y))}{2 \sinh (\frac{1}{2} (x-y))}$\\[12pt]
 $ \displaystyle h_{3}=\frac{2 \sin (\frac 12 x) \sin(\frac 12 y)}{\sin(\frac{1}{2}(x-y))}  $ &  $ \displaystyle h_{3} = \frac{xy}{x-y}$ & $ \displaystyle  h_{3}=\frac{2 \sinh (\frac 12 x) \sinh(\frac 12 y)}{\sinh(\frac{1}{2}(x-y))}    $ \\[12pt]
 $ \displaystyle  \omega = \frac{1}{4 \sin^{2}(\frac{1}{2} (x-y))}\, \dd x \wedge \dd y$&
$ \displaystyle \omega = \frac{1}{(x-y)^{2}}\, \dd x \wedge \dd y$&
$\displaystyle  \omega = \frac{1}{4 \sinh^{2}(\frac{1}{2} (x-y))}\, \dd x \wedge \dd y$\vspace*{0.2cm}\\[6pt]
\hline

\hline

\end{tabular}\hfill}
\end{table}


It is also worth emphasizing that the contraction parameter $\kk$ can alternatively be  interpreted as a perturbation parameter of
the Euclidean class I$_{4}$. In particular, if $\kk$ is considered to have a small value, we can take a power series expansion in $\kk$ of
the system  (\ref{eq:I4_Curv:system}). For instance, by truncating at the first-order in $\kk$, we obtain the following approximation:
\be
 \dv{x}{t} = b_{1}(t) + b_{2}(t)x + b_{3}(t)x^2-\frac 1{12}\kk  \left( 2 b_2(t) x^3+  b_3(t)x^4\right) + o[\kk^2],
 \label{aprox}
 \ee
(and similarly for the second equation in $y$) which can be interpreted as a fourth-order  perturbation of the Riccati equation (\ref{flatriccati}).
 In this sense, the introduction of $\kk$ can be regarded as the opposite process to the contraction procedure, i.e.~a classical deformation. This idea is quite similar to Poisson--Hopf deformations of LH systems, for which the quantum deformation parameter can always be thought of as a perturbation  (see, e.g.,~\cite[eq. (4.19)]{Ballesteros2021} for an example).


 \subsection{Constants of  the motion   of the curved I$_{4}$-class}
 \label{s31}

 We now proceed to compute $t$-independent constants of motion of the LH system $\X_{\kappa}$ \eqref{eq:I4_Curv:tvf}   by applying the so-called  {coalgebra formalism} \cite{Ballesteros2013,Ballesteros2021}. Let us consider the LH algebra $\cH_{\omega_{\kappa}}$ of the LH system $\X_{\kappa}$ expressed in a basis $\set{v_{1}, v_{2}, v_{3}}$ that formally satisfies the same Poisson brackets shown in \eqref{eq:I4_Curv:brackets}:
 \be
  \set{v_{1}, v_{2}} = - v_{1} + \tfrac{1}{2}\kk \, v_{3}, \qquad \set{v_{1}, v_{3}} = - 2 v_{2}, \qquad \set{v_{2}, v_{3}} = - v_{3}. \nonumber
  \ee
 There exists a non-trivial quadratic Casimir element given by
 \begin{equation}
 C= v_{1}v_{3} - v_{2}^{2} - \tfrac{1}{4} \kappa\, v_{3}^{2}, \qquad \set{C, \cdot} = 0.
 \label{eq:I4_Curv:Casimir}
 \end{equation}
 As a shorthand notation, let us denote by $\mathcal{M}_{[\kk]}$ the submanifold of $\S^{1}_{[\kappa]} \times \S^{1}_{[\kappa]}$ with $x\ne y$.  From \eqref{eq:I4_Curv:Ham} we construct the following Hamiltonian functions $( 1 \leq i \leq 3)$:
 \begin{equation}
 \begin{split}
h_{\kappa, i}^{(1)} &= h_{\kappa,i}(\x_{1}) \in C^{\infty}\bigl(\mathcal{M}_{[\kk]}\bigr), \\
h_{\kappa, i}^{(2)} &= h_{\kappa,i}(\x_{1}) + h_{\kappa, i}(\x_{2})  \in C^{\infty}\bigl(\mathcal{M}_{[\kk]}\times \mathcal{M}_{[\kk]}\bigr),\\
h_{\kappa, i}^{(3)} &= h_{\kappa,i}(\x_{1}) + h_{\kappa, i}(\x_{2}) + h_{\kappa, i}(\x_{3})  \in C^{\infty}\bigl(\mathcal{M}_{[\kk]}\times \mathcal{M}_{[\kk]}\times \mathcal{M}_{[\kk]}\bigr),
\end{split}
\label{eq:I4_Curv:Ham_pro}
 \end{equation}
 where $\x_{s} = (x_{s}, y_{s})$ denotes the coordinates in the $s^{\textup{th}}$-copy of  $\mathcal{M}_{[\kk]}$ within the product manifold. Note that each set of functions of \eqref{eq:I4_Curv:Ham_pro} so obtained satisfies the Poisson brackets \eqref{eq:I4_Curv:brackets} with respect to the Poisson bracket induced by taking the product symplectic structure $\omega_{\kappa}^{[3]}$ of \eqref{eq:I4_Curv:symplectic} to $\mathcal{M}_{[\kk]}^{3}\equiv \mathcal{M}_{[\kk]}\times \mathcal{M}_{[\kk]}\times \mathcal{M}_{[\kk]}$, that is,
 \be
 \omega_{\kappa}^{[3]} = \frac 14\sum_{s=1}^{3}\frac{1}{ \Sin_{\kappa}^{2} \bigl( \frac{1}{2} (x_s-y_s) \bigr)} \,\dd x_s\wedge \dd y_s .\nonumber
 \ee
  Now, using the Casimir \eqref{eq:I4_Curv:Casimir} we obtain the following constants of motion for the diagonal prolongation $\widetilde{\X}_{\kappa}^{3}$ of $\X_{\kappa}$ \eqref{eq:I4_Curv:tvf} to $\mathcal{M}_{[\kk]}^{3}$:
 \begin{equation}
 \begin{split}
 F_{\kappa}^{(1)} &= C\! \left( h_{\kappa,1}^{(1)}, h_{\kappa, 2}^{(1)}, h_{\kappa,3}^{(1)} \right) = - \frac{1}{4}, \\
  F_{\kappa}^{(2)} &= C\! \left( h_{\kappa,1}^{(2)}, h_{\kappa, 2}^{(2)}, h_{\kappa,3}^{(2)} \right)  =- \frac{\Sin_{\kappa}\bigl(\frac{1}{2}(x_{2} -y_{1}) \bigr) \Sin_{\kappa}\bigl(\frac{1}{2}(x_{1} -y_{2}) \bigr)}{\Sin_{\kappa}\bigl(\frac{1}{2}(x_{1} -y_{1}) \bigr)\Sin_{\kappa}\bigl(\frac{1}{2}(x_{2} -y_{2}) \bigr)} .
 \end{split}
 \label{eq:I4_Curv:constants}
 \end{equation}
Observe that  $F_{\kappa}^{(1)} <0$ for any value of $\kk$, so this corresponds  to $\mathcal C<0$ (\ref{casconstant}),  with $c=-1$  as shown in Table~\ref{table1} for the Euclidean class I$_{4}$. This fact will distinguish a specific LH system belonging to the \textit{curved I$_{4}$-LH class} from other within the  \textit{curved P$_{2}$-LH class}, with $\mathcal C>0$,  that we will develop in Section~\ref{s5}.

Note also that the well-known constant of motion of the Euclidean class I$_{4}$~\cite{Winternitz1983,Ballesteros2013,Blasco2015,Carinena2007} is recovered under the  contraction $\kk=0$:
\be
  F_{0}^{(2)}=- \frac{ (x_{2} -y_{1}  ) (x_{1} -y_{2})}{(x_{1} -y_{1})(x_{2} -y_{2}) } ,
  \label{eq:I4_Flat:constants}
\ee
corresponding to the set of coupled Riccati equations (\ref{flatriccati}).

Moreover,   $F_{\kappa}^{(2)}$ gives rise to  two additional constants of motion through the permutation $S_{ij}$ of the variables $(x_{i}, y_{i}) \leftrightarrow (x_{j}, y_{j})$; these are
\begin{equation}
\begin{split}
F_{\kappa,13}^{(2)} & = S_{13}\!\left(F_{\kappa}^{(2)} \right) =  - \frac{\Sin_{\kappa} (\frac{1}{2}(x_{2} -y_{3})) \Sin_{\kappa}(\frac{1}{2}(x_{3} -y_{2}))}{\Sin_{\kappa}(\frac{1}{2}(x_{3} -y_{3}))\Sin_{\kappa}(\frac{1}{2}(x_{2} -y_{2}))} ,\\
 F_{\kappa,23}^{(2)} &= S_{23}\!\left(F_{\kappa}^{(2)}\right) = - \frac{\Sin_{\kappa}(\frac{1}{2}(x_{3} -y_{1})) \Sin_{\kappa}(\frac{1}{2}(x_{1} -y_{3}))}{\Sin_{\kappa}(\frac{1}{2}(x_{1} -y_{1}))\Sin_{\kappa}(\frac{1}{2}(x_{3} -y_{3}))}.
\end{split}
\label{eq:I4_Curv:constantsB}
\end{equation}
The functions $h_{\kappa, i}^{(3)} $  (\ref{eq:I4_Curv:Ham_pro}) also allow to obtain another constant of motion which can be expressed in terms of (\ref{eq:I4_Curv:constants}) and (\ref{eq:I4_Curv:constantsB}) as
\be
F_{\kappa}^{(3)}  = C\! \left( h_{\kappa,1}^{(3)}, h_{\kappa, 2}^{(3)}, h_{\kappa,3}^{(3)} \right)  =F_{\kappa}^{(2)}+F_{\kappa,13}^{(2)} + F_{\kappa,23}^{(2)} +\frac 34 .
\label{eq:I4_Curv:constantsC}\nonumber
\ee


 \subsection{Superposition rules for the curved I$_{4}$-class}
 \label{s32}

So far we have obtained four $t$-independent constants of motion $\bigl\{F_{\kappa}^{(2)} ,
F_{\kappa,13}^{(2)},F_{\kappa,23}^{(2)},F_{\kappa}^{(3)} \bigr\}$ for the curved I$_{4}$-LH system $\X_{\kappa}$ \eqref{eq:I4_Curv:tvf}. To derive a superposition rule in this case, it is necessary to choose two functionally independent constants of motion, say $I_1$, $I_2$,  for the  diagonal prolongation $\widetilde{\X}_{\kappa}^{3}$ of $\X_{\kappa}$ to $\mathcal{M}_{[\kk]}^{3}$, so verifying the condition
\be
\frac {\partial(I_1,I_2)}   {\partial(x_{1}, y_{1})} \neq 0 .
\label{a1}
\ee
This ensures  that it is possible to express  the general solution $(x_{1}(t), y_{1}(t))$ of the LH system $\X_{\kappa}$ in terms of two particular solutions $(x_{2}(t), y_{2}(t))$ and $(x_{3}(t), y_{3}(t))$ and two significative constants, say $\mm_1$ and $\mm_2$,  by solving the equations $I_1=\mm_1$
and $I_2=\mm_2$  (see~\cite{Blasco2015,Carinena2007} for details).

There are several possible options for the constants $I_1$ and $I_2$. We take
$F_{\kappa}^{(2)} $ and $F_{\kappa,23}^{(2)}$ and then use $F_{\kappa,13}^{(2)}$ in order to simplify the final result, thus we set
\be
F_{\kappa}^{(2)}  = -\mm_1,\qquad F_{\kappa,23}^{(2)}=-\mm_2,\qquad F_{\kappa,13}^{(2)}=- \mm_3 .
\label{a2}
\ee
For the computations, we rewrite the expression for  $F_{\kappa}^{(2)} $ (\ref{eq:I4_Curv:constants})  (see the Appendix) as
\be
 F_{\kappa}^{(2)}  = -\left( \frac{  \Tan_{\kappa}\bigl(\frac{1}{2}x_{2} \bigr)  - \Tan_{\kappa}\bigl(\frac{1}{2}y_{1} \bigr)  }{ \Tan_{\kappa}\bigl(\frac{1}{2}x_{1} \bigr)  - \Tan_{\kappa}\bigl(\frac{1}{2}y_{1} \bigr) } \right)\left( \frac{  \Tan_{\kappa}\bigl(\frac{1}{2}x_{1} \bigr)  - \Tan_{\kappa}\bigl(\frac{1}{2}y_{2} \bigr)  }{ \Tan_{\kappa}\bigl(\frac{1}{2}x_{2} \bigr)  - \Tan_{\kappa}\bigl(\frac{1}{2}y_{2} \bigr) }\right) ,
 \label{a3}
\ee
and similarly for $F_{\kappa,13}^{(2)} $ and $F_{\kappa,23}^{(2)}$ (\ref{eq:I4_Curv:constantsB}).
Starting from the first equation in (\ref{a2}) we find that
 \be
 \Tan_\kk\bigl(\tfrac12 x_1\bigr)=\frac{\mm_1\! \left(\! \Tan_\kk\bigl(\tfrac12 x_2\bigr)-\Tan_\kk\bigl(\tfrac12 y_2\bigr) \right)\!\Tan_\kk\bigl(\tfrac12 y_1\bigr)-\left(\!\Tan_\kk\bigl(\tfrac12 x_2\bigr)-\Tan_\kk\bigl(\tfrac12 y_1\bigr)\right)\!\Tan_\kk\bigl(\tfrac12 y_2\bigr)}{\mm_1\!\left(\!\Tan_\kk\bigl(\tfrac12 x_2\bigr)-\Tan_\kk\bigl(\tfrac12 y_2\bigr)\right) -\left(\!\Tan_\kk\bigl(\tfrac12 x_2\bigr)-\Tan_\kk\bigl(\tfrac12 y_1\bigr)\right) } .
 \label{a4}
 \ee
 Substituting this result into the second equation  in (\ref{a2}) and  also introducing the third constant $\mm_3$  we arrive at the full superposition principle for  $ y_1$, namely
\be
\begin{split}
\Tan_\kk\bigl(\tfrac12 y_1^\pm\bigr)&=\frac 12 \left\{
\left(\! \Tan_\kk\bigl(\tfrac12 x_2\bigr)+\Tan_\kk\bigl(\tfrac12 x_3\bigr)\right)\left(\! \Tan_\kk\bigl(\tfrac12 y_2\bigr)-\Tan_\kk\bigl(\tfrac12 y_3\bigr)\right) \right.\\
&\qquad\quad +\mm_1\left(\! \Tan_\kk\bigl(\tfrac12 x_2\bigr)-\Tan_\kk\bigl(\tfrac12 y_2\bigr)\right ) \left(\! \Tan_\kk\bigl(\tfrac12 x_3\bigr)+\Tan_\kk\bigl(\tfrac12 y_3\bigr) \right)\\
&\qquad\quad  \left. -\mm_2\left(\! \Tan_\kk\bigl(\tfrac12 x_3\bigr)-\Tan_\kk\bigl(\tfrac12 y_3\bigr)\right)\left(\! \Tan_\kk\bigl(\tfrac12 x_2\bigr)+\Tan_\kk\bigl(\tfrac12 y_2\bigr)\right)\pm\sqrt{\Xi}\, \right\} \\
&\quad \times \left\{  \mm_1\left(\! \Tan_\kk\bigl(\tfrac12 x_2\bigr)-\Tan_\kk\bigl(\tfrac12 y_2\bigr) \right)- \mm_2\left(\! \Tan_\kk\bigl(\tfrac12 x_3\bigr)-\Tan_\kk\bigl(\tfrac12 y_3\bigr)\right)+\left(\! \Tan_\kk\bigl(\tfrac12 y_2\bigr)-\Tan_\kk\bigl(\tfrac12 y_3\bigr) \right) \right\}^{-1} \\[2pt]
\end{split}
\label{a5}
\ee
 (so with two  solutions according to the sign of the square root), where
\be
\begin{split}
\Xi&=\left(\! \Tan_\kk\bigl(\tfrac12 x_2\bigr)-\Tan_\kk\bigl(\tfrac12 x_3\bigr) \right)^2 \left(\! \Tan_\kk\bigl(\tfrac12 y_2\bigr)-\Tan_\kk\bigl(\tfrac12 y_3\bigr)\right)^2\\
&\qquad \quad+(\mm_1^2+\mm_2^2-2 \mm_1\mm_2\mm_3)\left(\! \Tan_\kk\bigl(\tfrac12 x_2\bigr)-\Tan_\kk\bigl(\tfrac12 y_2\bigr)\right)^2\left(\! \Tan_\kk\bigl(\tfrac12 x_3\bigr)-\Tan_\kk\bigl(\tfrac12 y_3\bigr)\right)^2\\
&\qquad \quad   -2 (\mm_1+\mm_2-\mm_1\mm_2)\left(\! \Tan_\kk\bigl(\tfrac12 x_2\bigr)-\Tan_\kk\bigl(\tfrac12 y_2\bigr)\right)\left(\! \Tan_\kk\bigl(\tfrac12 x_3\bigr)-\Tan_\kk\bigl(\tfrac12 y_3\bigr)\right)\\
&\qquad \qquad\times \left(\! \Tan_\kk\bigl(\tfrac12 x_2\bigr)-\Tan_\kk\bigl(\tfrac12 x_3\bigr)\right)\left(\! \Tan_\kk\bigl(\tfrac12 y_2\bigr)-\Tan_\kk\bigl(\tfrac12 y_3\bigr)\right).
\end{split}
\label{a6}\nonumber
\ee
Note that the result (\ref{a5}) is well-defined whenever the function $\Xi$ is non-negative, which requires some restrictions on the parameters $\mu_i$  determined by the specific particular solutions under consideration. Hereafter, this fact will be assumed when a square root appears in a superposition rule (as in Subsection~\ref{s52}). Finally, 
inserting the above expressions into (\ref{a4}), we obtain the complete superposition rule for the curved I$_{4}$-LH system $\X_{\kappa}$ \eqref{eq:I4_Curv:tvf} as the general solution $(x_1^\pm(t),y_1^\pm(t))$  with the following dependence
\be
x_1^\pm(x_2,y_2,x_3,y_3,\mm_1,\mm_2,\mm_3),\qquad y_1^\pm(x_2,y_2,x_3,y_3,\mm_1,\mm_2,\mm_3),\qquad
\Xi(x_2,y_2,x_3,y_3,\mm_1,\mm_2,\mm_3),\nonumber
\ee
 such that $\mm_3=\mm_3(x_2,y_2,x_3,y_3)$ through $F_{\kappa,13}^{(2)}  = S_{13}(F_{\kappa}^{(2)})=-\mm_3$ (\ref{a3}).

 The superposition rule for the curved I$_{4}$-LH class is rather simplified  for the  system with $\kk=0$, yielding
 \be
\begin{split}
y_1^\pm&=\frac{(x_2+x_3)(y_2-y_3)+\mm_1(x_2-y_2)(x_3+y_3)-\mm_2(x_3-y_3)(x_2+y_2)\pm\sqrt{\Xi}}
{2\bigl(\mm_1(x_2-y_2)- \mm_2(x_3-y_3)+(y_2-y_3) \bigr)},\\[2pt]
\Xi&=(x_2-x_3)^2 (y_2-y_3)^2+(\mm_1^2+\mm_2^2-2 \mm_1\mm_2\mm_3)(x_2-y_2)^2(x_3-y_3)^2\\
&\qquad\qquad  -2 (\mm_1+\mm_2-\mm_1\mm_2)(x_2-y_2)(x_3-y_3)(x_2-x_3)(y_2-y_3),\\[2pt]
 x_1^\pm&=\frac{\mm_1(x_2-y_2)y_1^\pm-(x_2-y_1^\pm)y_2}{\mm_1(x_2-y_2) -(x_2-y_1^\pm) },
\end{split}
\label{a8}
\ee
  where $\mm_3=\mm_3(x_2,y_2,x_3,y_3)$ now via $F_{0,13}^{(2)}  = S_{13}\bigl(F_{0}^{(2)} \bigr) =-\mm_3$ (\ref{eq:I4_Flat:constants}).

 It is worth observing that, to the best of our knowledge, not only the curved LH systems constructed throughout  this section together with their constants of  the motion  and superposition principles  constitute  new results, but also the superposition rule (\ref{a8}) for the  Euclidean coupled Riccati equations (\ref{flatriccati}), thus completing previous  works (see \cite{Ballesteros2013} and references therein).

    Furthermore, recall that a single Riccati equation, take only the first one in (\ref{flatriccati})    with   variable $ x$, is a Lie system with VG Lie algebra isomorphic to $ \sl(2, \R)$, but it is not a LH system.  This system appears in the classification of Lie systems on the plane~\cite{Ballesteros2015} as the class I$_{3}$ for which there does not exist an associated symplectic form, so that the application of the coalgebra formalism to obtain constants of motion and superposition principles is precluded.
  Nevertheless,  the well-known superposition rule for the Riccati equation~\cite{Winternitz1983,Carinena2007} can easily be derived from the last equation in (\ref{a8}). Let us denote by $\X^{\rm Ric}$ the $t$-dependent vector field  for  the Riccati equation spanned by the projection $y=0$ of the conformal vector fields (\ref{eq:I4_Euc:vf})   with variable $\tilde x\in \R$.  Under the following identification
\be
x_1\equiv \tilde x,\qquad     y_1\equiv  \tilde x_3,\qquad
 x_2\equiv  \tilde x_2,\qquad  y_2\equiv  \tilde x_1 ,
 \label{a9}
\ee
 we retrieve from (\ref{a8}) the general solution of the Riccati equation $\tilde x(t)$   expressed in terms of three particular solutions $(\tilde x_1(t),\tilde x_2(t),\tilde x_3(t))$ and a single constant $\mm_1$ as
\be
\tilde x =\frac{\mm_1(\tilde x_2-\tilde x_1)\tilde x_3+(\tilde x_3-\tilde x_2)\tilde x_1}{\mm_1(\tilde x_2-\tilde x_1) +(\tilde x_3-\tilde x_2) }.
 \label{a10}
\ee
This result is, in fact, provided by $ F_{0}^{(2)}=-\mm_1$ (\ref{eq:I4_Flat:constants}) which is now understood as a constant of motion for the diagonal prolongation $\widetilde{\X}^{\rm Ric}$    of $\X^{\rm Ric}$   to  $\R^4$.

 Likewise, we will call the first equation of  ({\ref{eq:I4_Curv:system}) the {\em curved  Riccati equation}, with variable $x\in\S^{1}_{[\kappa]}$, whose corresponding $t$-dependent vector field $\X_\kk^{\rm Ric}$
 is spanned by the projection $y=0$ of (\ref{eq:I4_Curv:vf}) (see~(\ref{eq:Conf:CK1})). This is again  a Lie system but not a LH one, since no symplectic form on $\S^{1}_{[\kappa]}$ exists.  However, a superposition rule for $\X_\kk^{\rm Ric}$  is provided by (\ref{a4}) under the identification (\ref{a9}),  determining the general solution $\tilde x(t)$:
  \be
 \Tan_\kk\bigl(\tfrac12 \tilde x\bigr)=\frac{\mm_1\! \left(\! \Tan_\kk\bigl(\tfrac12 \tilde x_2\bigr)-\Tan_\kk\bigl(\tfrac12 \tilde x_1\bigr) \right)\!\Tan_\kk\bigl(\tfrac12 \tilde x_3\bigr)+\left(\!\Tan_\kk\bigl(\tfrac12 \tilde x_3\bigr)-\Tan_\kk\bigl(\tfrac12 \tilde x_2\bigr)\right)\!\Tan_\kk\bigl(\tfrac12 \tilde x_1\bigr)}{\mm_1\!\left(\!\Tan_\kk\bigl(\tfrac12 \tilde x_2\bigr)-\Tan_\kk\bigl(\tfrac12 \tilde x_1\bigr)\right) +\left(\!\Tan_\kk\bigl(\tfrac12 \tilde x_3\bigr)-\Tan_\kk\bigl(\tfrac12 \tilde x_2\bigr)\right) } ,\nonumber
 \ee
which is reduced to (\ref{a10}) for the  contraction $\kk=0$. In this curved case, $F_{\kappa}^{(2)} =-\mm_1$ (\ref{a3})  is  a constant of the motion for the diagonal prolongation $\widetilde{\X}_\kk^{\rm Ric}$    of $\X_\kk^{\rm Ric}$   to  $\bigl(\S^{1}_{[\kappa]} \bigr) ^4$.


 \section{Applications of the curved I$_{4}$-class}
  \label{s4}

The Euclidean I$_{4}$-LH class is known to include the following relevant systems \cite{Blasco2015}: the coupled Riccati equations, the split-complex Riccati equation, a planar diffusion Riccati system, and the Kummer--Schwarz and Ermakov equations for $c < 0$ (\ref{casconstant}). The curved counterpart of the  coupled Riccati equations has already been studied, since they  appear directly from the conformal vector fields (\ref{eq:I4_Curv:vf}).  Consequently, by making use of the curved I$_{4}$-LH class   presented previously,  we   generalize in this section the remaining particular LH systems to the 2D   spaces $\S^{1}_{[\kappa]} \times \S^{1}_{[\kappa]}$.  In this way,   the    known Euclidean LH systems are extended to the  flat $2$-torus $\T^{2} = \S^{1} \times \S^{1}$ ($\kk>0$) and the
 product of two hyperbolic lines $\H^{1} \times \H^{1}$ ($\kk<0$).

The procedure requires to obtain an appropriate  local diffeomorphism between the specific LH system on $\S^{1}_{[\kappa]} \times \S^{1}_{[\kappa]}$ under consideration and the curved I$_{4}$-LH class of Section~\ref{s3}. Recall that, as local diffeomorphisms between open subsets of $\S^{1}_{[\kappa]} \times \S^{1}_{[\kappa]}$ are induced from local diffeomorphisms between open subsets of the domain of the parametrization \eqref{CK1}, the curved analogues of the systems of the Euclidean I$_{4}$-LH class can be constructed by applying local diffeomorphisms of the Euclidean case to the domain of the parametrization \eqref{CK1}.

 Once this is achieved, the corresponding constants of the motion and  the superposition principle can be written directly from the results of Subsections~\ref{s31} and \ref{s32}, respectively, although for the sake of brevity we omit these expressions. Nevertheless, we point out that  in all applications the first-order constant of motion is $F_{\kappa}^{(1)}   = - \frac{1}{4}$ (\ref{eq:I4_Curv:constants}), which is a direct consequence of taking a negative value for the constant $c$ (\ref{casconstant}) (see Table~\ref{table1}).


 \subsection{Curved split-complex Riccati equation}
   \label{s41}

On the Euclidean plane $\E^{2}$ with coordinates $(u, v)$, consider a  hypercomplex  unit $\iota$ which commutes with real numbers and such that $\iota^{2} \in \set{-1, +1, 0}$ and define the hypercomplex number $z := u + \iota v$~\cite{Yaglom1979,Gromov1990,Kisil2012}.
  The differential equation
\be
\dv{z}{t} = b_{1}(t) + b_{2}(t)z + b_{3}(t) z^{2}
\label{b1}
\ee
 is known as the {\em Cayley--Klein Riccati equation}~\cite{Blasco2015, Estevez2016}, and is equivalent to the following first-order system of differential equations on $\E^{2}$:
 \begin{equation}
 \dv{u}{t} = b_{1}(t) + b_{2}(t) u + b_{3}(t) \bigl(u^{2} + \iota^{2} v^{2}\bigr), \qquad  \dv{v}{t} = b_{2}(t) v + 2b_{3}(t) uv.
 \label{eq:CKRiccati_Euc}
 \end{equation}

If $\iota^{2} = +1$, we obtain the so-called split-complex numbers $z\in \mathbb C'$  and  $\iota$  is   usually referred to  as the hyperbolic,   double or Clifford unit~\cite{Yaglom1979,Kisil2012}.  In this case, the system \eqref{eq:CKRiccati_Euc}  reads
 \begin{equation}
 \dv{u}{t} = b_{1}(t) + b_{2}(t) u + b_{3}(t) \bigl(u^{2} +  v^{2}\bigr), \qquad \dv{v}{t} = b_{2}(t) v +2 b_{3}(t) uv,
 \label{eq:SCRiccati:Euc:system}
 \end{equation}
which comes from the so-called {\em split-complex Riccati equation} (\ref{b1}). It is associated with the $t$-dependent vector field  $\X = b_{1}(t) \X_{1} + b_{2}(t) \X_{2} +b_{3}(t) \X_{3}$,
where
\be
 \X_{1} = \pdv{u}, \qquad \X_{2} = u \pdv{u} + v \pdv{v}, \qquad \X_{3} = \bigl(u^{2} + v^{2}\bigr) \pdv{u} + 2 uv \pdv{v}  ,\nonumber
 \ee
satisfying the  commutation relations (\ref{sl2}). In addition, these are Hamiltonian vector fields  relative to the symplectic form given by
\begin{equation}
\omega=-\frac{1}{ 2 v^2}  \,{\rm d} u \wedge {\rm d} v,\nonumber
\end{equation}
and their associated Hamiltonian functions are
\begin{equation}
 h_1=\frac{1}{2 v} ,\qquad
h_2=  \frac{u}{2 v}  ,\qquad
h_3=\frac{u^2-v^2}{2 v}.\nonumber
\end{equation}
All the above expressions are  mapped into those of the Euclidean I$_{4}$-LH class displayed in Table~\ref{table1}  via the change of variables~\cite{Blasco2015}
\begin{equation}
x := u +v, \qquad y := u-v, \qquad u = \tfrac{1}{2}(x+y), \qquad v = \tfrac{1}{2}(x-y).
\label{eq:SCRiccati:change}
\end{equation}
Observe that the domain for each set of variables is ${\R}^2_{x\ne y} $  and ${\R}^2_{v\ne 0}$.

Within the framework  established in Section~\ref{s3}, we apply the transformation \eqref{eq:SCRiccati:change} to the curved I$_{4}$-LH
class (see Table~\ref{table2}),  so now defined on $ \S^{1}_{[\kappa]} \times \S^{1}_{[\kappa]}$, thus obtaining the $t$-dependent vector field \eqref{eq:I4_Curv:tvf}   with    vector fields, coming from (\ref{eq:I4_Curv:vf}),    given by
 \begin{equation}
 \begin{split}
  & \X_{\kappa,1} =  \pdv{u}, \qquad \X_{\kappa,2} = \Cos_{\kappa}(v) \Sin_{\kappa}(u) \pdv{u} + \Cos_{\kappa}(u) \Sin_{\kappa}(v) \pdv{v}, \\[2pt]
  & \X_{\kappa, 3} = 2\left(  \frac{1-\Cos_{\kappa}(u) \Cos_{\kappa}(v)}{\kk}\right) \pdv{u} +  2  \Sin_{\kappa}(u) \Sin_{\kappa}(v) \pdv{v} .
 \end{split}
 \label{eq:SCRiccati:Curv:vf}
 \end{equation}
 They span a VG Lie algebra $V^{X_{\kappa}} \simeq \so(2, 1)$, since they commutators are those of \eqref{eq:I4_Curv:comm}. The corresponding first-order system of ODEs on $\S^{1}_{[\kappa]} \times \S^{1}_{[\kappa]}$  turns out to be
 \begin{equation}
 \begin{split}
 \dv{u}{t} &= b_{1}(t) + b_{2}(t) \Cos_{\kappa}(v) \Sin_{\kappa}(u) + 2 b_{3}(t) \left(  \frac{1-\Cos_{\kappa}(u) \Cos_{\kappa}(v)}{\kk}\right)  , \\
 \dv{v}{t} &= b_{2}(t) \Cos_{\kappa}(u) \Sin_{\kappa}(v) +2 b_{3}(t)    \Sin_{\kappa}(u) \Sin_{\kappa}(v) .
 \end{split}
 \label{eq:SCRiccati:Curv:system}
 \end{equation}
The vector fields \eqref{eq:SCRiccati:Curv:vf} are Hamiltonian vector fields relative to the symplectic form
 \begin{equation}
 \omega_{\kappa} = - \frac{1}{2 \Sin_{\kappa}^{2}(v)}\, \dd u \wedge \dd v ,
 \label{eq:SCRiccati:Curv:symplectic}
 \end{equation}
and their associated Hamiltonian functions, fulfilling (\ref{condition}), read
 \begin{equation}
 h_{\k, 1} = \frac{1}{2 \Tan_{\kappa}(v)}, \qquad h_{\kappa, 2} = \frac{\Sin_{\kappa}(u)}{2 \Sin_{\kappa}(v)}, \qquad h_{\kappa, 3} = \frac{ \Cos_{\kappa}(v) -\Cos_{\kappa}(u) }{\kk \Sin_{\kappa}(v)} .
 \label{eq:SCRiccati:Curv:Ham}
 \end{equation}
They span a LH algebra $\cH_{\omega_{\kappa}} \simeq \so(2, 1)$  with respect to the Poisson bracket $\set{\cdot, \cdot}_{\omega_{\kappa}}$ induced by the symplectic form \eqref{eq:SCRiccati:Curv:symplectic}, as their brackets are the same as in \eqref{eq:I4_Curv:brackets}.

The vector fields, the symplectic structure and the Hamiltonian functions  of the split-complex Riccati equation \eqref{eq:SCRiccati:Euc:system}  on $\E^{2}$ are recovered under the limit $\kappa \to 0$ from \eqref{eq:SCRiccati:Curv:vf}, \eqref{eq:SCRiccati:Curv:symplectic} and \eqref{eq:SCRiccati:Curv:Ham}, respectively. We thus call the system \eqref{eq:SCRiccati:Curv:system} the \textit{curved split-complex Riccati equation}.


 \subsection{A curved diffusion Riccati system}
   \label{s42}

  Let us consider the so-called {\em planar diffusion Riccati system} on  $\E^{2}$ introduced in~\cite{Suazo2011} and \cite{Suazo2014} (see \cite[Subsection~6.4]{Blasco2015} and set $c_0=1$):
 \begin{equation}
 \dv{u}{t} = - b(t) + 2c(t) u + 4a(t) u^{2} + a(t) v^{4}, \qquad \dv{v}{t} = c(t)v + 4a(t)u v,
 \label{eq:PDRiccati:Euc:system_c0}
 \end{equation}
 where $a(t), b(t)$ and $c(t)$ are arbitrary $t$-dependent functions.   This system is associated with the $t$-dependent vector field
\be
 \X = - 2 b(t) \X_{1} + 2 c(t) \X_{2}+ 2a(t) \X_{3} ,
 \label{ca}
 \ee
 where the vector fields
 \begin{equation}
  \X_{1} =\frac 12 \pdv{u}, \qquad \X_{2} =  u \pdv{u} + \frac 12v \pdv{v}, \qquad \X_{3} =\frac 12 (4u^{2} + v^{4}) \pdv{u} + 2uv \pdv{v}  ,
 \label{eq:PDRiccati:Euc:vf}\nonumber
 \end{equation}
 verifying the  commutation rules (\ref{sl2}),
 are Hamiltonian vector fields  with respect  to the symplectic form given by~\cite{Blasco2015}
\begin{equation}
\omega=-\frac{2 }{   v^3}  \, {\rm d} u \wedge {\rm d} v, \nonumber
\end{equation}
with associated Hamiltonian functions
\begin{equation}
 h_1=\frac{1}{2 v^2} ,\qquad
h_2=  \frac{u}{ v^2}  ,\qquad
h_3=2 \frac{u^2}{v^2}-\frac 12 v^2. \nonumber
\end{equation}
The (Hamiltonian) vector fields of the Euclidean I$_{4}$-class in the form shown in Table~\ref{table1} are reproduced after  the change of variables defined by
 \begin{equation}
 x:= 2 u + v^{2}, \qquad y:= 2 u - v^{2}, \qquad u = \frac{1}{4}(x + y), \qquad v = \frac{\pm 1}{\sqrt{2}}\, \sqrt{|x-y|} ,
 \label{eq:PDRiccati:change}\nonumber
 \end{equation}
with domains  ${\R}^2_{x\ne y} $  and ${\R}^2_{v\ne 0}$.

Then, we apply the above transformation on $\S^{1}_{[\kappa]} \times \S^{1}_{[\kappa]}$ to the curved I$_{4}$-LH class, giving rise to the curved $t$-dependent vector field counterpart $\X_{\kappa} $ of (\ref{ca}),   where the vector fields $\X_{\kappa,i}$ read as
 \begin{equation}
 \begin{split}
  & \X_{\kappa,1} =\frac 12  \pdv{u},\qquad
  \X_{\kappa,2} = \frac{1}{2}  \Cos_{\kappa}(  v^{2})   \Sin_{\kappa}(2 u  )  \pdv{u} + \frac{1}{2v}  \Cos_{\kappa}(2 u  )  \Sin_{\kappa}( v^{2})   \pdv{v}, \\[2pt]
  & \X_{\kappa, 3} =  \left(\frac{1-\Cos_{\kappa}(2 u  )  \Cos_{\kappa}(  v^{2}) }{\kk}  \right) \pdv{u} + \frac{1}{v}  \Sin_{\kappa}(2 u  )  \Sin_{\kappa}( v^{2} ) \pdv{y} ,
 \end{split}
 \label{eq:PDRiccati:Curv:vf}
 \end{equation}
closing onto the commutation relations (\ref{eq:I4_Curv:comm}). The first-order system of ODEs on $\S^{1}_{[\kappa]} \times \S^{1}_{[\kappa]}$  associated with $\X_{\kappa} $  is given by
  \begin{equation}
 \begin{split}
 \dv{u}{t} &= - b(t) +  {c(t)}  \Cos_{\kappa}(  v^{2})   \Sin_{\kappa}(2 u  ) +2 a(t)\left(\frac{1-\Cos_{\kappa}(2 u  )  \Cos_{\kappa}(  v^{2}) }{\kk}  \right) ,\\
 \dv{v}{t} &=c(t) \frac{1}{v}   \Cos_{\kappa}(2 u  )  \Sin_{\kappa}( v^{2}) + 2 a(t)   \frac{1}{v}  \Sin_{\kappa}(2 u  )  \Sin_{\kappa}( v^{2} ) .
 \end{split}
 \label{eq:PDRiccati:Curv:system}
 \end{equation}
 The vector fields \eqref{eq:PDRiccati:Curv:vf} turn out to be Hamiltonian vector fields relative to the symplectic form
  \begin{equation}
  \omega_{\kappa} = - \frac{2 v}{\Sin^2_{\kappa}(v^{2})}\, \dd u \wedge \dd v .
  \label{eq:PDRiccati:Curv:symplectic}
  \end{equation}
 Their associated Hamiltonian functions read
  \begin{equation}
  h_{\kappa, 1} = \frac{1}{2\Tan_{\kappa}(v^{2})}, \qquad h_{\kappa, 2} = \frac{\Sin_{\kappa}(2u)}{2 \Sin_{\kappa}(v^{2})}, \qquad h_{\kappa, 3} = \frac{  \Cos_{\kappa}(  v^{2}) -\Cos_{\kappa}(2 u  ) }{\kk\Sin_{\kappa}(v^{2})},
  \label{eq:PDRiccati:Curv:Ham}
  \end{equation}
and  fulfil  the Poisson brackets (\ref{eq:I4_Curv:brackets}). The vector fields,  symplectic form and  Hamiltonian functions of the planar diffusion Riccati system  on $\E^{2}$ \eqref{eq:PDRiccati:Euc:system_c0} are recovered from \eqref{eq:PDRiccati:Curv:vf}, \eqref{eq:PDRiccati:Curv:symplectic} and \eqref{eq:PDRiccati:Curv:Ham} after the contraction $\kappa \to 0$. We therefore call the system \eqref{eq:PDRiccati:Curv:system} the \textit{curved Riccati diffusion system}.


 \subsection{Curved Kummer--Schwarz equation for $c <0$}
\label{s43}

We now consider the second-order Kummer--Schwarz equation on $\R$ studied in~\cite{Ballesteros2013,Ballesteros2015,Blasco2015,Berkovich2007,Carinena2012,deLucas2013}:
 \[ \dv[2]{u}{t} = \frac{3}{2u} \left( \dv{u}{t} \right)^{2} - 2 c\, u^{3} + 2 \eta(t) u, \]
 where $\eta(t)$ is an arbitrary $t$-dependent function and $c$ is a constant. This  equation is equivalent to the following first-order system of ODEs on the Euclidean plane $\E^{2}$
   \begin{equation}
\dv{u}{t} = v, \qquad \dv{v}{t} = \frac{3}{2} \frac{v^{2}}{u} - 2c\, u^{3} + 2\eta(t)u,
 \label{eq:KS:Euc:system}
 \end{equation}
which is related to the $t$-dependent vector field
 \begin{equation}
 \X  = \X_{3} + \eta(t) \X_{1},
 \label{eq:KS:Euc:tvf}
 \end{equation}
  where the   vector fields  given by
  \begin{equation}
  \X_{1} = 2 u \pdv{v}, \qquad \X_{2} = u \pdv{u} + 2v \pdv{v}, \qquad \X_{3} =  v \pdv{u} + \left( \frac{3}{2} \frac{v^{2}}{u} - 2 c\, u^{3} \right) \pdv{v}
  \label{eq:KS:Euc:vf}
  \end{equation}
  verify the  commutation relations (\ref{sl2}). Whenever $c<0$, we are dealing with the
  Euclidean I$_{4}$-class. Let us denote the constant $c$ as
  \be
  c=-\frac 1{ 4 \lambda^{2}} ,\qquad \lambda\in\R- \set{0} .\nonumber
  \ee
The vector fields  \eqref{eq:KS:Euc:vf} are Hamiltonian vector fields relative to the symplectic form
 \be
 \omega= \frac{\lambda }{2   u^3} \, {\rm d} u \wedge {\rm d} v ,\nonumber
 \ee
 with associated Hamiltonian functions
 \begin{equation}
  h_1=\frac{\lambda}{u} ,\qquad
h_2=  \frac{\lambda v}{ 2 u^2}  ,\qquad
h_3=- \frac{u}{4 \lambda}+\frac {\lambda v^2}{4 u^3}.\nonumber
 \end{equation}
  The (Hamiltonian) vector fields  are mapped  into those encompassing the Euclidean I$_{4}$-class (see Table~\ref{table1}) via the change of variables defined by~\cite{Ballesteros2015}
 \begin{equation}
 x:=\frac{ v  }{2u }+ \frac{u  }{2 \lambda}, \qquad   y:=\frac{  v  }{2u  }- \frac{u   }{2  \lambda}, \qquad   u = \lambda (x-y), \qquad  v = \lambda (x^{2} - y^{2})  ,
 \label{eq:KSI4:change}\nonumber
 \end{equation}
according to the domains  ${\R}^2_{x\ne y} $  and ${\R}^2_{u\ne 0}$.

Then, we apply  the above transformation on $\S^{1}_{[\kappa]} \times \S^{1}_{[\kappa]}$ to the curved I$_{4}$-LH class in order to  use  the results obtained in Section~\ref{s3}. This yields   the curved analogue $ \X_{\kappa}$ of (\ref{eq:KS:Euc:tvf})  where the vector fields
   \begin{equation}
 \begin{split}
   \X_{\kappa,1} &=  2u \pdv{v},\\[2pt]
   \X_{\kappa,2} &= 2 \lambda   \Sin_{\kappa}\!\left( \frac{u}{2\lambda}  \right)\! \Cos_{\kappa} \!\left( \frac{v}{2u} \right)\pdv{u}   + 2   \left\{     u    \Cos_{\kappa}\!\left( \frac{u}{2\lambda}  \right)\! \Sin_{\kappa} \!\left( \frac{v}{2u} \right) +
 \frac {\lambda v}  u    \Sin_{\kappa}\!\left( \frac{u}{2\lambda}  \right)\! \Cos_{\kappa} \!\left( \frac{v}{2u} \right)  \right\}\pdv{v} , \\[2pt]
   \X_{\kappa,3} &=4 \lambda   \Sin_{\kappa}\!\left( \frac{u}{2\lambda}  \right)\! \Sin_{\kappa} \!\left( \frac{v}{2u} \right)\pdv{u} + 4   \left\{  u\left(  \! \frac{1 - \Cos_{\kappa}\!\left( \frac{u}{2\lambda}  \right)\! \Cos_{\kappa} \!\left( \frac{v}{2u} \right) }{\kk}   \! \right)  \! +  \frac{\lambda v}{u}  \Sin_{\kappa}\!\left( \frac{u}{2\lambda}  \right)\! \Sin_{\kappa} \!\left( \frac{v}{2u} \right)\right\}  \!\pdv{v}   ,
 \end{split}
 \label{eq:KSI4:Curv:vf}
 \end{equation}
 have commutation relations \eqref{eq:I4_Curv:comm}. The  first-order system of ODEs on $\S^{1}_{[\kappa]} \times \S^{1}_{[\kappa]}$ associated with $ \X_{\kappa}$ is given by
 \begin{equation}
 \begin{split}
\dv{u}{t} &= 4 \lambda \Sin_{\kappa}\!\left( \frac{u}{2\lambda}  \right)\! \Sin_{\kappa} \!\left( \frac{v}{2u} \right),\\[2pt]
\dv{v}{t} &= 4 u\left( \frac{1 - \Cos_{\kappa}\!\left( \frac{u}{2\lambda}  \right)\! \Cos_{\kappa} \!\left( \frac{v}{2u} \right) }{\kk}\right) + 4\,\frac{ \lambda v}{u}  \Sin_{\kappa}\!\left( \frac{u}{2\lambda}  \right)\! \Sin_{\kappa} \!\left( \frac{v}{2u} \right)+ 2\eta(t) u .
 \end{split}
 \label{eq:KSI4:Curv:system}
 \end{equation}
The vector fields \eqref{eq:KSI4:Curv:vf} are Hamiltonian vector fields relative to the symplectic form
 \begin{equation}
 \omega_{\kappa} = \frac{1}{8 \lambda u \Sin_{\kappa}^{2}\bigl( \frac{u}{2 \lambda} \bigr)}\, \dd u \wedge \dd v ,
 \label{eq:KSI4:Curv:symplectic}
 \end{equation}
 and the Hamiltonian functions associated with the vector fields \eqref{eq:KSI4:Curv:vf} read
  \begin{equation}
  h_{\kappa, 1} = \frac{1}{2 \Tan_{\kappa} (\frac{u}{2 \lambda}  )}, \qquad h_{\kappa,2} = \frac{\Sin_{\kappa} ( \frac{v}{2u} )}{2 \Sin_{\kappa}( \frac{u}{2\lambda}  )}, \qquad h_{\kappa, 3} = \frac{\Cos_{\kappa}( \frac{u}{2\lambda}  ) - \Cos_{\kappa} ( \frac{v}{2u})  }{\kk\Sin_{\kappa}(\frac{u}{2 \lambda}  )} ,
  \label{eq:KSI4:Curv:Ham}
  \end{equation}
  obeying the Poisson brackets (\ref{eq:I4_Curv:brackets}).  The vector fields spanning the VG Lie algebra, symplectic form and  Hamiltonian functions of the Kummer--Schwarz equation for $c <0$ \eqref{eq:KS:Euc:system} on $\E^{2}$ are recovered from \eqref{eq:KSI4:Curv:vf}, \eqref{eq:KSI4:Curv:symplectic} and \eqref{eq:KSI4:Curv:Ham}, respectively, after the contraction $\kappa \to 0$. Therefore, we call the system \eqref{eq:KSI4:Curv:system} the \textit{curved Kummer--Schwarz equations} for $c <0$.


 \subsection{Curved Ermakov equation for $c <0$}
 \label{s44}

 The Ermakov equation \eqref{eq:Er_Euc:system}, outlined in Section~\ref{s1}, with $t$-dependent vector field $ \X $ \eqref{eq:Er_Euc:tvf} and vector fields \eqref{eq:Er_Euc:vf}  belongs to the Euclidean I$_{4}$-LH class whenever $c < 0$ \cite{Ballesteros2015,Blasco2015}. For the computations, we  introduce a non-zero real constant $\lambda$ as
 \be
  c=-\frac {\lambda^{4}}{ 4 } ,\qquad \lambda\in \R - \set{0} .
  \label{lamm}
  \ee
The   change of coordinates on $\E^{2}$ defined by
\begin{equation}
 x:= -\frac{v }{u}+ \frac{ \lambda^{2} }{2u^{2}}, \qquad y:=- \frac{v }{u}- \frac{ \lambda^{2} }{2u^{2}}, \qquad u = \pm\frac{\lambda}{\sqrt{ \abs{x-y}}}, \qquad v = \mp \frac{\lambda(x+y)}{2 \sqrt{ \abs{x-y}}},
\label{eq:ErI4:change}
\end{equation}
  maps the vector fields \eqref{eq:Er_Euc:vf} into the vector fields spanning the Euclidean I$_{4}$-LH class as shown in Table~\ref{table1}.  Hence the domains for both set of coordinates are ${\R}^2_{x\ne y} $  and ${\R}^2_{u\ne 0}$.    Then,  we apply the transformation \eqref{eq:ErI4:change} on $\S^{1}_{[\kappa]} \times \S^{1}_{[\kappa]}$ to the curved I$_{4}$-LH class. This leads to   the $t$-dependent vector field
$
\X_{\kappa} = \X_{\kappa, 3} + \Omega^{2}(t) \X_{\kappa, 1},
$
where the vector fields given by
   \begin{equation}
 \begin{split}
   \X_{\kappa,1} &=  -u \pdv{v},\\
   \X_{\kappa,2} &= - \frac{u^{3}}{\lambda^{2}}\Cos_{\kappa} \! \left( \frac{v}{u} \right)  \!\Sin_{\kappa}  \!\left( \frac{\lambda^{2}}{2u^{2}} \right) \!  \pdv{u} + \left\{   {u} \Sin_{\kappa} \! \left( \frac{v}{u} \right)\!  \Cos_{\kappa}\! \left( \frac{\lambda^{2}}{2u^{2}} \right)- \frac{u^2 v}{\lambda^{2}} \Cos_{\kappa} \! \left( \frac{v}{u} \right) \! \Sin_{\kappa}\! \left( \frac{\lambda^{2}}{2u^{2}} \right)  \right\} \! \pdv{v}, \\
     \X_{\kappa, 3} &=  \frac{2u^{3}}{\lambda^{2}}  \Sin_{\kappa} \!\left(\frac{v}{u} \right) \! \Sin_{\kappa} \!\left(\frac{\lambda^{2}}{2u^{2}} \right)\!  \pdv{u} +\left\{   \frac{2 u^2 v}{\lambda^2} \Sin_{\kappa} \! \left( \frac{v}{u} \right)\!  \Sin_{\kappa}\!\left( \frac{\lambda^{2}}{2u^{2}} \right)-2 u \left(\frac{1- \Cos_{\kappa} \!\left(\frac{v}{u} \right)\!  \Cos_{\kappa} \!\left(\frac{\lambda^{2}}{2u^{2}} \right)}{\kk} \right)   \right\} \! \pdv{v},
 \end{split}
 \label{eq:ErI4:Curv:vf}
 \end{equation}
 satisfy the commutation relations \eqref{eq:I4_Curv:comm}. The first-order system of ODEs on $\S^{1}_{[\kappa]} \times \S^{1}_{[\kappa]}$ associated with  $\X_{\kappa} $ reads
  \begin{equation}
 \begin{split}
\dv{u}{t} &= \frac{2u^{3}}{\lambda^{2}}  \Sin_{\kappa} \!\left(\frac{v}{u} \right) \! \Sin_{\kappa} \!\left(\frac{\lambda^{2}}{2u^{2}} \right), \\
\dv{v}{t} &= \frac{2 u^2 v}{\lambda^2} \Sin_{\kappa} \! \left( \frac{v}{u} \right) \! \Sin_{\kappa}\!\left( \frac{\lambda^{2}}{2u^{2}} \right)-2 u \left(\frac{1- \Cos_{\kappa} \!\left(\frac{v}{u} \right)\! \Cos_{\kappa} \!\left(\frac{\lambda^{2}}{2u^{2}} \right)}{\kk} \right) - \Omega^{2}(t) u.
 \end{split}
 \label{eq:ErI4:Curv:system}
 \end{equation}
 Moreover, the vector fields \eqref{eq:ErI4:Curv:vf} are Hamiltonian vector fields relative to the symplectic form
 \begin{equation}
 \omega_{\kappa} = \frac{\lambda^{4}}{4 u^4 \Sin_{\kappa}^{2}\! \left( \frac{\lambda^{2}}{2u^{2}} \right)}  \, \dd u \wedge \dd v .
 \label{eq:ErI4:Curv:symplectic}
 \end{equation}
Their associated Hamiltonian functions, determined by the condition  (\ref{condition}), turn out to be
 \begin{equation}
 h_{\kappa, 1} = \frac{\lambda^{2}}{4 \Tan_{\kappa} \!\left( \frac{\lambda^{2}}{2u^{2}} \right)}, \qquad h_{\kappa, 2} = -\frac{\lambda^{2} \Sin_{\kappa} \! \left( \frac{v}{u} \right)}{4 \Sin_{\kappa}\!\left( \frac{\lambda^{2}}{2u^{2}} \right)}, \qquad
 h_{\kappa, 3} =\lambda^{2}\,  \frac {    \Cos_{\kappa}\!\left(  \frac{ \lambda^{2}  }{ 2u^{2}  }\right) -
 \Cos_{\kappa}\!\left(   \frac{ v  }{u  } \right)       }     { 2\kk \Sin_{\kappa}\!\left( \frac{\lambda^{2}}{2u^{2}} \right)},
  \label{eq:ErI4:Curv:Ham}
 \end{equation}
 which verify  the commutation relations \eqref{eq:I4_Curv:brackets}.

The vector fields \eqref{eq:Er_Euc:vf}, the symplectic form (\ref{canE}) and the Hamiltonian functions \eqref{eq:Er_Euc:Ham} of the Ermakov equation \eqref{eq:Er_Euc:system} for $c < 0$ on $\E^{2}$ are recovered from \eqref{eq:ErI4:Curv:vf}, \eqref{eq:ErI4:Curv:symplectic} and \eqref{eq:ErI4:Curv:Ham}, respectively, after the contraction $\kappa \to 0$. We therefore call the system \eqref{eq:ErI4:Curv:system} the \textit{curved Ermakov equation} for $c <0$. In addition, such a curved system can alternatively be regarded as the curved Milne--Pinney equation or time-dependent curved Smorodinsky--Winternitz  system on $\S^{1}_{[\kappa]} \times \S^{1}_{[\kappa]}$. As in the previous applications, the corresponding $t$-independent constants of the motion and superposition rules can   be easily derived from the results of Section~\ref{s3}, now via the change of variables (\ref{eq:ErI4:change}). Due to their mathematical and physical relevance,  let us write, for example,   the constants of the motion (\ref{eq:I4_Curv:constants}) for  \eqref{eq:ErI4:Curv:system}:
\be
 F_{\kappa}^{(1)} = - \frac{\lambda^4}{16}= \frac {c}{4},\qquad
  F_{\kappa}^{(2)} =\frac{\lambda^4}{8}\,\frac{  \Cos_{\kappa}\!\left(  \frac{ \lambda^{2}  (u_1^{2} +u_2^{2}   )   }{ 2u_1^{2} u_2^{2}  }\right)   -\Cos_{\kappa}\!\left(  \frac{  u_1 v_2-u_2 v_1   }{  u_1  u_2   }\right)  }{\kk  \Sin_{\kappa}\!\left( \frac{\lambda^{2}}{2u_1^{2}} \right) \!\Sin_{\kappa}\!\left( \frac{\lambda^{2}}{2u_2^{2}} \right)} .\nonumber
 \ee
 The    contraction $\kk\to 0$   leads to the invariant of the Milne--Pinney equation or Smorodinsky--Winternitz  system exactly in the form already obtained in~\cite{Ballesteros2013,LucasPLA}, namely
\be
  F_{0}^{(2)} =\frac 14 (u_1 v_2-u_2 v_1)^2-\frac{\lambda^4}{16} \, \frac{ \bigl(u_1^2+u_2^2 \bigr)^2}{ u_1^2u_2^2}= \frac 14 (u_1 v_2-u_2 v_1)^2+\frac{c}{4} \left(\frac{u_1^2}{u_2^2}+  \frac{u_2^2}{u_1^2} \right)+\frac{c}{2}  ,\nonumber
\ee
 which was the cornerstone for deriving a superposition rule in those works. In our case, this would be recovered from (\ref{a8}) and the one corresponding to the curved system (\ref{eq:ErI4:Curv:system}) would be provided by (\ref{a5}) both through the map (\ref{eq:ErI4:change}).

 Finally, remind, as already commented in Section~\ref{s3} (see eq.~(\ref{aprox})), that any curved LH system can be interpreted as a perturbation on the contraction parameter $\kk$ of the initial Euclidean system.
In particular,  for the curved Ermakov system  (\ref{eq:ErI4:Curv:system}),  we obtain at the first-order in $\kk$ the following  approximation:
 \begin{equation}
 \begin{split}
\dv{u}{t} &=  v +\frac \kk 6 \left( \frac {c\,v}{  u^4} - \frac {v^3}{u^2}\right) + o[\kk^2],\\
\dv{v}{t} &= - \Omega^{2}(t) u +\frac {c}{u^3}+ \frac \kk {12} \left( \frac{c^2}{u^7}-\frac {4 c\, v^2}{u^5}-\frac {v^4}{u^3} \right) + o[\kk^2] ,
 \end{split}
 \label{eq:ErI4:Curv:systemA}
 \end{equation}
to be compared with (\ref{eq:Er_Euc:system}) and where we have introduced the parameter $c<0$ instead of $\lambda$ (\ref{lamm}}).


 \section{The class P$_{2}$ of Lie--Hamilton systems  on curved spaces}
 \label{s5}

 The class P$_{2}$ of LH systems on the Euclidean plane $\E^{2}$ (see Table~\ref{table1}) is spanned by the vector fields
 \begin{equation}
 \X_{1} = \pdv{x}, \qquad \X_{2} = x \pdv{x} + y \pdv{y}, \qquad \X_{3} = (x^{2}- y^{2}) \pdv{x} + 2xy \pdv{y},
 \label{eq:P2_Euc:vf}
 \end{equation}
 which give rise to a subalgebra $\so(2, 1)\simeq \sl(2, \R)$, with commutation rules  (\ref{sl2}),  of the conformal algebra $\conf_{0,+} \simeq \so(3, 1)$ of   $\E^{2}\equiv \S^{2}_{[0], +}$. Therefore, they   have a very concrete geometrical meaning  as   conformal symmetries. From (\ref{eq:Conf:specificB}) with $\kk_2=+1$, it is found that they correspond, in this order, to the generators of translations along  the basic geodesic $l_1$, dilations and specific conformal transformations  related to   $l_1$ on  $\E^{2}$, that is,
  $\X_1= -\P_1$, $\X_2=-\D$ and $\X_3=2\G_1$.

    In order to generalize the Euclidean class P$_{2}$, let us consider the curved counterparts  \eqref{eq:Conf:specific} of the vector fields \eqref{eq:P2_Euc:vf}, thus yielding   the following conformal vector fields on the 2D CK spaces $\S^{2}_{[\kappa_{1}], \kappa_{2}}$:
\begin{equation}
\begin{split}
 \X_{\k, 1} &= \pdv{x}, \qquad \X_{\k, 2} = \frac{\Sin_{\kappa_{1}}(x)}{\Cos_{\kappa_{1} \kappa_{2}}(y)} \pdv{x} + \Cos_{\kappa_{1}}(x) \Sin_{\kappa_{1} \kappa_{2}}(y) \pdv{y}, \\
  \X_{\k, 3} &= \frac{2}{\Cos_{\kappa_{1} \kappa_{2}}(y)}\, \bigl( \Ver_{\kappa_{1}}(x) - \kappa_{2} \Ver_{\kappa_{1} \kappa_{2}}(y) \bigr) \pdv{x} + 2 \Sin_{\kappa_{1}}(x) \Sin_{\kappa_{1} \kappa_{2}}(y) \pdv{y},
\end{split}
\label{eq:P2_Curv:vf}
\end{equation}
 where     we have used the shorthand notation $\k= (\kappa_{1}, \kappa_{2})$ introduced in Section~\ref{s2}. Their commutation relations read  (see (\ref{cCK}))
\begin{equation}
  [\X_{\k, 1}, \X_{\k, 2}] = \X_{\k, 1} - \tfrac{1}{2} \kappa_{1} \X_{\k, 3}, \qquad [\X_{\k,1}, \X_{\k,3}] = 2 \X_{\k, 2}, \qquad [\X_{\k,2}, \X_{\k,3}] = \X_{\k,3}.
 \label{eq:P2_Curv:comm}
 \end{equation}
 The $t$-dependent vector field
 \begin{equation}
 \X_{\k} = b_{1}(t) \X_{\k, 1} + b_{2}(t) \X_{\k, 2} + b_{3}(t) \X_{\k,3} ,
 \label{eq:P2_Curv:tvf}
 \end{equation}
 where $b_{i}(t)$ are arbitrary $t$-dependent functions, is associated with the following first-order system of ODEs on $\S^{2}_{[\kappa_{1}], \kappa_{2}}$:
 \begin{equation}
  \begin{split}
 \dv{x}{t} &= b_{1}(t) + b_{2}(t) \frac{\Sin_{\kappa_{1}}(x)}{\Cos_{\kappa_{1} \kappa_{2}}(y)}  +  {2 b_{3}(t)} \, \frac{  \Cos_{\kappa_{1} \kappa_{2}}(y) - \Cos_{\kappa_{1}}(x)  }{ \kappa_{1}\Cos_{\kappa_{1} \kappa_{2}}(y)}, \\
 \dv{y}{t} &= b_{2}(t)\Cos_{\kappa_{1}}(x) \Sin_{\kappa_{1} \kappa_{2}}(y) + 2 b_{3}(t) \Sin_{\kappa_{1}}(x) \Sin_{\kappa_{1} \kappa_{2}}(y).
 \end{split}
 \label{eq:P2_Curv:system}
 \end{equation}
 Consequently,  $\X_{\k}$ is a Lie system whose VG Lie algebra $V^{X_{\k}}$ spanned by the vector fields \eqref{eq:P2_Curv:vf} is isomorphic to $\so(2, 1)$. In addition, $\X_{\k}$  is also a LH system
since the relation (\ref{LieDer})  provides a  symplectic form
 \begin{equation}
 \omega_{\k} = \frac{\Cos_{\kappa_{1} \kappa_{2}}(y)}{\Sin_{\kappa_{1} \kappa_{2}}^{2}(y)}\, \dd x \wedge \dd y ,
 \label{eq:P2_Curv:symplectic}
 \end{equation}
turning the
 vector fields \eqref{eq:P2_Curv:vf}  into    Hamiltonian vector fields via the condition (\ref{condition}); these are
 \begin{equation}
 \begin{split}
 h_{\k, 1} &= - \frac{1}{\Sin_{\kappa_{1} \kappa_{2}}(y)}, \qquad h_{\k, 2} = - \frac{\Sin_{\kappa_{1}}(x)}{\Tan_{\kappa_{1} \kappa_{2}}(y)}, \\
 h_{\k, 3} &=\frac{2}{\Sin_{\kappa_{1}\kappa_{2}}(y)}\, \bigl( \kappa_{1} \kappa_{2} \Ver_{\kappa_{1}}(x) \Ver_{\kappa_{1} \kappa_{2}}(y) - \kappa_{2} \Ver_{\kappa_{1} \kappa_{2}}(y) - \Ver_{\kappa_{1}}(x) \bigr) \\
  &=      2 \, \frac{  \Cos_{\kappa_{1}}(x) \Cos_{\kappa_{1} \kappa_{2}}(y) -1 }{\kappa_{1}\Sin_{\kappa_{1} \kappa_{2}}(y)}  .
  \end{split}
 \label{eq:P2_Curv:Ham}
 \end{equation}
 Their  Poisson brackets $\set{\cdot, \cdot}_{\omega_{\k}}$ induced by  $\omega_{\k} $ \eqref{eq:P2_Curv:symplectic} read as
  \begin{equation}
 \set{h_{\k,1}, h_{\k, 2}}_{\omega_{\k}} = - h_{\k, 1} + \tfrac{1}{2} \kappa_{1} h_{\k, 3}, \qquad \set{h_{\k, 1}, h_{\k,3}}_{\omega_{\k}} = - 2 h_{\k,2}, \qquad \set{h_{\k, 2}, h_{\k, 3}}_{\omega_{\k}} = - h_{\k, 3},
 \label{eq:P2_Curv:brackets}
 \end{equation}
 showing that $\X_{\k}$ is a LH system with LH algebra $\cH_{\omega_{\k}}  \simeq \so(2, 1)$.

 Summing up, we call the LH system $\X_{\k}$ the \textit{curved P$_{2}$-LH class} as it
   actually constitutes a generalization of the Euclidean    P$_{2}$-LH class to the 2D curved spaces $\S^{2}_{[\kappa_{1}], \kappa_{2}}$, recovering   the Euclidean LH structure, shown in Table~\ref{table1}, from \eqref{eq:P2_Curv:vf}, \eqref{eq:P2_Curv:symplectic} and \eqref{eq:P2_Curv:Ham} under the contraction $\kappa_{1} \to 0$ and setting  $\kappa_{2} = +1$.

  \begin{table}[htbp]
{\footnotesize
 \noindent
\caption{{\small  The curved P$_{2}$-LH class on the nine 2D CK spaces $\S^{2}_{[\kappa_{1}], \kappa_{2}}$ according to the `normalized'  values     $\kappa_{i} \in\{1, 0, -1\}$. For each   space   it is shown, in geodesic parallel coordinates $(x,y)$ \eqref{eq:CK2:coord}, the    metric $\dd s^2_{\k}$ \eqref{eq:CK2:metric_curv}, domain of the variables,  vector fields $\X_{\k,i}$ \eqref{eq:P2_Curv:vf},
Hamiltonian functions $h_{\k,i}$ \eqref{eq:P2_Curv:Ham}, and   symplectic form $\omega_{\k}$ \eqref{eq:P2_Curv:symplectic}.
 For the sake of clarity, we drop the index $\k =(\kappa_{1},\kappa_{2})$. }}
\label{table3}

\medskip
\!\!\!\!\!\! \!\!\! \!\!\!\!\!\!
\begin{tabular}{lll}
\hline

\hline
\\[-4pt]

$\bullet$ Sphere  $\S^2$& $\bullet$ Euclidean plane $\E^2$
&$\bullet$ Hyperbolic  plane $\H^2$  \\[2pt]

$\S^2_{[+],+}=\mathrm{\mathrm{SO}}(3)/\mathrm{\mathrm{SO}}(2)$&$ \S^2_{[0],+}= \mathrm{ISO}(2)/\mathrm{\mathrm{SO}}(2)$&
$ \S^2_{[-],+}= \mathrm{\mathrm{SO}}(2,1)/\mathrm{\mathrm{SO}}(2)$\\[4pt]

$\dd s^2=\cos^2 \!y \,  \dd x^2+ \dd y^2$&
$\dd s^2= \dd x^2+ \dd y^2$&
$\dd s^2=\cosh^2\! y \, \dd x^2+ \dd y^2$\\[2pt]
$x\in (-\pi,\pi],\  y \in [-\tfrac \pi 2, \tfrac \pi 2] ,\ y\ne 0$&
$x\in \R,\ y\in \R,\ y\ne 0$&
$x\in\R,\ y\in \R,\ y\ne 0$
\\[4pt]

$  \X_1={\partial}_x  $ & $\X_1={\partial}_x   $ & $\X_1={\partial}_x   $\\[4pt]
$\X_{2}=\displaystyle{ \frac{\sin x }{ \cos y} }\,{\partial_x} + \cos x   \sin y \,\partial_{y} $ & $\X_{2} = x \partial_{x} + y {\partial_y} $ & $\X_{2}= \displaystyle{ \frac{\sinh x }{ \cosh y} }\, \partial_{x} + \cosh x   \sinh y \, \partial_{y} $\\[8pt]
$\X_{3}= 2\,\displaystyle{  \frac{ \cos y- \cos x}{ \cos y} } \, \partial_{x} + 2 \sin x   \sin y \, \partial_{y}  $ & $\X_{3} = (x^{2} - y^{2})   \partial_{x} + 2xy  \partial_{y}$ & $\X_{3}=  2\,\displaystyle{  \frac{  \cosh x-\cosh y}{ \cosh y} } \,    \partial_{x} + 2 \sinh x  \sinh y \, \partial_{y}    $
\\[4pt]

 $\displaystyle  h_{1}= -\frac 1{ \sin y }\qquad h_{2}= -\frac{ \sin x}{\tan y}$ & $  h_{1}=  \displaystyle - \frac{1}{y} \qquad h_{2}= -  \frac{x}{y} $ & $\displaystyle  h_{1}= -\frac 1{ \sinh y} \qquad h_{2}=-  \frac{ \sinh x}{\tanh y}$\\[8pt]
  $\displaystyle h_{3}=   2\,\frac{  \cos x  \cos y -1 }{ \sin y} $ & $\displaystyle  h_{3} =- \frac{x^{2} + y^{2}}{y}$ & $\displaystyle  h_{3} =  2\,\frac{ 1 - \cosh x   \cosh y}{\sinh y}$ \\[8pt]
 $\displaystyle  \omega =\frac{\cos y}{\sin^2 \! y}   \, \dd x \wedge \dd y$&
$ \displaystyle \omega = \frac{1}{y^{2}}\, \dd x \wedge \dd y $&
$\displaystyle \omega =\frac{\cosh y}{\sinh^2\! y} \, \dd x \wedge \dd y$\\[14pt]

$\bullet$ Oscillating NH  space ${\bf NH}_+^{1+1}$ & $\bullet$ Galilean plane ${\bf
G}^{1+1}$
&$\bullet$ Expanding NH  space ${\bf NH}_-^{1+1}$   \\[2pt]

$\S^2_{[+],0}=\mathrm{ISO}(2)/ \R$&$ \S^2_{[0],0}=\mathrm{IISO}(1)/\R$&
$\S^2_{[-],0}=\mathrm{ISO}(1,1)/\R$\\[4pt]
$ \dd s^2= \dd x^2   $&
$  \dd s^2= \dd  x^2  $&
$   \dd s^2= \dd x^2   $\\[2pt]
$x\in (-\pi,\pi],\  y\in \R,\ y\ne 0$&
$x\in \R,\ y\in \R,\ y\ne 0$&
$x\in \R,\ y\in \R,\ y\ne 0$
\\[4pt]

 $  \X_1={\partial}_x  $ & $  \X_1={\partial}_x   $ & $ \X_1={\partial}_x   $\\[4pt]
$ \X_{2}=  \sin x \, \partial_{x} + y   \cos x \, \partial_{y}$ & $\X_{2} = x \partial_{x} + y \partial_{y}  $ & $\X_{2}=  \sinh x \, \partial_{x} + y  \cosh x \, \partial_{y}$\\[4pt]
$\X_{3}= 2( 1 - \cos x) \partial_{x} + 2 y \sin x \, \partial_{y}$ & $\X_{3} =    x^{2} \partial_{x} + 2xy \partial_{y}$ & $ \X_{3}=  2( \cosh x - 1) \partial_{x} + 2 y \sinh x \, \partial_{y}$
\\[4pt]

 $ \displaystyle h_{1}=  - \frac{1}{y} \qquad h_{2}=- \frac{\sin x}{y} $ & $ \displaystyle  h_{1}=  - \frac{1}{y}  \qquad h_{2}=-  \frac{x}{y} $ &  $ \displaystyle  h_{1}=  - \frac{1}{y} \qquad h_{2}=- \frac{\sinh x}{y} $\\[8pt]
    $ \displaystyle h_{3}= 2\, \frac{  \cos x-1}{y}   $ & $ \displaystyle h_{3} =- \frac{x^{2}}{y}$ & $ \displaystyle h_{3}= 2\, \frac{ 1 - \cosh x}{y}$ \\[8pt]
     $ \displaystyle \omega =\frac{1}{y^{2}}  \, \dd x \wedge \dd y $&
$\displaystyle \omega = \frac{1}{y^{2}}  \, \dd x \wedge \dd y $ &
 $\displaystyle \omega = \frac{1}{y^{2}}  \, \dd x \wedge \dd y $ \\[14pt]

$\bullet$ Anti-de Sitter space ${\AdS}^{1+1}$ & $\bullet$ Minkowskian plane ${\M}^{1+1}$
&$\bullet$ De Sitter  space ${\dS}^{1+1}$  \\[2pt]

$\S^2_{[+],-}=\mathrm{SO}(2,1)/\mathrm{SO}(1,1)$&$\S^2_{[0],-}= \mathrm{ISO}(1,1)/\mathrm{SO}(1,1)$&
$\S^2_{[-],-}=\mathrm{SO}(2,1)/\mathrm{SO}(1,1)$\\[4pt]
$\dd s^2=\cosh^2\!y  \,  \dd x^2- \dd y^2$&
$\dd s^2= \dd x^2- \dd y^2$&
$\dd s^2=\cos^2 \! y \,\dd x^2-\dd y^2$\\[2pt]
$x\in (-\pi,\pi],\  y\in \R,\ y\ne 0$&
$x\in\R,\ y\in \R,\ y\ne 0$&
$x\in\R,\ y\in (-\pi,\pi),\ y\ne 0$
\\[4pt]

 $  \X_1={\partial}_x  $ & $\X_1={\partial}_x   $ & $\X_1={\partial}_x   $\\[4pt]
$\displaystyle \X_{2}=\frac{\sin x}{ \cosh y}\, {\partial_x} +\cos x  \sinh y \,{\partial_y} $ & $ \X_{2} = x \partial_{x} + y \partial_{y}$ & $ \displaystyle \X_{2}=\frac{ \sinh x}{ \cos y}\, {\partial_x} +\cosh x \sin y \,{\partial_y} $\\[8pt]
$\displaystyle \X_{3}=2\,\frac{ \cosh y-\cos x}{ \cosh y} \, \partial_{x} + 2 \sin x  \sinh y \, \partial_{y}$ & $ \X_{3}  = (x^{2} +y^{2}) \partial_{x} + 2xy \partial_{y}$ & $ \displaystyle \X_{3}=2\, \frac{\cosh x-\cos y}{\cos y}  \,\partial_{x} + 2 \sinh x  \sin y \, \partial_{y}$
\\[4pt]

 $\displaystyle h_{1}=-\frac 1{\sinh y} \qquad  h_{2}=-\frac{\sin x}{\tanh y}$ & $   \displaystyle h_{1}=  - \frac{1}{y} \qquad h_{2}=-  \frac{x}{y}$ & $ \displaystyle  h_{1}=-\frac 1{\sin y}\qquad h_{2}=-\frac{\sinh x}{\tan  y} $\\[8pt]
 $ \displaystyle h_{3}=2\, \frac{ \cos x  \cosh y-1}{\sinh y }    $ & $ \displaystyle h_{3} =- \frac{x^{2} - y^{2}}{y}$ & $\displaystyle h_{3}= 2\, \frac{1 - \cosh x   \cos y }{\sin y}   $ \\[8pt]
 $\displaystyle \omega =\frac{\cosh y}{\sinh^2 \! y}  \, \dd x \wedge \dd y$&
$ \displaystyle \omega = \frac{1}{y^{2}}\,  \dd x \wedge \dd y$&
$\displaystyle \omega =\frac{\cos y}{\sin^2 \! y} \, \dd x \wedge \dd y$\\[12pt]
\hline

\hline

\end{tabular}\hfill}
\end{table}


  We  explicitly display in Table~\ref{table3} the curved P$_{2}$-LH class on each of the nine 2D CK spaces. In all cases, the VG Lie algebra $V^{X_{\k}}\simeq \so(2, 1)$ \eqref{eq:P2_Curv:comm} and the LH algebra $\cH_{\omega_{\k}}\simeq \so(2, 1)$ \eqref{eq:P2_Curv:brackets}. The domain of the variables $(x,y)$  guarantees a well-defined symplectic form and  Hamiltonian functions.

Recall that the six spaces with $\kk_2\le 0$ can be regarded as $(1+1)$D kinematical spacetimes
considering the relationships (\ref{luz}), where $x$ becomes the time coordinate and $y$ the spatial one.
 Under this interpretation, the vector fields $\X_{\k, i}$  (\ref{eq:P2_Curv:vf})  correspond to the generators of time translations   along the time-like   geodesic $l_1$,   dilations and specific conformal transformations also related to $l_1$ on $\S^{2}_{[\kappa_{1}], \kappa_{2}}$. Therefore,  the resulting LH systems shown in Table~\ref{table3} are of `time-like' type.  In this respect, observe that although the `main' time-like metric \eqref{eq:CK2:metric_curv} is degenerate in the Newtonian spaces with $\kk_2=0$ $(c=\infty)$, with foliation determined by (\ref{metricN}), our results allow the construction of  Newtonian LH systems in a consistent way, as it was already performed considering the isometries on $\S^{2}_{[\kappa_{1}], \kappa_{2}}$  from the Euclidean  P$_{1}$-LH class in~\cite{Tobolski2017}.

Furthermore, we stress that, in this kinematical framework, it would be possible to construct `space-like' LH systems by starting from the generator of spatial translations $\P_2$  along the geodesic $l_2$ instead of $\P_1=\partial_{x}$ (\ref{eq:Conf:specific}). In this case, the well adapted variables would be the geodesic parallel coordinates of type II $(x', y')$ (\ref{eq:CK2:coord}). In particular, for such an analogous `space-like' approach,
 one should consider   the vector fields $\X_{\k, 1} = - \P_{2}$, $\X_{\k, 2} = - \D$ and $\X_{\k, 3} = 2 \G_{2}$, which in terms of $(x', y')$ are given by~\cite{Herranz2002}
\begin{equation}
\begin{split}
&\X_{\k, 1} =  \pdv{y'}, \qquad \X_{\k, 2} = \Sin_{\kappa_{1}}(x') \Cos_{\kappa_{1} \kappa_{2}}(y') \pdv{x'} + \frac{\Sin_{\kappa_{1} \kappa_{2}}(y')}{\Cos_{\kappa_{1}}(x')} \pdv{y'},\\
& \X_{\k, 3} = 2 \kappa_{2} \Sin_{\kappa_{1}}(x') \Sin_{\kappa_{1} \kappa_{2}}(y') \pdv{x'} - \frac{2}{\Cos_{\kappa_{1}}(x')} \,\bigl( \Ver_{\kappa_{1}}(x') - \kappa_{2} \Ver_{\kappa_{1} \kappa_{2}}(y') \bigr) \pdv{y'} ,\nonumber
\end{split}
\end{equation}
fulfilling the following commutation relations (see (\ref{cCK}))
\begin{equation}
  [\X_{\k, 1}, \X_{\k, 2}] = \X_{\k, 1} - \tfrac{1}{2} \kappa_{1} \X_{\k, 3}, \qquad [\X_{\k,1}, \X_{\k,3}] = 2 \kk_2\X_{\k, 2}, \qquad [\X_{\k,2}, \X_{\k,3}] = \X_{\k,3},
 \label{eq:P2_Curv:commB}\nonumber
 \end{equation}
and to be compared with (\ref{eq:P2_Curv:comm}).
 Hence, they now correspond to the generators of spatial translations   along the space-like   geodesic $l_2$ (see Figure~\ref{fig:coord_curved}),   dilations and specific conformal transformations   related to $l_2$ on $\S^{2}_{[\kappa_{1}], \kappa_{2}}$. Clearly, such `space-like' LH systems would be diffeomorphic to the previous  `time-like' ones only for the three Riemannian spaces with $\kk_2>0$, under an appropriate change of variables, but they would provide new LH systems for the six models of spacetimes.

Then, as we have already commented in Section~{\ref{s3}} (see eq.~(\ref{aprox})),
the contraction parameters $\k =(\kappa_{1},\kappa_{2})$ can be seen as `integrable' perturbation parameters from the initial `flat' system. In  the curved P$_{2}$-LH class, the most contracted LH system is just the Galilean one, shown in the center of Table~\ref{table3}, leading to  following system
 \begin{equation}
  \begin{split}
 \dv{x}{t} = b_{1}(t) + b_{2}(t) x +  {  b_{3}(t)}  x^2, \qquad
 \dv{y}{t}  = b_{2}(t ) y+ 2 b_{3}(t)  xy,
 \end{split}
 \label{Galilei}
 \end{equation}
which is formed by a Riccati equation in $x$ coupled with another equation in $y$.   From it we can express  the curved P$_{2}$-LH class (\ref{eq:P2_Curv:system}) as a perturbation at the first-order in both $(\kappa_{1},\kappa_{2})$, namely,
 \begin{equation}
  \begin{split}
 \dv{x}{t} &=   b_{1}(t) + b_{2}(t) x +  {  b_{3}(t)}  x^2 -\frac 1{12}\kk_1\bigl(2 b_2(t) x^3 + b_3(t) x^4 \bigr)- \kk_2 b_3(t) y^2+ o[\kk_1^2,\kk_2^2,\kk_1\kk_2], \\[2pt]
 \dv{y}{t} &=  b_{2}(t ) y+ 2 b_{3}(t)  xy-\frac 16 \kk_1\bigl(3 b_2(t) x^2 y + 2 b_3(t) x^3 y \bigr)  + o[\kk_1^2,\kk_2^2,\kk_1\kk_2] .
 \end{split}
 \label{GalileiPerturbation}
 \end{equation}
Other perturbations can be constructed starting from another contracted LH system, that is, from the Euclidean, Minkowkian, and both Newton--Hooke ones.


\subsection{Constants of  the motion of the curved P$_{2}$-class}
\label{s51}

Analogously to the curved I$_{4}$-class   (see Section~\ref{s31}), we now compute the $t$-independent constants of the motion of the LH system  $\X_{\k}$ (\ref{eq:P2_Curv:system}). Since the LH algebra $\cH_{\omega_{\k}}$  \eqref{eq:P2_Curv:brackets} of the curved P$_{2}$-class and the LH algebra of the curved I$_{4}$-class \eqref{eq:I4_Curv:brackets} formally satisfy the same brackets, provided that $\kk\equiv \kk_1$ (regardless of the value of $\kk_2$), the Casimir \eqref{eq:I4_Curv:Casimir} also holds for $\cH_{\omega_{\k}}$.
 From \eqref{eq:P2_Curv:Ham} we construct the following Hamiltonian functions $( 1 \leq i \leq 3)$:
 \begin{equation}
 \begin{split}
h_{\k, i}^{(1)} &= h_{\k,i}(\x_{1}) \in C^{\infty} \bigl(\S^{2}_{[\kappa_{1}], \kappa_{2}} \bigr), \\
h_{\k, i}^{(2)} &= h_{\k,i}(\x_{1}) + h_{\k, i}(\x_{2})  \in C^{\infty} \bigl(\S^{2}_{[\kappa_{1}], \kappa_{2}} \times \S^{2}_{[\kappa_{1}], \kappa_{2}} \bigr),\\
h_{\k, i}^{(3)} &= h_{\k,i}(\x_{1}) + h_{\k, i}(\x_{2})+ h_{\k, i}(\x_{3})  \in C^{\infty} \bigl(\S^{2}_{[\kappa_{1}], \kappa_{2}} \times \S^{2}_{[\kappa_{1}], \kappa_{2}} \times \S^{2}_{[\kappa_{1}], \kappa_{2}} \bigr),
\end{split}
\label{eq:P2_Curv:Ham_pro}
 \end{equation}
 where $\x_{s} = (x_{s}, y_{s})$ denotes the coordinates in the $s^{\textup{th}}$-copy of $\S^{2}_{[\kappa_{1}], \kappa_{2}}$ within the product manifold. Each set of Hamiltonian  functions  verifies the Poisson brackets \eqref{eq:P2_Curv:brackets} with respect to the Poisson bracket induced by taking the product symplectic structure $\omega_{\k}^{[3]}$ of \eqref{eq:P2_Curv:symplectic} to $\bigl(\S^{2}_{[\kappa_{1}], \kappa_{2}} \bigr)^{3}$:
  \be
 \omega_{\k}^{[3]} = \sum_{s=1}^{3}  \frac{\Cos_{\kappa_{1} \kappa_{2}}(y_s)}{\Sin_{\kappa_{1} \kappa_{2}}^{2}(y_s)} \,\dd x_s\wedge \dd y_s .\nonumber
 \ee

 Now, using the Casimir \eqref{eq:I4_Curv:Casimir}, with $\kk\equiv \kk_1$,  we obtain the following constants of the motion for the diagonal prolongation $\widetilde{\X}_{\k}^{3}$ of $\X_{\k}$ to $\bigl(\S^{2}_{[\kappa_{1}], \kappa_{2}} \bigr)^{3}$:
 \begin{equation}
 \begin{split}
 F_{\k}^{(1)} = C\! \left( h_{\k,1}^{(1)}, h_{\k, 2}^{(1)}, h_{\k,3}^{(1)} \right) &= \kappa_{2}, \\
  F_{\k}^{(2)} = C\! \left( h_{\k,1}^{(2)}, h_{\k, 2}^{(2)}, h_{\k,3}^{(2)} \right)  &= 2 \left( \kappa_{2} + \frac{1- \Cos_{\kappa_{1}}(x_{1} - x_{2}) \Cos_{\kappa_{1} \kappa_{2}}(y_{1}) \Cos_{\kappa_{1} \kappa_{2}}(y_{2})}{\kappa_{1} \Sin_{\kappa_{1} \kappa_{2}}(y_{1}) \Sin_{\kappa_{1} \kappa_{2}}(y_{2})}  \right) .
 \end{split}
 \label{eq:P2_Curv:constants}
 \end{equation}
 We illustrate the CK contraction approach by performing the flat limit $\kk_1\to 0$ and the non-relativistic limit $\kk_2\to 0$ of $ F_{\k}^{(2)}$, yielding the constants of the motion for the LH systems shown in the second column and second row in Table~\ref{table3}:
 \be
 F_{(0,\kk_2)}^{(2)} = \frac{(x_1-x_2)^2 + \kk_2 (y_1+y_2)^2}{y_1 y_2},\qquad
 F_{( \kk_1,0)}^{(2)} =2 \left( \, \frac{ 1 - \Cos_{\kappa_{1}}(x_{1} - x_{2})}{\kk_1 y_1 y_2} \right).
 \label{contractF}
 \ee
Both reduce to the constant of the motion for  the (most contracted) Galilean system with $\k=(0,0)$:
\be
 F_{(0,0)}^{(2)} = \frac{(x_1-x_2)^2}{y_1 y_2}.\nonumber
\ee
Recall that the constant of motion $ F_{(0,\kk_2)}^{(2)} $ leads to the one corresponding to the   Euclidean P$_{2}$-class   for $\kk_2=+1$ obtained in~\cite{Blasco2015}.

Furthermore,  $F_{\k}^{(2)}$ provides two other constants of the motion by means of the permutation $S_{ij}$ of the variables $(x_{i}, y_{i}) \leftrightarrow (x_{j}, y_{j})$:
\begin{equation}
\begin{split}
F_{\k,13}^{(2)}  = S_{13}\!\left(F_{\k}^{(2)}\right) &=   2 \left( \kappa_{2} +  \frac{1- \Cos_{\kappa_{1}}(x_{3} - x_{2}) \Cos_{\kappa_{1} \kappa_{2}}(y_{2}) \Cos_{\kappa_{1} \kappa_{2}}(y_{3})}{\kappa_{1} \Sin_{\kappa_{1} \kappa_{2}}(y_{2}) \Sin_{\kappa_{1} \kappa_{2}}(y_{3})} \right) , \\[2pt]
  F_{\k,23}^{(2)}  = S_{23}\!\left(F_{\k}^{(2)} \right) &=   2 \left( \kappa_{2} +  \frac{1- \Cos_{\kappa_{1}}(x_{1} - x_{3}) \Cos_{\kappa_{1} \kappa_{2}}(y_{1}) \Cos_{\kappa_{1} \kappa_{2}}(y_{3})}{\kappa_{1} \Sin_{\kappa_{1} \kappa_{2}}(y_{1}) \Sin_{\kappa_{1} \kappa_{2}}(y_{3})} \right) .
  \label{eq:P2_Curv:constantsB}
\end{split}
\end{equation}
Finally, the functions $h_{\k, i}^{(3)} $  (\ref{eq:P2_Curv:Ham_pro}) give rise to  another constant of the motion that can be written in terms of (\ref{eq:P2_Curv:constants}) and (\ref{eq:P2_Curv:constantsB}) in the form
\be
F_{\k}^{(3)}  = C\! \left( h_{\k,1}^{(3)}, h_{\k, 2}^{(3)}, h_{\k,3}^{(3)} \right)  =F_{\k}^{(2)}+F_{\k,13}^{(2)} + F_{\k,23}^{(2)} - 3 \kk_2 .
\label{P2C}\nonumber
\ee


 \subsection{Superposition rules for the curved P$_{2}$-class}
 \label{s52}

Similarly to the curved I$_{4}$-class, addressed in Section~\ref{s32}, we have deduced
four $t$-independent constants of the motion $\bigl\{F_{\k}^{(2)} ,
F_{\k,13}^{(2)},F_{\k,23}^{(2)},F_{\k}^{(3)} \bigr\}$ for the curved P$_{2}$-LH system $\X_{\k}$ (\ref{eq:P2_Curv:system}), so holding for any choice of $\k$. They allow the derivation of a
superposition rule    for the  diagonal prolongation $\widetilde{\X}_{\k}^{3}$ of $\X_{\k}$ to $\bigl(\S^{2}_{[\kappa_{1}], \kappa_{2}} \bigr)^{3}$. Again, it is necessary to select two functionally independent constants of the motion, $I_1$ and $I_2$,   fulfilling the condition (\ref{a1}). In this case, we can set $F_{\k}^{(2)}=\mu_1$ (\ref{eq:P2_Curv:constants}) and $F_{\k,23}^{(2)}=\mu_2$ (\ref{eq:P2_Curv:constantsB}), and   try to deduce the general solution $(x_{1}(t), y_{1}(t))$ of the LH system in terms of two particular solutions $(x_{2}(t), y_{2}(t))$ and $(x_{3}(t), y_{3}(t))$ along with the constants $\mu_1$ and $\mu_2$.  Nevertheless, the computations are very cumbersome, precluding to express a general superposition rule in an explicit way for  an arbitrary $\k=(\kappa_1,\kappa_2)$. However, for the contracted cases, with some $\kk_i=0$, it is possible to obtain such a rule in an explicit  form.
Therefore, we shall restrict ourselves to present a  superposition principle for the three flat spaces $ \S^{2}_{[0], \kappa_{2}}  $ and for the three non-relativistic ones $ \S^{2}_{[\kk_1], 0}  $, separately.

 Let us consider the flat LH system $\X_{(0,\kk_2)}$ (\ref{eq:P2_Curv:system}) defined on the Euclidean  $ {\bf \E^2}\equiv \S^{2}_{[0], +}$, Galilean ${\bf G}^{1+1}\equiv \S^{2}_{[0], 0}$ and Minkowskian ${\bf
M}^{1+1}\equiv \S^{2}_{[0], -}$ planes (displayed in the second column of Table~\ref{table3}). Under the contraction $\kk_1\to 0$, the symplectic form (\ref{eq:P2_Curv:symplectic}) and Hamiltonian functions (\ref{eq:P2_Curv:Ham}), determining $\X_{(0,\kk_2)}$,  reduce to
\be
\omega_{(0,\kk_2)} = \frac{1}{ y^2}\, \dd x \wedge \dd y ,\quad \  h_{(0,\kk_2), 1} = - \frac{1}{y}, \quad \ h_{(0,\kk_2), 2}  = - \frac{x}{y}, \quad\ h_{(0,\kk_2), 3}  = - \frac{x^2+\kk_2 y^2}{y}.\nonumber
\ee
Then we take the constant of the motion $ F_{(0,\kk_2)}^{(2)} $ (\ref{contractF})  together with its permutations  $F_{(0,\kk_2),13}^{(2)} $,  $F_{(0,\kk_2),23}^{(2)} $, and set
 \be
 \begin{split}
 F_{(0,\kk_2)}^{(2)} &= \frac{(x_1-x_2)^2 + \kk_2 (y_1+y_2)^2}{y_1 y_2}=\mu_1,\quad\  F_{(0,\kk_2),23}^{(2)}= \frac{(x_1-x_3)^2 + \kk_2 (y_1+y_3)^2}{y_1 y_3}=\mu_2,\\
  F_{(0,\kk_2),13}^{(2)}&= \frac{(x_2-x_3)^2 + \kk_2 (y_2+y_3)^2}{y_2 y_3}=\mu_3 .
 \end{split}
 \label{contractF2}
 \ee
From the first equation we find that
\be
x_1^{\pm}(y_1,x_2,y_2,\mu_1)=  x_2\pm \sqrt{\mu_1 y_1 y_2 - \kk_2 (y_1+y_2)^2} .
\label{x1A}
\ee
Substituting this result in the second relation in ({\ref{contractF2}), we obtain
 \be
 \begin{split}
& y_1^{\pm}(x_2,y_2,x_3,y_3,\mu_1,\mu_2)\\
&\qquad =\left\{  y_2y_3\bigl(  (\mu_1\mu_3 +A) y_2+    (\mu_2\mu_3 +A) y_3 \bigr) \pm 2  \sqrt{ y_2^2 y_3^2 B\bigl( \kk_2 (y_2+y_3)^2 - \mu_3 y_2y_3 \bigr) }      \right\}
 \\
  &\qquad\qquad\quad \times  \left\{ \mu_1( \mu_1-4 \kk_2) y_2^2 + \mu_2( \mu_2-4 \kk_2) y_3^2 - 2 (\mu_1\mu_2+ A) y_2 y_3 \right\}^{-1},\\
  A& =  - 2 \kk_2 (\mu_1+\mu_2+\mu_3) +8\kk_2^2,\\
 B&=-\mu_1\mu_2\mu_3 + \kk_2 (\mu_1+\mu_2+\mu_3)^2 - 8 \kk_2^2 (\mu_1+\mu_2+\mu_3)+ 16 \kk_2^3,
 \end{split}
 \label{contractF3}\nonumber
 \ee
  reminding that  $\mu_3=\mu_3(x_2,y_2,x_3,y_3)$ via the third relation in  ({\ref{contractF2}). Finally, introducing this last expression in (\ref{x1A}) we arrive at $x_1^{\pm}(x_2,y_2,x_3,y_3,\mu_1,\mu_2)$, thus  completing the superposition principle for  the flat LH system $\X_{(0,\kk_2)}$.

As a second subfamily within the curved P$_{2}$-LH system $\X_{\k}$ (\ref{eq:P2_Curv:system}), let us focus on the three non-relativistic systems $\X_{(\kk_1,0)}$, so with underlying   oscillating NH  $ {\bf \NH_+^{1+1}}\equiv \S^{2}_{[+], 0}$, Galilean ${\bf G}^{1+1}\equiv \S^{2}_{[0] , 0}$ (again) and  expanding NH ${\bf
NH}_-^{1+1}\equiv \S^{2}_{[-], 0}$ $(1+1)$D spacetimes (displayed in the second row of Table~\ref{table3}). Observe that under the non-relativistic limit   $\kk_2\to 0$ $(c\to \infty)$, the symplectic form (\ref{eq:P2_Curv:symplectic}) and Hamiltonian functions (\ref{eq:P2_Curv:Ham}), defining $\X_{(\kk_1,0)}$, lead to
\be
\omega_{(\kk_1,0)} = \frac{1}{ y^2}\, \dd x \wedge \dd y ,\quad \  h_{(\kk_1,0), 1} = - \frac{1}{y},  \quad \  h_{(\kk_1,0), 2} = - \frac{ \Sin_{\kappa_{1}}(x ) }{y}, \quad \ h_{(\kk_1,0), 3}  = 2\, \frac{ \Cos_{\kappa_{1}}(x ) -1 }{\kk_1y}.\nonumber
\ee
We express the constant of the motion  $F_{( \kk_1,0)}^{(2)} $ (\ref{contractF}) in the form
\be
F_{( \kk_1,0)}^{(2)} = \frac{4\Sin^2_{\kappa_{1}}( \frac 12(x_1-x_2) )  }{y_1y_2} =\mu_1,\nonumber
\ee
so that
\be
y_1(x_1,x_2,y_2,\mu_1)=\frac{4\Sin^2_{\kappa_{1}}( \frac 12(x_1-x_2) ) }{\mu_1 y_2} .
\label{supery1}
\ee
We introduce this result in  $F_{( \kk_1,0),23}^{(2)}  = S_{23}\!\left(F_{( \kk_1,0)}^{(2)}\right)=\mu_2$ and, after some  computations, we get the following equation
 \be
 \mu_1\,\frac{y_2}{y_3}\left( \frac{  \Tan_{\kappa_{1}} \bigl(\frac 12 x_1 \bigr )  \Cos_{\kappa_{1}} \bigl(\frac 12 x_3 \bigr ) - \Sin_{\kappa_{1}} \bigl(\frac 12 x_3 \bigr )  }{\Tan_{\kappa_{1}} \bigl(\frac 12 x_1 \bigr )  \Cos_{\kappa_{1}} \bigl(\frac 12 x_2 \bigr ) - \Sin_{\kappa_{1}} \bigl(\frac 12 x_2 \bigr ) } \right)^2=\mu_2.\nonumber
 \ee
This gives rise to $x_1(x_2,y_2,x_3,y_3,\mu_1,\mu_2)$ in the form
\be
 \Tan_{\kappa_{1} } \bigl(\tfrac 12 x_1^{\pm} \bigr )= \frac{ \sqrt{ \mu_1 }  \Sin_{\kappa_{1}} \bigl(\tfrac 12 x_3 \bigr)  \sqrt{  y_2}  \pm  \sqrt{ \mu_2 }  \Sin_{\kappa_{1}} \bigl(\tfrac 12 x_2 \bigr)   \sqrt{  y_3} }{ \sqrt{ \mu_1 }  \Cos_{\kappa_{1}} \bigl(\tfrac 12 x_3 \bigr)  \sqrt{  y_2}  \pm  \sqrt{ \mu_2 }  \Cos_{\kappa_{1}} \bigl(\tfrac 12 x_2 \bigr)  \sqrt{  y_3}   } ,\nonumber
\ee
which, by substitution in (\ref{supery1}) yields $y_1(x_2,y_2,x_3,y_3,\mu_1,\mu_2)$  completing the full superposition rule for the  non-relativistic LH systems $\X_{(\kk_1,0)}$.

Consequently, we have deduced two equivalent superposition rules  for LH systems based on the
the Galilean ${\bf G}^{1+1}\equiv \S^{2}_{[0] , 0}$ space, both leading to the following  expressions
\be
 \begin{split}
 x_1^{\pm}(x_2,y_2,x_3,y_3,\mu_1,\mu_2)&= \frac{\mu_1 x_3 y_2- \mu_2  x_2  y_3  \pm (x_2 - x_3) \sqrt{\mu_1\mu_2 y_2 y_3} }{\mu_1 y_2 - \mu_2 y_3} ,\\
  y_1^{\mp}(x_2,y_2,x_3,y_3,\mu_1,\mu_2)&= \frac{ (x_2-x_3)^2\bigl(     \mu_1  y_2+ \mu_2   y_3 \mp 2     \sqrt{\mu_1\mu_2 y_2 y_3}\,  \bigr) }{(\mu_1 y_2 - \mu_2 y_3)^2}.\nonumber
  \end{split}
 \ee


 \section{Applications of the curved P$_{2}$-class}
\label{s6}

 The Euclidean P$_{2}$-LH class is known to include the following relevant systems \cite{Blasco2015}: the complex Riccati equation,  the Kummer--Schwarz and Ermakov equations, both for $c>0$ (see (\ref{casconstant}) and Table~\ref{table1}). By making use of the results of the previous section on the curved P$_{2}$-LH class we generalize such Euclidean systems to the nine CK spaces $ \S^{2}_{[\kappa_{1}], \kappa_{2}}  $.


 \subsection{Curved complex Riccati equation}
 \label{s61}

 The  complex Riccati equation on $ {\bf \E^2}\equiv \S^{2}_{[0], +}$ corresponds to the case
    in  (\ref{b1}) when    $z$ is a complex number such that   $\iota^{2}  =-1$. This equation
   has been widely studied from different points of view  (see, e.g.,~\cite{Campos1997,Zoladek2000,Egorov2007,Wilczynski2008,Suazo2011,Ortega2012,Schuch2012} and references therein) and leads to the following   first-order system on  $\E^{2}$:
   \begin{equation}
 \begin{split}
 \dv{x}{t} &= b_{1}(t) + b_{2}(t) x + b_{3}(t) \bigl(x^{2} - y^{2}\bigr), \\
 \dv{y}{t} &= b_{2}(t)y + 2 b_{3}(t) xy,
 \end{split}
 \label{eq:CRiccati:Euc:system}\nonumber
 \end{equation}
 provided that $z=x+{\rm i} y$. This system is just recovered from the general    $t$-dependent vector field     $\X_{\k}$ (\ref{eq:P2_Curv:tvf})  with equations (\ref{eq:P2_Curv:system}) for the particular case with $\kk_1\to 0$ and $\kk_2=+1$. Therefore, we shall call the LH system \eqref{eq:P2_Curv:system} the \textit{curved complex Riccati equation} on $\S^{2}_{[\kk_1], \kk_2}$, being endowed with  the VG Lie algebra \eqref{eq:P2_Curv:comm}, symplectic form \eqref{eq:P2_Curv:symplectic} and LH algebra  \eqref{eq:P2_Curv:brackets}.

 Observe that for the three   flat LH systems $\X_{(0,\kk_2)}$ (\ref{eq:P2_Curv:system}), covering
    $ {\bf \E^2}\equiv \S^{2}_{[0], +}$,   ${\bf G}^{1+1}\equiv \S^{2}_{[0], 0}$ and   ${\bf
M}^{1+1}\equiv \S^{2}_{[0], -}$, it is found that
   \begin{equation}
 \begin{split}
 \dv{x}{t} &= b_{1}(t) + b_{2}(t) x + b_{3}(t) \bigl(x^{2} - \kk_2 y^{2} \bigr), \\
 \dv{y}{t} &= b_{2}(t)y + 2 b_{3}(t) xy,
 \end{split}
 \label{eq:CRiccati:Euc:systemB}
 \end{equation}
whose superpostion rule has been deduced in Section~\ref{s52}. Hence, in the Galilei case with $\kk_2=0$ we recover the system (\ref{Galilei}), formed by  the usual Riccati equation in $x$ coupled with a second equation in $y$.

\medskip
It is worth mentioning that the system \eqref{eq:CRiccati:Euc:systemB} can be solved without using the superposition rule. Indeed, solving the second equation for $x$ leads to
\begin{equation*}
x(t)=\frac{\dv{y}{t} - b_2(t)y}{2b_3(t)y}
\end{equation*}
and a second-order ODE in $y(t)$, which, after the change of variables $y(t)=\xi(t)^{-2}$, adopts the form
\begin{equation}\label{red}
\frac{{\rm d}^2 \xi}{{\rm d}t^2}=\frac{{\rm d}}{{\rm d}t}{\rm log}(|b_3(t)|)\dv{\xi}{t} - A(t) \xi +\frac{b_3(t)^2 \kappa_2}{\xi^3},
\end{equation}
 where
 \begin{equation*}
A(t)=b_1(t)b_3(t)-\frac{1}{4}b_2(t)^2+\frac{1}{2}\dv{b_2(t)}{t}-\frac{b_2(t)\dv{b_3(t)}{t}}{2b_3(t)}.
\end{equation*}
The ODE \eqref{red} admits a 3D Lie point symmetry algebra, incidentally isomorphic to $\mathfrak{sl}(2,\mathbb{R})$, implying that it can be linearized via a nonpoint transformation \cite{Leac,C136}. However, as this transformation also involves the independent variable, it alters the physical significance of the solution.


 \subsection{Curved Kummer--Schwarz equation for $c > 0$}
  \label{s62}

 The Kummer--Schwarz  equation \eqref{eq:KS:Euc:system}~\cite{Berkovich2007,Carinena2012,deLucas2013}, with variables $(u,v)$,   $t$-dependent vector field \eqref{eq:KS:Euc:tvf} and vector fields \eqref{eq:KS:Euc:vf}, belongs to the Euclidean P$_{2}$-LH class whenever the constant $c > 0$~\cite{Ballesteros2013,Ballesteros2015,Blasco2015}, that we shall express as
\be
c=\frac 1{  \lambda^{2}} ,\qquad \lambda\in\R - \{0\}.\nonumber
\ee
  In this case, the change of coordinates on $\E^{2}$ defined by
 \begin{equation}
 x:= \frac{v}{2u}, \qquad y := \frac{u}{\lambda}, \qquad u = \lambda y, \qquad v = 2 \lambda x y ,
 \label{eq:KSP2:change}
 \end{equation}
 maps the vector fields \eqref{eq:KS:Euc:vf} into \eqref{eq:P2_Euc:vf} spanning the Euclidean P$_{2}$-LH class with domains ${\R}^2_{u\ne 0}$     and ${\R}^2_{y\ne 0} $, respectively.  This shows (see Table~\ref{table1}) that the vector fields \eqref{eq:KS:Euc:vf} are Hamiltonian vector fields with respect to the symplectic form
\begin{equation}
\omega = - \frac{\lambda}{2u^{3}}\, \dd u \wedge \dd v, \nonumber
\end{equation}
with associated Hamiltonian functions
\begin{equation}
h_{1} =  - \frac{\lambda}{u}, \qquad h_{2} = -\frac{\lambda v}{2u^{2}}, \qquad h_{3} = - \frac{u}{\lambda} - \frac{\lambda v^{2}}{4u^{3}}. \nonumber
\end{equation}

 In order to generalize the Kummer--Schwarz equation \eqref{eq:KS:Euc:system} for $c > 0$ to the nine CK spaces $\S^{2}_{[\kappa_{1}], \kappa_{2}}$ using the results obtained in Section~\ref{s5}, we apply the transformation \eqref{eq:KSP2:change} to the curved P$_{2}$-class. This  procedure yields   the  $t$-dependent vector field
$ \X_{\k} = \X_{\k, 3} + \eta(t) \X_{\k, 1} $ as the curved counterpart of the Euclidean system
\eqref{eq:KS:Euc:tvf}, where the vector fields
 \begin{equation}
 \begin{split}
 \X_{\k, 1} & = 2 u \pdv{v}, \\[2pt]
 \X_{\k, 2} & = \lambda \Cos_{\kappa_{1}} \!\left( \frac{v}{2u} \right)\! \Sin_{\kappa_{1} \kappa_{2}} \!\left( \frac{u}{\lambda} \right) \pdv{u}  + \left\{ 2 u \frac{\Sin_{\kappa_{1}}\! \left( \frac{v}{2u} \right)}{\Cos_{\kappa_{1} \kappa_{2}}  \!\left( \frac{u}{\lambda} \right) }+ \frac{\lambda v}{u} \Cos_{\kappa_{1}} \! \left( \frac{v}{2u} \right) \! \Sin_{\kappa_{1} \kappa_{2}} \!\left( \frac{u}{\lambda} \right)   \right\} \pdv{v},\\[2pt]
\X_{\k, 3} &=   2 \lambda \Sin_{\kappa_{1}}  \!\left( \frac{v}{2u} \right)  \!\Sin_{\kappa_{1} \kappa_{2}}  \!\left( \frac{u}{\lambda} \right)   \pdv{u} \\[2pt]
&\qquad +\left\{ 4u\left(  \frac{  \Cos_{\kappa_{1} \kappa_{2}}  \!\left( \frac{u}{\lambda} \right)  - \Cos_{\kappa_{1}} \! \left( \frac{v}{2u} \right)  }{\kk_1 \Cos_{\kappa_{1} \kappa_{2}}  \!\left( \frac{u}{\lambda} \right)} \right)+  \frac{2 \lambda v}{u} \Sin_{\kappa_{1}}   \!\left( \frac{v}{2u} \right) \! \Sin_{\kappa_{1} \kappa_{2}}  \!\left( \frac{u}{\lambda} \right)\right\} \pdv{v},
 \end{split}
 \label{eq:KSP2:Curv:vf}
 \end{equation}
 obey the commutation relations \eqref{eq:P2_Curv:comm}. Hence, we obtain the  following first-order system of differential equations on $\S^{2}_{[\kappa_{1}], \kappa_{2}}$, so generalizing (\ref{eq:KS:Euc:system}):
 \begin{equation}
 \begin{split}
 \dv{u}{t} &=   2 \lambda \Sin_{\kappa_{1}}  \!\left( \frac{v}{2u} \right)  \!\Sin_{\kappa_{1} \kappa_{2}}  \!\left( \frac{u}{\lambda} \right),  \\[2pt]
 \dv{v}{t} &= 4u\left(  \frac{  \Cos_{\kappa_{1} \kappa_{2}}  \!\left( \frac{u}{\lambda} \right)  - \Cos_{\kappa_{1}} \! \left( \frac{v}{2u} \right)  }{\kk_1 \Cos_{\kappa_{1} \kappa_{2}}  \!\left( \frac{u}{\lambda} \right)} \right)+  \frac{2 \lambda v}{u} \Sin_{\kappa_{1}}   \!\left( \frac{v}{2u} \right) \! \Sin_{\kappa_{1} \kappa_{2}}  \!\left( \frac{u}{\lambda} \right)+ 2 \eta(t) u .
 \end{split}
 \label{eq:KSP2:Curv:system}
 \end{equation}

 In addition, the vector fields \eqref{eq:KSP2:Curv:vf} are Hamiltonian vector fields relative to the symplectic form
 \begin{equation}
 \omega_{\k} = - \frac{\Cos_{\kappa_{1} \kappa_{2}}   \!\left( \frac{u}{\lambda} \right)}{2 \lambda u \Sin_{\kappa_{1} \kappa_{2}}^{2}   \!\left( \frac{u}{\lambda} \right)}\,  \dd u \wedge \dd v ,
 \label{eq:KSP2:Curv:symplectic}
 \end{equation}
 with associated Hamiltonian functions given by
  \begin{equation}
 \begin{split}
 h_{\k, 1} &= - \frac{1}{\Sin_{\kappa_{1} \kappa_{2}}   \!\left( \frac{u}{\lambda} \right)}, \qquad h_{\k, 2} = - \frac{\Sin_{\kappa_{1}}   \!\left( \frac{v}{2u} \right)}{\Tan_{\kappa_{1} \kappa_{2}}  \! \left( \frac{u}{\lambda} \right)}, \qquad
 h_{\k, 3} = 2\, \frac{\Cos_{\kappa_{1}} \!\left( \frac{v}{2u} \right) \! \Cos_{\kappa_{1} \kappa_{2}}  \!\left( \frac{u}{\lambda} \right)-1}{\kappa_{1}\Sin_{\kappa_{1} \kappa_{2}} \left( \frac{u}{\lambda} \right)}  .
 \end{split}
 \label{eq:KSP2:Curv:Ham}
 \end{equation}
 The corresponding Poisson brackets $\set{\cdot, \cdot}_{\omega_{\k}}$, induced by the symplectic form \eqref{eq:KSP2:Curv:symplectic}, are identical to those in \eqref{eq:P2_Curv:brackets}.  Observe that, as expected, the system (\ref{eq:KSP2:Curv:system}) can also  be obtained from   the Hamilton equations of the time-dependent Hamiltonian $h_{\k} = h_{\k, 3} + \eta(t) h_{\k, 1}$.

The vector fields spanning the corresponding VG Lie algebra,  symplectic form and  Hamiltonian functions  of the Kummer--Schwarz equation \eqref{eq:KS:Euc:system} for $c > 0$ on $\E^{2}$ are recovered from \eqref{eq:KSP2:Curv:vf}, \eqref{eq:KSP2:Curv:symplectic} and \eqref{eq:KSP2:Curv:Ham} after the contraction $\kappa_{1} \to 0$ with $ \kappa_{2} =+1$. Thus, we call the system \eqref{eq:KSP2:Curv:system} the \textit{curved Kummer--Schwarz equation} for $c >0$.


 \subsection{Curved Ermakov equation for $c > 0$}
   \label{s63}

As last application, let us consider the   Ermakov equation \eqref{eq:Er_Euc:system} with positive constant $c > 0$ which corresponds to the Euclidean P$_{2}$-LH class~\cite{Ballesteros2015,Blasco2015}. If we set
\be
c= \lambda^{4},\qquad \lambda\in\R -\{0\},\nonumber
\ee
 it can be shown that  the change of coordinates on $\E^{2}$ given by
\begin{equation}
x:= - \frac{v}{u}, \qquad y := \pm \frac{\lambda^{2}}{u^{2}}, \qquad u = \frac{\lambda}{\sqrt{\abs{y}}}, \qquad v = - \frac{\lambda x}{\sqrt{\abs{y}}},
\label{eq:ErP2:change}
\end{equation}
maps the vector fields \eqref{eq:Er_Euc:vf} into those  determining the Euclidean P$_{2}$-LH class  \eqref{eq:P2_Euc:vf} with domains ${\R}^2_{u\ne 0}$ and  ${\R}^2_{y\ne 0} $, respectively. \!

 Applying again the transformation \eqref{eq:ErP2:change} to the curved P$_{2}$-LH class   constructed in Section~\ref{s5}, we are led to the following curved analogue
 \be
 \X_{\k} = \X_{\k, 3} + \Omega^{2}(t) \X_{\k, 1}
 \label{zzaa}
,
\ee
on  the nine   spaces $\S^{2}_{[\kappa_{1}], \kappa_{2}}$  of the  $t$-dependent Ermakov  vector field   $ \X $ \eqref{eq:Er_Euc:tvf}, where the `curved' vector fields now read
 \begin{equation}
 \begin{split}
 \X_{\k, 1} & = - u \pdv{v}, \\[2pt]
 \X_{\k, 2} & = - \frac{u^{3}}{2 \lambda^{2}} \Cos_{\kappa_{1}}  \!\left( \frac{v}{u} \right) \!  \Sin_{\kappa_{1} \kappa_{2}}  \!\left( \frac{\lambda^{2}}{u^{2}} \right) \pdv{u} + \left\{ \frac{u \Sin_{\kappa_{1}} \!  \left( \frac{v}{u} \right) }{\Cos_{\kappa_{1} \kappa_{2}} \! \left( \frac{\lambda^{2}}{u^{2}} \right) } - \frac{u^{2}v}{2 \lambda^{2}} \Cos_{\kappa_{1}}  \! \left( \frac{v}{u} \right)  \! \Sin_{\kappa_{1} \kappa_{2}}  \!\left( \frac{\lambda^{2}}{u^{2}} \right)\right\} \pdv{v}, \\[2pt]
 \X_{\k, 3} &= \frac{u^{3}}{\lambda^{2}} \Sin_{\kappa_{1}}  \! \left( \frac{v}{u} \right)  \! \Sin_{\kappa_{1} \kappa_{2}}  \! \left( \frac{\lambda^{2}}{u^{2}} \right) \pdv{u} \\[2pt]
 &\qquad +\left\{  2u \left( \frac{\Cos_{\kappa_{1}}  \!\left( \frac{v}{u} \right)- \Cos_{\kappa_{1} \kappa_{2}}  \!\left( \frac{\lambda^{2}}{u^{2}} \right) }{\kk_1\Cos_{\kappa_{1} \kappa_{2}}\! \left( \frac{\lambda^{2}}{u^{2}} \right)}   \right)  +  \frac{u^{2}v}{\lambda^{2}} \Sin_{\kappa_{1}} \! \left( \frac{v}{u} \right) \!\Sin_{\kappa_{1} \kappa_{2}} \!\left( \frac{\lambda^{2}}{u^{2}} \right)  \right\}\pdv{v}  .
 \end{split}
 \label{eq:ErP2:Curv:vf}
 \end{equation}
 These satisfy  the commutation relations \eqref{eq:P2_Curv:comm}. Therefore, we obtain  the following first-order system of differential equations on $\S^{2}_{[\kappa_{1}], \kappa_{2}}$ associated with  $\X_{\k} $ (\ref{zzaa}):
\begin{equation}
 \begin{split}
 \dv{u}{t} &=  \frac{u^{3}}{\lambda^{2}} \Sin_{\kappa_{1}}  \! \left( \frac{v}{u} \right)  \! \Sin_{\kappa_{1} \kappa_{2}}  \! \left( \frac{\lambda^{2}}{u^{2}} \right) , \\[2pt]
 \dv{v}{t} &=2u \left( \frac{\Cos_{\kappa_{1}}  \!\left( \frac{v}{u} \right)- \Cos_{\kappa_{1} \kappa_{2}}  \!\left( \frac{\lambda^{2}}{u^{2}} \right) }{\kk_1\Cos_{\kappa_{1} \kappa_{2}}\! \left( \frac{\lambda^{2}}{u^{2}} \right)}   \right)  +  \frac{u^{2}v}{\lambda^{2}} \Sin_{\kappa_{1}} \! \left( \frac{v}{u} \right) \!\Sin_{\kappa_{1} \kappa_{2}} \!\left( \frac{\lambda^{2}}{u^{2}} \right)   - \Omega^{2}(t) u.
 \end{split}
 \label{eq:ErP2:Curv:system}
 \end{equation}

 Furthermore, the vector fields \eqref{eq:ErP2:Curv:vf} are Hamiltonian vector fields relative to the symplectic form
 \begin{equation}
 \omega_{\k} = \frac{\lambda^{4} \Cos_{\kappa_{1} \kappa_{2}} \! \left( \frac{\lambda^{2}}{u^{2}} \right)}{u^{4}\Sin_{\kappa_{1} \kappa_{2}}^{2} \!  \left( \frac{\lambda^{2}}{u^{2}} \right)}\, \dd u \wedge \dd v,
 \label{eq:ErP2:Curv:symplectic}\nonumber
 \end{equation}
 such that their associated Hamiltonian functions turn out to be
  \begin{equation}
 \begin{split}
 h_{\k, 1} &= \frac{\lambda^{2} }{ 2\Sin_{\kappa_{1} \kappa_{2}} \!  \left( \frac{\lambda^{2}}{u^{2}} \right)}, \qquad h_{\k, 2}  = -\frac{\lambda^{2} \Sin_{\kappa_{1}}  \!\left( \frac{v}{u} \right)}{2 \Tan_{\kappa_{1} \kappa_{2}} \! \left( \frac{\lambda^{2}}{u^{2}} \right)}, \qquad
 h_{\k, 3} =\lambda^{2}   \,  \frac{ 1-\Cos_{\kappa_{1}}  \!\left( \frac{v}{u} \right)  \!\Cos_{\kappa_{1} \kappa_{2}} \!\left( \frac{\lambda^{2}}{u^{2}} \right)}{\kappa_{1}\Sin_{\kappa_{1} \kappa_{2}}  \!\left(  \frac{\lambda^{2}}{u^{2}} \right)}  ,
 \end{split}
 \label{eq:ErP2:Curv:Ham}\nonumber
 \end{equation}
which close on the Poisson brackets   \eqref{eq:P2_Curv:brackets}. The differential equations (\ref{eq:ErP2:Curv:system}) can alternatively be obtained as the Hamilton equations of the time-dependent Hamiltonian  $h_{\k} = h_{\k, 3} + \Omega^2(t) h_{\k, 1}$.

 Finally, since the   Ermakov equation for $c > 0$ on $\E^{2}$  \eqref{eq:Er_Euc:system} is recovered   from the above expressions on
  $\S^{2}_{[\kappa_{1}], \kappa_{2}}$ for the particular case when $\kk_1\to 0$ and $\kk_2=+1$,   we  call the LH system \eqref{eq:ErP2:Curv:system} the \textit{curved Ermakov equation} for $c >0$.


\section{Conclusions and final remarks}
\label{s7}

In this work, the isometry-based formalism proposed in \cite{Tobolski2017} has been completed and expanded from a conformal-based point of view. As observed in Section~\ref{s1}, the curved LH systems obtained in \cite{Tobolski2017} do not include any physically relevant applications, as no remarkable system appart from the complex Bernouilli equation belongs to the Euclidean I$_{2}$-LH class. Nevertheless, the conformal symmetries of the 1D and 2D CK spaces lead to a natural generalization of the Euclidean classes I$_{4}$ and P$_{2}$  to curved spaces: the so-called curved I$_{4}$ and P$_{2}$-LH classes. The former is defined on $\S^{1}_{[\kappa]} \times \S^{1}_{[\kappa]}$, the product of two 1D CK spaces, while the latter is defined on the proper 2D CK spaces $\S^{2}_{[\kappa_{1}], \kappa_{2}}$.

The curved I$_{4}$-LH class, obtained in Section~\ref{s3}, yields   novel LH systems on two different spaces (see Table~\ref{table2}), being the Euclidean I$_{4}$-LH class recovered under the contraction $\kappa \to 0$.
From this, the curved counterparts of  systems belonging to the Euclidean I$_{4}$-LH class arise in a natural way (see Section~\ref{s4}): curved coupled Riccati equations, a curved split-complex Riccati equation, a curved diffusion Riccati system, a curved Kummer--Schwarz equation and a curved Ermakov equation, being the well-known Euclidean systems retrieved under the contraction $\kappa \to 0$. In particular,  the superposition rule for this curved class produces, after contraction $\kappa \to 0$, a  new superposition rule for the Euclidean coupled Riccati equations \eqref{flatriccati}, from which the well-known superposition rule for the Riccati equation is derived after projection. This example suggests the possibility that, although a Lie system on a manifold $M$ is not necessarily a LH one, it may happen that the Lie system on $M \times M$ generated by the sum of the realizations spanning the VG Lie algebra of the system on $M$ is a LH system. In this case, a superposition rule for the system on $M$ can be obtained by projecting a superposition rule for the LH system on $M \times M$, the latter being more feasible to be obtained by means of the coalgebra formalism.  Also, as pointed out in the expressions \eqref{aprox} and \eqref{eq:ErI4:Curv:systemA}, we stress that the parameter $\kappa$ can be considered as an integrable deformation of the initial Euclidean LH system. Following a similar ansatz, the curved P$_{2}$-LH class is presented in Section~\ref{s5}, from which new LH systems on the 2D CK spaces emerge (see Table~\ref{table3}), particularly recovering the Euclidean P$_{2}$-LH systems under the contraction $\kappa_{1} \to 0$, with $\kappa_{2} = +1$. Within this curved LH class,   the curved analogues of a complex Riccati equation, of a Kummer--Schwarz equation and of a Ermakov equation are obtained in Section~\ref{s6}. As for the curved P$_{2}$-LH class is concerned, the graded contraction parameters $\kappa_{1}$ and $\kappa_{2}$ (corresponding now to the curvature and the signature of the metric \eqref{eq:CK2:metric_curv}), can be considered as integrable deformation parameters (see (\ref{GalileiPerturbation})). Although the approximate systems obtained by truncation of the series in $\kappa_i$ are generally not related to a VG algebra, and therefore do not inherit a definite geometrical structure, their solutions may be inferred using either the Lie symmetry method \cite{Ovs} or techniques of perturbation theory \cite{Kum}, taking into account that their solution must provide, in the limit $\kappa_i\rightarrow 0$, the solution of the non-perturbed system.

In this context,  some natural open problems emerge, which are be expected to be (at least partially) solved by applying the same geometrical formalism.
For instance, the higher-dimensional extension of the present work can be carried for the so-called 2$N$D CK spaces \cite{Herranz98}, for which the conformal symmetries can be derived following the same geometrical ansatz proposed in \cite{Herranz2002}. At this point, we must recall that, unlike   the 2D case, not all the 2$N$D CK spaces are symplectic manifolds (the only sphere with a symplectic structure is the 2D one). Thus, it would be interesting to study those 2$N$D CK spaces admitting a symplectic form turning its conformal symmetries into Hamiltonian vector fields, from which new higher-dimensional (curved) LH classes would emerge. It is also worthy to be mentioned that the so-called rank-two CK spaces seem to possess a natural symplectic structure \cite{Herranz98}. Nevertheless, its precise structure has not been found yet.
Finally, it would also be interesting to determine explicit solutions for those curved LH systems based on the book Lie algebra $\mathfrak{b}_{2}$, which constitutes a distingushed subalgebra of $\mathfrak{sl}(2,\mathbb{R})$, as explicit solutions for the $\mathfrak{b}_{2}$-LH systems on the Euclidean plane have been found recently \cite{Campoamor2023}.

Work along these various lines is currently in progress.

\section*{Acknowledgements}

\phantomsection
\addcontentsline{toc}{section}{Acknowledgments}

This work has been supported by Agencia Estatal de Investigaci\'on (Spain)   under  the grant PID2023-148373NB-I00 funded by MCIN/AEI/10.13039/501100011033/FEDER, UE.  F.J.H.~acknowledges support  by the  Q-CAYLE Project  funded by the Regional Government of Castilla y Le\'on (Junta de Castilla y Le\'on, Spain) and by the Spanish Ministry of Science and Innovation (MCIN) through the European Union funds NextGenerationEU (PRTR C17.I1).  O.C. acknowledges a fellowship (grant C15/23) supported by Universidad Complutense de Madrid and Banco   Santander.  The authors also acknowledge the contribution of RED2022-134301-T funded by MCIN/AEI/10.13039/ 501100011033 (Spain).


\section*{Appendix. Fundamental relations of $\kk$-trigonometric functions}
\setcounter{equation}{0}
\renewcommand{\theequation}{A.\arabic{equation}}

\phantomsection
\addcontentsline{toc}{section}{Appendix. Fundamental relations of $\kk$-trigonometric functions}

Consider the  $\kappa$-dependent trigonometric functions: $\kappa$-cosine $\!\Cos_{\kappa} (u)$,  $\kappa$-sine $\!\Sin_{\kappa} (u)$,
  {$\kappa$-tangent} $\!\Tan_{\kappa} (u)$ and   {$\kappa$-versed sine}   $\!\Ver_{\kappa} (u)$ defined in   \eqref{eq:kdep:cos_sin}  and \eqref{eq:kdep:tan_ver}.  Assume that $u$ and $v$ are two arbitrary real arguments. Observe that in Sections~\ref{s5} and \ref{s6}, the label $\kk$    corresponds   to $\kk_1$ or to the product $\kk_1\kk_2$.

Their derivatives   are given by
\begin{equation}
\begin{split}
&\dv{u} \Cos_{\kappa} (u) = - \kappa \Sin_{\kappa} (u), \qquad \dv{u} \Sin_{\kappa}(u) = \Cos_{\kappa} (u), \\
& \dv{u} \Tan_{\kappa}(u) = \frac{1}{\Cos_{\kappa}^{2}(u)}, \qquad\quad \dv{u} \Ver_{\kappa}(u) = \Sin_{\kappa}(u).
 \end{split}
\nonumber
\end{equation}

The main identities for such $\kappa$-functions read  (see \cite{Herranz2000} for details):
\begin{equation}
\begin{split}
& \Cos_{\kappa}^{2}(u) + \kappa \Sin_{\kappa}^{2}(u) = 1. \\
 \end{split}\nonumber
\end{equation}
\begin{equation}
\begin{split}
\Cos_\kappa(2u)&=\Cos^2_\kappa(u)- \kappa\Sin^2_\kappa(u) ,\\
\Sin_\kappa(2u)&=2\Sin_\kappa(u)\Cos_\kappa(u)  ,\\
\Ver_\kappa(2u)&=2\Sin^2_\kappa(u)=4\Ver_\kappa(u)-2\kappa\Ver^2_\kappa(u),\\
\Tan_\kappa(2u)&={{2\Tan_\kappa(u)}\over {1- \kappa\Tan^2_\kappa(u)}}.
 \end{split}\nonumber
\end{equation}
\begin{equation}
\begin{split}
\Cos^2_\kappa\biggl({u\over 2}\biggr)&={{\Cos_\kappa(u)+1}\over 2}
=1-\kappa{\Ver_\kappa(u)\over 2},\\
\Sin^2_\kappa\biggl({u\over 2}\biggr)&={{1-\Cos_\kappa(u)}\over 2\kappa}={1\over 2}\Ver_\kappa(u),\\
\Tan_\kappa\biggl({u\over 2}\biggr)&={{1-\Cos_\kappa(u)}\over \kappa\Sin_\kappa(u)}=
{{\Sin_\kappa(u)}\over {\Cos_\kappa(u)+1}}= {{\Ver_\kappa(u)}\over
{\Sin_\kappa(u)}}.
\end{split}
\nonumber
 \end{equation}
\begin{equation}
\begin{split}
\Cos_\kappa(u)&={{1-\kappa\Tan^2_\kappa\bigl({u\over 2}\bigr)}\over {1+\kappa\Tan^2_\kappa\bigl({u\over
2}\bigr)}},\qquad
\Sin_\kappa(u)={{2\Tan_\kappa\bigl({u\over 2}\bigr)}\over {1+\kappa\Tan^2_\kappa\bigl({u\over
2}\bigr)}},\\
\Ver_\kappa(u)&={{2\Tan^2_\kappa\bigl({u\over 2}\bigr)}\over {1+\kappa\Tan^2_\kappa\bigl({u\over
2}\bigr)}},\qquad
\Tan_\kappa(u)={{2\Tan_\kappa\bigl({u\over 2}\bigr)}\over {1-\kappa\Tan^2_\kappa\bigl({u\over
2}\bigr)}}.
\end{split}
\nonumber
 \end{equation}
\begin{equation}
\begin{split}
\Cos_\kappa(u\pm v)&=\Cos_\kappa(u)\Cos_\kappa(v)\mp \kappa\Sin_\kappa(v)\Sin_\kappa(u),\\
\Sin_\kappa(u\pm v)&=\Sin_\kappa(u)\Cos_\kappa(v)\pm \Sin_\kappa(v)\Cos_\kappa(u),\\
\Ver_\kappa(u\pm v)&=\Ver_{\kappa}(u)+\Ver_{\kappa}(v)-\kappa \Ver_{\kappa}(u)\Ver_{\kappa}(v)\pm
\Sin_\kappa(u)\Sin_\kappa(v),\\
\Tan_\kappa(u\pm v)&={{\Tan_\kappa(u)\pm \Tan_\kappa(v)}\over {1\mp
\kappa\Tan_\kappa(u)\Tan_\kappa(v)}}.
\end{split}
\nonumber
 \end{equation}
\begin{equation}
\begin{split}
\Cos_\kappa(u+v)+\Cos_\kappa(u-v)&=2\Cos_\kappa(u)\Cos_\kappa(v),\\
\Cos_\kappa(u+v)-\Cos_\kappa(u-v)&=-2\kappa\Sin_\kappa(u)\Sin_\kappa(v),\\
\Sin_\kappa(u+v)+\Sin_\kappa(u-v)&=2\Sin_\kappa(u)\Cos_\kappa(v),\\
\Sin_\kappa(u+v)-\Sin_\kappa(u-v)&=2\Cos_\kappa(u)\Sin_\kappa(v),\\
\Ver_\kappa(u+v)+\Ver_\kappa(u-v)&=2\bigl\{\Ver_\kappa(u)+\Ver_\kappa(v)-\kappa\Ver_\kappa(u)\Ver_\kappa(v)\bigr\},\\
\Ver_\kappa(u+v)-\Ver_\kappa(u-v)&=2\Sin_\kappa(u)\Sin_\kappa(v).
\end{split}
\nonumber
 \end{equation}


\end{document}